\documentclass[a4paper,11pt]{article}

\usepackage{jheppub}
\author[a]{Bob Knighton,}
\author[b]{Vit Sriprachyakul,}
\author[c]{Jakub Vo\v{s}mera}
\affiliation[a]{Department of Applied Mathematics \& Theoretical Physics, University of Cambridge,\\
Wilberforce Road, Cambridge CB3 0WA, United Kingdom}
\affiliation[b]{Institut f\"ur Theoretische Physik,
ETH Z\"urich,\\
Wolfgang-Pauli-Stra{\ss}e 27,
8093 Z\"urich, Switzerland}
\affiliation[c]{Institut de Physique Théorique,\\
CNRS, CEA, Université Paris-Saclay, Orme des Merisiers,\\ 
Gif-sur-Yvette, 91191 CEDEX, France}
\emailAdd{rik23@cam.ac.uk}
\emailAdd{vsriprachyak@phys.ethz.ch}
\emailAdd{jakub.vosmera@ipht.fr}

\usepackage{bm} % Bold Greek letters
\usepackage[dvipsnames]{xcolor}
\definecolor{detail}{RGB}{110,110,110}
\usepackage{amsmath} % Mathematics
\usepackage{nccmath} % Mathematics
\usepackage{amssymb} % \mathbb{•}
\usepackage{amsthm} % For proof environment
\usepackage{mathtools} % To be able to define bra, ket and braket
\usepackage[utf8]{inputenc} % For special characters in an Inspire reference
\usepackage{braket}
\usepackage{enumerate}
\usepackage[inline]{enumitem}
\usepackage{comment}
\usepackage{soul} % To strikeout text
\usepackage{tensor}

\usepackage{caption}
\usepackage{subcaption}
\usepackage[mathscr]{euscript}

\usepackage{tikz}

\usetikzlibrary{calc,decorations.markings}
\usetikzlibrary{decorations.pathmorphing}
\usetikzlibrary{shapes,backgrounds}
\usetikzlibrary{fadings}
\usetikzlibrary{cd}
\usetikzlibrary{decorations.pathreplacing,calligraphy}
\usetikzlibrary{arrows, decorations.pathreplacing, 3d}

\newcommand\SmallMatrix[1]{{%
  \tiny\arraycolsep=0.3\arraycolsep\ensuremath{\begin{bmatrix}#1\end{bmatrix}}}}

\tikzset{
partial ellipse/.style args={#1:#2:#3}{
insert path={+ (#1:#3) arc (#1:#2:#3)}
}
}
\newcommand{\p}{\partial}

\usepackage{stackengine}

\def\be{\begin{equation}}
\def\ee{\end{equation}}

\title{Topological defects and tensionless holography}

\begingroup\allowdisplaybreaks
\begin{document}

\abstract{
We study topological defect lines in the symmetric-product orbifolds $\mathrm{Sym}^N(X)$ for a generic seed CFT $X$. We focus on the defects which preserve the maximum of the $S_N$ symmetry. When $X$ is taken to describe the free theory of four fermions and four bosons on a $\mathbb{T}^4$, we construct holographically dual backgrounds describing propagation of tensionless closed strings in the presence of spacetime defects wrapping the conformal boundary. We find a precise match between the spectra of local on-shell closed-string vertex operators in the bulk and the spectra of non-local disorder fields in the spacetime theory.
}

\maketitle

\section{Introduction}

Over the recent years, the duality  \cite{Eberhardt:2018ouy} between the symmetric-product orbifold $\mathrm{Sym}^N(\mathbb{T}^4)$ CFT and the tensionless closed string theory on the $\rm AdS_3\times S^3\times\mathbb{T}^4$ background with $k=1$ unit of pure NS-NS flux has become a very well-studied and carefully understood example of the $\mathrm{AdS}/\mathrm{CFT}$ correspondence \cite{Maldacena:1997re}. This was possible due to the fact that both sides of this particular instance of holographic duality come out as exactly tractable theories \cite{Maldacena:2000hw,Maldacena:2000kv,Maldacena:2001km,Gaberdiel:2018rqv,Giribet:2018ada,Eberhardt:2019qcl,Eberhardt:2019niq,Eberhardt:2019ywk,Dei:2019osr,Eberhardt:2020akk,Eberhardt:2020bgq,Dei:2020zui,Gaberdiel:2020ycd,Knighton:2020kuh,Gaberdiel:2021njm,Gaberdiel:2022als,Gaberdiel:2022oeu,Fiset:2022erp,Dei:2023ivl} (see \cite{Knighton:2023xzg,Demulder:2023bux} for reviews). 
At the same time, this makes it an ideal playground for testing and possibly extending the standard holographic dictionary \cite{Gubser:1998bc,Aharony:1999ti}. Such a paradigm was demonstrated in \cite{Gaberdiel:2021kkp} where, based on exact worldsheet calculations, a duality was proposed between certain species of D-branes in $\mathrm{AdS}_3$ and boundary states in the symmetric-product orbifold theory. It is therefore natural to follow in this vein and try to learn more useful lessons towards understanding various aspects of holography.

The main goal of this paper is to study \emph{topological defect lines} in the symmetric product orbifold CFTs and investigate the effects of their presence on the physics of the tensionless strings propagating in the bulk.  

Topological defect operators provide an indispensable tool for realizing the concept of generalized symmetry in quantum field theory \cite{Gaiotto:2014kfa,Bhardwaj:2017xup,Chang:2018iay,Lin:2019hks,Thorngren:2019iar,Thorngren:2021yso,Komargodski:2020mxz,Choi:2021kmx,Kaidi:2021xfk,Huang:2021zvu,Cordova:2022ieu,Roumpedakis:2022aik,Heckman:2022xgu} (see \cite{Cordova:2022ruw,Schafer-Nameki:2023jdn,Shao:2023gho} for reviews).
Specifically in 2d conformal field theory, these are represented by insertions of codimension-one interfaces into correlators. These can be arbitrarily deformed, as long as they do not cross punctures with operator insertions \cite{Oshikawa:1996dj,Petkova:2000ip,Bachas:2001vj,Recknagel:2002qq,Quella:2006de,Frohlich:2004ef,Frohlich:2006ch,Frohlich:2009gb,Fuchs:2007tx,Brunner:2007qu,Petkova:2013yoa}. The defect operators enact global symmetries (invertible or not) on the CFT Hilbert space. Moving topological interfaces through CFT correlators then implements the conservation laws associated with the symmetries represented by the defect. Performing such moves in the case of non-invertible defects, one is typically left behind with correlators involving defect lines joining insertions of non-local \emph{disorder (or defect-ending) fields} -- the fields on which a defect may terminate -- and defect junctions. This may give rise to Kramers-Wannier-like order-disorder dualities for CFT correlators \cite{Frohlich:2004ef}. 

Symmetric product orbifold theories $\mathrm{Sym}^N(X)$ can be constructed as examples of well-defined two-dimensional conformal field theories given any consistent 2d CFT $X$ which is to be used as a seed. As a consequence, the program of writing down topological defects in symmetric product orbifolds proceeds as in any other 2d CFT. In particular, one has to ensure that any candidate topological defect solves the standard set of bootstrap conditions, starting with the defect Cardy condition at genus one \cite{Petkova:2000ip}. 

On the other hand, the question of making topological defects part of the conventional string perturbation theory is much more delicate. On the positive side, as a topological defect defines a map on the Hilbert space of local bulk CFT operators, it should be clear what is meant by locally wrapping a defect loop around a single isolated puncture with an insertion of a local vertex operator on the worldsheet. However, this stays far from exhausting the possibilities one has for writing down correlators in a defect CFT: these may include complicated defect-line networks involving insertions of non-local defect-ending fields. In particular, the partition functions of the defect-ending fields are not constrained by modular T-invariance, so that the spins of these fields are not restricted to be integer. Such worldsheet insertions would be problematic even in \emph{off-shell} string perturbation theory as the level matching condition is an important ingredient ensuring consistency of interactions in standard formulations of string field theory.\footnote{See \cite{Erbin:2022cyb,Okawa:2022mos} for a recent discussion on the possibility of formulating string field theory without the level matching condition.} Moreover, allowing topological defects to wrap non-trivially around cycles of higher-genus worldsheet correlators would generally infringe on their modular properties and therefore spoil the standard prescription on evaluating amplitudes by integrating over the moduli space. Generic defect configurations would therefore seem hard to make sense of on the worldsheet.

In light of these comments, given the existence of the duality between the symmetric-product orbifold theory $\mathrm{Sym}^N(\mathbb{T}^4)$
and the $k=1$ string theory on $\mathrm{AdS}_3\times \mathrm{S}^3\times\mathbb{T}^4$ which is amenable to an exact worldsheet description, one may hope to learn useful lessons on how to make topological defects come to terms with string perturbation theory by characterizing bulk duals of topological defects in the spacetime CFT. Furthermore, such a program may shed light on realizations of non-invertible symmetries on the worldsheet \cite{Heckman:2024obe} by following the holographic correspondence starting from spacetime generalized symmetries encoded by non-invertible defects in the $\mathrm{Sym}^N(\mathbb{T}^4)$ theory.\footnote{See \cite{Benini:2022hzx,Apruzzi:2022rei,GarciaEtxebarria:2022vzq,Heckman:2022muc,Heckman:2022xgu,Heckman:2022xgu,Antinucci:2022vyk,Yu:2023nyn,Dierigl:2023jdp,Cvetic:2023plv,Bah:2023ymy,Heckman:2024oot,Heckman:2024obe} for recent works discussing topological defects and generalized symmetry in the context of holography and string theory.}

Operators in the twisted sector symmetric product orbifold, even without the presence of defect operators, are inherently non-local, as they have branch cuts causing fundamental fields to become multi-valued. A standard trick in computing correlation functions of such operators is to pass to a covering space on which the fundamental fields in the path integral are single-valued \cite{Arutyunov:1997gt,Lunin:2000yv}. In the large-$N$ limit, correlators of the symmetric orbifold $\text{Sym}^N(X)$ admit a genus expansion in $1/\sqrt{N}$, where the genus is identified with the genus of the covering space contributing at that order. Holographically, this has a very clean interpretation: the covering spaces in the symmetric orbifold are worldsheets in $\text{AdS}_3$ \cite{Pakman:2009zz}.

The dictionary between covering spaces and worldsheets is an extremely powerful one conceptually, since it allows one to identify what a bulk worldsheet calculation should look like, even if one does not have a direct definition of the bulk theory. For example, if the seed theory $X$ is formulated on a manifold with boundary, then the covering spaces arising in the calculation of correlation functions will also have boundaries. This allows one to immediately identify end-of-the-world branes as the holographic dual of certain classes of boundary states in the symmetric orbifold \cite{Gaberdiel:2021kkp}.

Following this logic for the computation of symmetric orbifold correlators with topological defect lines, we are lead to a natural guess, namely that the bulk duals of these defects should be given by a construction involving topological defects on the string worldsheet. From a bulk spacetime perspective, these topological defects should be thought of as brane-like objects extending the spacetime defects into the bulk \cite{GarciaEtxebarria:2022vzq,Heckman:2022muc,Heckman:2022xgu,Dierigl:2023jdp,Cvetic:2023plv}. The worldsheet theory will see a defect line wherever the worldsheet intersects one of these branes. Crucially, these branes are not to be thought of as giving rise to open-string degrees of freedom as was the case in \cite{Gaberdiel:2021kkp}. Instead, we will see that they are to be associated with modified \emph{closed-string} backgrounds.

In this paper, we explore precisely how this duality plays out. After constructing various classes of topological defect lines in the symmetric orbifold CFT, we construct bulk/worldsheet backgrounds which are holographically dual to the topological defects of the spacetime boundary theory. We do so by exhibiting the corresponding modular-invariant worldsheet partition function which arises upon \emph{gauging} a suitable defect on the worldsheet which inherits the global symmetry action from the spacetime defect, in line with the observations of \cite{Harlow:2018tng}. Specifically, we identify the states which are dual to twisted-sector disorder fields in the symmetric product CFT. These states belong to the twisted sectors of the (generalized) orbifold which arises upon gauging the aforementioned worldsheet defect, and live in spectrally-flowed sectors of the worldsheet WZW model. In all cases we consider, we find that the worldsheet theory naturally supports topological defect lines identical to those which are `lifted' to the covering space in the symmetric orbifold. Specializing to the symmetric orbifold of the $\mathbb{T}^4$ sigma model (the case of interest holographically), we find a precise match between the spectrum of disorder fields for an explicit example of maximally-symmetric defect in the $\mathrm{Sym}^N(\mathbb{T}^4)$ theory and the worldsheet partition function constructed by gauging a worldsheet defect involving the corresponding seed $\mathbb{T}^4$ defect as a factor. 
We also comment on a slightly more exotic class of defects in the symmetric orbifold (the `sign' defects), and interpret them from the worldsheet perspective as well.

\vspace{0.5cm}

\noindent This paper is organized as follows. In Section \ref{sec:sym-defects}, we construct topological defects in the symmetric orbifold theory which preserve the $S_N$ symmetry. In Section \ref{sec:worldsheet}, we specialize to the symmetric orbifold of $\mathbb{T}^4$ and construct a dual worldsheet background in $\text{AdS}_3\times\text{S}^3\times\mathbb{T}^4$ whose spectrum precisely matches that of the dual CFT. We end our exposition in Section \ref{sec:discussion} with a brief summary and a list of potential future directions. Finally, we include various appendices reviewing topological defects in 2D CFTs in general, as well as providing various examples in well-known theories.

\vspace{0.25cm}

\noindent\textbf{Note added:} During the completion of this manuscript, the work \cite{Gutperle:2024vyp} appeared which independently covers the topic of defects in symmetric-product orbifold CFTs and their holographic duals.

\section{Topological defects in the symmetric product orbifold}\label{sec:sym-defects}

In this section we will discuss the structure of the symmetric product orbifolds of topological defect CFTs. While we are primarily interested in the case of $\mathbb{T}^4$, we will not insist on fixing a particular seed CFT for much of this discussion. See Appendix \ref{section2} for a short review of topological defect lines in 2D CFTs and Appendices \ref{defects} and \ref{defectsFree} for a number of explicit examples of TDLs in various CFTs.

\subsection{Symmetric product without defects}

First, let us recall the definition of the symmetric orbifold CFT (see \cite{Eberhardt:2019kyt,Roumpedakis:2018tdb} for reviews). Let $X$ be a CFT with central charge $c$. We can construct a new CFT by considering the tensor product of $N$ copies of $X$ with itself, and orbifolding the $S_N$ permutation symmetry. The resulting CFT is the symmetric orbifold of $X$, written as
\begin{equation}
\text{Sym}^N(X)=X^{\otimes N}/S_N\,,
\end{equation}
and has central charge $cN$. 

\subsubsection{Spectrum and partition function}

As in any orbifold CFT, the Hilbert space of $\text{Sym}^N(X)$ decomposes into twisted sectors, labeled by conjugacy classes of $S_N$. Specifically, we write
\begin{equation}\label{eq:sym-hilbert-space}
\mathcal{H}=\bigoplus_{[g]}\text{Inv}_{S_N}\left(\mathcal{H}_{[g]}\right)\,,
\end{equation}
where the sum is over conjugacy classes $[g]$ of the symmetric group. The notation $\text{Inv}_{S_N}$ means that we only keep states which are invariant under the action of $S_N$. Thus, states in the symmetric orbifold fall into twisted sectors labeled by a permutation $[\sigma]$, as well as $r$ representations of the Virasoro algebra with central charge $c$. We will use this fact below in constructing topological defects in $\text{Sym}^N(X)$. 

The partition function of the orbifold theory is found by computing the trace over the spectrum of $\mathcal{H}_{[g]}$ with the insertion of the projection operator $\frac{1}{N!}\sum_{h}h$, which projects onto the space of states invariant under $S_N$, as in \eqref{eq:sym-hilbert-space}. Geometrically, this trace computes the partition function of $X^{\otimes N}$ such that the fields are twisted by the permutation $g$ along the space cycle and by $h$ along the time cycle. This is only a consistent boundary condition if $g$ and $h$ commute, and so the partition function can be reduced to a sum over pairs $h,g$ such that $hg=gh$, that is
\begin{equation}
Z_N(t)=\frac{1}{N!}\sum_{\substack{g,h\in S_N\\ gh=hg}}\text{Tr}_{\mathcal{H}_{[g]}}[h\,q^{L_0-\frac{c}{24}}\bar{q}^{\bar{L}_0-\frac{c}{24}}]\,,\label{eq:ZNt}
\end{equation}
where $q=e^{2\pi i t}$. Here the sum over $g$ runs over the twisted sectors while the sum over $h$ projects onto $S_{N}$ invariant states.
In order to proceed with finding a more explicit expression for the symmetric-product partition function, it is first convenient to express the projected traces over twisted sectors in terms of the seed CFT partition function $Z(t,\bar{t})$ as \cite{Bantay:1997ek,Bantay:1999us,Haehl:2014yla}
\begin{align}
    \text{Tr}_{\mathcal{H}_{[g]}}[h\,q^{L_0-\frac{c}{24}}\bar{q}^{\bar{L}_0-\frac{c}{24}}] = \prod_{\xi\in \mathcal{O}(g,h)} {Z}(t_\xi)\,,\label{eq:TrZ}
\end{align}
where $\mathcal{O}(g,h)$ denotes the set of orbits of the subgroup generated by the two commuting elements $g,h\in S_N$ acting on the set $\{1,2,\ldots,N\}$. Also, the modular parameter $t_\xi$ is defined as
\begin{align}
    t_\xi = \frac{\mu_\xi t+\kappa_\xi}{\lambda_\xi}\label{eq:txi}
\end{align}
where $\mu_\xi$ is the number of $g$-orbits in $\xi$, $\lambda_\xi$ is the common length of the $g$-orbits in $\xi$ and $\kappa_\xi$ is the smallest non-negative integer such that the relation $h^{\mu_\xi} = g^{\kappa_\xi}$ holds on all elements of $\xi$. In particular, it follows that $\kappa_\xi\in \{0,\ldots, \lambda_\xi-1\}$. Substituting the trace \eqref{eq:TrZ} into \eqref{eq:ZNt}, the sum over $g,h\in S_N$ can be reorganized as a sum over integer partitions of $N$ to give
\begin{align}
Z_N(t)=\sum_{\substack{\{m_l\}\\\sum_l m_l =N}}\prod_{l} \frac{1}{m_l!}\, T_{m_l} {Z}(t)  \,,  
\end{align}
where the Hecke operator $T_k$ acts on the seed CFT partition function as
\begin{align}
      T_k Z(t)=\frac{1}{k}\sum_{w|k}\sum_{b=0}^{w-1}  {Z}\bigg(\frac{\frac{k}{w}t+b}{w}\bigg)\,.
\end{align}
Instead of fixing the number $N$ of the copies of the seed CFT, one may decide to work at a fixed fugacity $p$ for $N$. The corresponding grandcanonical partition function 
then exponentiates as
\begin{align}
    {\mathfrak{Z}}(p;t)=1+\sum_{N=1}^\infty p^N Z_N(t) = \exp\bigg(\sum_{k=1}^\infty p^k {{T}}_k Z(t)\bigg)\,.
\end{align}
Finally, in order to exhibit the interpretation of multi-cycle twisted sectors in terms of multiparticle states, let us note that one can rewrite the grandcanonical partition function in the DMVV form\footnote{Here and in the following, we will be suppressing the omnipresent shift $\frac{c}{24 n}(n^2-1)$ in the ground state energy.} \cite{Dijkgraaf:1996xw}
\begin{align}
   \mathfrak{Z}(p;t)&=\prod_{w=1}^\infty  \prod_{\substack{h,\bar{h}\\ h-\bar{h}\in w\mathbb{Z}}}\Big(1-p^{w}  {q}^{\frac{h}{w}}\bar{{q}}^{\frac{\bar{h}}{w}}\Big)^{-{c(h,\bar{h})}}\,,\label{eq:Zsp}
\end{align}
where the positive integers $c(h,\bar{h})$ denote the number of states with dimensions $(h,\bar{h})$ in the spectrum. This can be recognized as a partition function of a Fock space, where each triplet $(h,\bar{h},w)$ contributes $c(h,\bar{h})$ bosonic creation operators.
The single-particle spectrum is then encoded in the partition function \cite{Eberhardt:2018ouy}
\begin{align}
    Z^{\mathrm{s.p.}}(t)=\sum_{w=1}^\infty Z^\prime\bigg(\frac{t}{w}\bigg)\,,
\end{align}
where the $^\prime$ was introduced as a reminder that the projection onto states with $h-\bar{h}\in w\mathbb{Z}$ is in place.

\subsubsection{Fractional Virasoro algebra}

The theory $X^{\otimes N}$ contains $N$ copies $L^{(i)}_n$ of the Virasoro algebra. In the symmetric orbifold $\text{Sym}^N(X)$, the structure of the Virasoro algebra depends on the precise twisted sector one lives in. The conjugacy classes $[\sigma]$ of $S_N$ are labeled by the cycle type $\ell_1,\ldots,\ell_r$ of their permutations. As explained in \cite{Borisov:1997nc,Jevicki:1998bm}, the conjugacy class of cycle type $\ell_1,\ldots,\ell_r$ contains $r$ fractional Virasoro algebras of the form
\begin{equation}
L^{(j)}_{p}\,,\quad p\in\frac{\mathbb{Z}}{\ell_j}\,,\quad j=1,\ldots,r\,.
\end{equation}
These generators satisfy the fractional Virasoro algebra
\begin{equation}
[L_p^{(j)},L_q^{(j)}]=(p-q)L_{p+q}^{(j)}+\frac{c\ell_j}{12}p(p^2-1)\delta_{p+q,0}\,.
\end{equation}
Moreover, the representations of the fractional Virasoro algebra are isomorphic to those of the seed theory Virasoro algebra, since we can make the redefinition
\begin{equation}
\hat{L}_{n}^{(j)}=L^{(j)}_{n/p}+\frac{c(\ell_j^2-1)}{24\ell_j}\delta_{n,0}\,,\quad n\in\mathbb{Z}
\end{equation}
The scaled generators $\hat{L}_n^{(j)}$ satisfy the original, non-fractional, Virasoro algebra with central charge $c$.\footnote{If $X$ admits further symmetries $W_n$, we can also define a fractional $\mathcal{W}$-algebra on $X^{\otimes N}/S_N$, see \cite{Borisov:1997nc}.}

\subsection{Lifting defects to the symmetric orbifold}

Let us consider a topological defect line $\mathcal{L}$ in the seed theory $X$. As explained in detail in Appendix \ref{section2}, we can think of $\mathcal{L}$ as a linear operator $\mathcal{L}:\mathcal{H}\to\mathcal{H}$ which commutes with the Virasoro generators.
Let $i$ be an index labelling the Verma modules of the Virasoro algebra on $X$. Since $\mathcal{L}$ commutes with the Virasoro generators on $X$, we can in general write
\begin{equation}\label{eq:defect-decomp-general}
\mathcal{L}=\sum_{i}D_{i}\Pi_{i}\,,
\end{equation}
where $\Pi_{i}$ projects onto the Verma modules and $D_i$ is the eigenvalue of $\mathcal{L}$ in that module. These operators implement the action of generalized symmetries on the Hilbert space $\mathcal{H}$ of the theory $X$.

\subsubsection{Defect operators}

In lifting the defects to the symmetric orbifold theory, we want to consider linear operators $\boldsymbol{\mathcal{L}}:\mathcal{H}\to\mathcal{H}$ which commute with the (fractional) Virasoro algebra of the orbifold CFT in each twisted sector. As a consequence, we can decompose $\boldsymbol{\mathcal{L}}$ as a sum over projectors on each twisted sector $\mathcal{H}_{[\sigma]}$. Putting everything together, we can decompose $\boldsymbol{\mathcal{L}}$ as
\begin{equation}
\boldsymbol{\mathcal{L}}=\sum_{[\sigma]}\sum_{i_1,\ldots,i_r}D_{i_1,\ldots,i_r,[\sigma]}\Pi_{i_1,\ldots,i_r}\Pi_{[\sigma]}\,,
\end{equation}
where $\Pi_{[\sigma]}$ is a projection onto the twisted sector $\mathcal{H}_{[\sigma]}$, and $\Pi_{i_1,\ldots,i_r}$ projects onto the Verma modules $i_j$ of the Virasoro generators $\smash{\hat{L}^{(j)}_n}$ (and where $r$ is the number of disjoint cycles in $[\sigma]$).

As was done for the boundary states in \cite{Gaberdiel:2021kkp} (see also \cite{Belin:2021nck} for a more general analysis), we will focus exclusively on the case of \textit{maximally-fractional} topological defects. These are the defects which can be thought of as an uplift of a single defect $\mathcal{L}$ of the seed CFT $X$. If $\mathcal{L}$ has a decomposition of the form $\eqref{eq:defect-decomp-general}$, then we can define a defect in the symmetric orbifold as
\begin{equation}\label{eq:maximally-fractional-defect}
\boldsymbol{\mathcal{L}}_\rho=\sum_{[\sigma]}\sum_{i_1,\ldots,i_r}D_{i_1}\cdots D_{i_r}\chi_{\rho}([\sigma])\Pi_{i_1,\ldots,i_r}\Pi_{[\sigma]}\,.
\end{equation}
Here, $\chi_{\rho}$ is the character of some $S_N$ representation $\rho$, which we allow to be generic. Defects of the form \eqref{eq:maximally-fractional-defect} represent a large class of defects (and, correspondingly, generalized symmetries) in $\text{Sym}^N(X)$ which can be constructed from a pair $(\mathcal{L},\rho)$ of a topological defect $\mathcal{L}$ in $X$ and a representation $\rho$ of $S_N$. If one takes $\rho$ to be the trivial representation, the defect \eqref{eq:maximally-fractional-defect} gives a direct uplift of any generalized symmetry of the seed theory $X$ into the symmetric-product orbifold theory $\text{Sym}^N(X)$.

\subsubsection{Fusion rules and invertibility} 
\label{subsec:SymFusion}

Let us consider two topological defects $\mathcal{L}_1,\mathcal{L}_2$ in the seed theory $X$. The fusion of these defects gives a third defect $\mathcal{L}_1\mathcal{L}_2$ with decomposition
\begin{equation}
\mathcal{L}_1\mathcal{L}_2=\sum_{i}D_i(\mathcal{L}_1)D_i(\mathcal{L}_2)\Pi_i\,.
\end{equation}
Now, let us consider two topological defects $\boldsymbol{\mathcal{L}}_1,\boldsymbol{\mathcal{L}_2}$ in the symmetric orbifold. We take $\boldsymbol{\mathcal{L}}_1$ to be constructed from the pair $(\mathcal{L}_1,\rho_1)$, and similarly for $\boldsymbol{\mathcal{L}}_2$. We have
\begin{subequations}
\begin{align}
\boldsymbol{\mathcal{L}}_1\boldsymbol{\mathcal{L}}_2&=\sum_{[\sigma]}\sum_{i_1,\ldots,i_r}D_{i_1}(\mathcal{L}_1)D_{i_1}(\mathcal{L}_2)\cdots D_{i_r}(\mathcal{L}_1)D_{i_r}(\mathcal{L}_2)\chi_{\rho_1}([\sigma])\chi_{\rho_2}([\sigma])\Pi_{i_1,\ldots,i_r}\Pi_{[\sigma]}\\
&=\sum_{[\sigma]}\sum_{i_1,\ldots,i_r}D_{i_1}(\mathcal{L}_1\mathcal{L}_2)\cdots D_{i_r}(\mathcal{L}_1\mathcal{L}_2)\chi_{\rho_1\otimes\rho_2}([\sigma])\Pi_{i_1,\ldots,i_r}\Pi_{[\sigma]}\,.
\end{align}
\end{subequations}
That is, the fusion of maximally-symmetric defects in $\text{Sym}^N(X)$ is inherited from the fusion of the seed defects in $X$ and the representation category of $S_N$.

Given a maximally-symmetric defect $\boldsymbol{\mathcal{L}}_\rho$ in $\text{Sym}^N(X)$, the above discussion tells us that $\boldsymbol{\mathcal{L}}_\rho$ is invertible if and only if both the corresponding seed defect and the representation $\rho$ are invertible. This gives us precisely two cases of invertible maximally-fractional defects in $\text{Sym}^N(X)$:
\begin{itemize}

    \item $\mathcal{L}$ is an invertible defect in $X$ and $\rho$ is the trivial representation. This is the case we will be most interested in below and we will simply denote it by $\boldsymbol{\mathcal{L}}_0$. It defines an uplift of any invertible symmetry of the seed theory $X$ into the symmetric-product orbifold $\text{Sym}^N(X)$.

    \item $\mathcal{L}$ is an invertible defect in $X$ and $\rho$ is the alternating representation $\chi_{\rho}([\sigma])=\text{sgn}(\sigma)$. In this case, we have the symmetric orbifold defect
    \begin{equation}
    \sum_{[\sigma]}\text{sgn}(\sigma)\sum_{i_1,\ldots,i_r}D_{i_1}(\mathcal{L})\cdots D_{i_r}(\mathcal{L})\Pi_{i_1,\ldots,i_r}\Pi_{[\sigma]}\,.
    \end{equation}
    Without loss of generality, we can always take $\mathcal{L}$ to be the trivial defect with $D_{i}=1$, and recover the more general case by fusing with $\boldsymbol{\mathcal{L}}_0$. With this in mind, we define the `sign defect'
    \begin{equation}
    \boldsymbol{\mathcal{L}}_{\text{sign}}=\sum_{[\sigma]}\text{sgn}(\sigma)\Pi_{[\sigma]}\,.
    \end{equation}
    The sign defect simply maps the state $\ket{\psi}^{([\sigma])}$ in the $[\sigma]$-twisted sector to the state ${\mathrm{sgn}(\sigma)}\ket{\psi}^{([\sigma])}$. Blowing up a defect loop $\boldsymbol{\mathcal{L}}_\mathrm{sign}$ inside any $\text{Sym}^N(X)$ $n$-point correlation function on the sphere which involves insertions of single-cycle twist fields with cycle lengths $w_i$ and shrinking it around the punctures, this yields the selection rule \cite{Lunin:2000yv}
    \begin{align}
        \sum_{i=1}^n (w_i +1) \in 2\mathbb{Z}\label{eq:wSelection}
    \end{align}
    in order for the correlator to be non-zero.
\end{itemize}
Also, given a non-invertible duality defect in the seed theory (that is, a defect which fuses with its conjugate to yield a superposition of invertible group-like defects), the corresponding uplift $\boldsymbol{\mathcal{L}}_\rho$ into $\text{Sym}^N(X)$ is a duality defect provided that $\rho$ has invertible fusion rules in $S_N$, i.e. it fuses with the conjugate representation $\rho^\ast$ to a superposition of the trivial and the sign representations.

\subsection{Defect correlators on a torus}

Given a maximally-fractional defect $\boldsymbol{\mathcal{L}}_\rho$, the focus of our interest will now be the Hilbert space $\mathcal{H}(\boldsymbol{\mathcal{L}}_\rho)$ of states in $\text{Sym}^N(X)$ twisted by the defect, i.e.\ the space of defect-ending fields. 
The content of $\mathcal{H}(\boldsymbol{\mathcal{L}}_\rho)$ is encoded in the torus correlation function $(Z_N)_{0,1}(t)$ involving a single insertion of $\boldsymbol{\mathcal{L}}_\rho$ along the temporal cycle of the base torus of the symmetric product orbifold CFT \cite{Petkova:2000ip}. Positive integrality of the multiplicities entering this torus correlator will serve as an important consistency check of the defect lines \eqref{eq:maximally-fractional-defect}. The spectrum of the \emph{non-local} bulk fields encoded by this defect partition function generally differs from the spectrum of local bulk fields dictated by the modular invariant partition function \eqref{eq:ZNt}. In particular, as the defect partition function is not required to be invariant under the modular T-transformation (as explained in Appendix \ref{section2}, it instead has to obey the transformation rules \eqref{eq:modular} under $\mathrm{PSL}(2,\mathbb{Z})$), it allows for fields with non-integer spin.
We will be especially interested into the spectra of \emph{disorder fields} (the fields on which invertible group-like defects can end) as these typically enter statements of order-disorder dualities, which in turn arise as a consequence of moving non-invertible duality defects inside correlators of local bulk fields. More generally, one can be interested in correlators $(Z_N)_{\alpha,\beta}(t)$ involving an instertion of a defect which winds $\alpha$ times around the spatial cycle and $\beta$ times around the temporal cycle.  These are related by modular transformations as the labels $(\alpha,\beta)$ transform as a \emph{modular doublet} under $\mathrm{PSL}(2,\mathbb{Z})$ (see \eqref{eq:modular}).

We start with an insertion of the defect line around the \emph{spatial} cycle of the torus which is determined simply by inserting the defect operator \eqref{eq:maximally-fractional-defect} into the trace over bulk-CFT Hilbert space. Doing this, one obtains
\begin{align}
    (Z_N)_{1,0}(t) = \frac{1}{N!}\sum_{\substack{g,h\in S_N\\ gh=hg}}\chi_\rho([g])\prod_{\xi\in \mathcal{O}(g,h)} {Z}_{\mu_\xi,0 }(t_\xi)\,,\label{eq:ZSNdefClosed}
\end{align}
where $Z_{m,n}(t)$ are the defect torus correlators in the seed theory where the defect $\mathcal{L}$ winds $m$ times around the spatial cycle and $n$ times around the temporal cycle. The modular parameter $t_\xi$ was defined in \eqref{eq:txi}.

\subsubsection{A useful result}

In order to perform the modular S-transformation on \eqref{eq:ZSNdefClosed}, we make use of the following result. Let $k$ be a fixed integer and let $w,b$ be such that $w|k$ and $b\in \mathbb{Z}/w\mathbb{Z}$. Given $w,b$ we can define a transformation $(w,b)\to (\tilde{w},\tilde{b})$ where $\tilde{w}$ and $\tilde{b}$ are given as
\begin{subequations}
\label{eq:abtransf}
    \begin{align}
        \tilde{w}&=\frac{k}{\mathrm{gcd}(w,b)}\,,\\
        \tilde{b}&=\frac{k}{w}\,\gamma(w,b)\,.
    \end{align}
\end{subequations}
Here $\gamma(w,b)$ is the modular inverse of $-\frac{b}{\mathrm{gcd}(w,b)}$ in $\mathbb{Z}/(\frac{w}{\mathrm{gcd}(w,b)}\mathbb{Z})$, that is
\begin{align}
    -\gamma(w,b)\frac{b}{\mathrm{gcd}(w,b)}\equiv 1\ \mathrm{mod}\ \frac{w}{\mathrm{gcd}(w,b)}\,.
\end{align}
It is not difficult to see that $\tilde{w}|k$ and $\tilde{b}\in \mathbb{Z}/\tilde{w}\mathbb{Z}$. Furthermore, let $M$ be the modular transformation which maps
\begin{align}
    \frac{\frac{k}{w}t+b}{w}\ \overset{M}{\longmapsto}\   \frac{\frac{k}{\tilde{w}}\tilde{t}+\tilde{b}}{\tilde{w}}\,,
\end{align}
where $\tilde{t}=-\frac{1}{t}$. Then one can run the Euclidean algorithm to establish that $M$ acts on modular doublets as
\begin{align}
   \big(\tfrac{k}{\tilde{w}}\alpha-\tilde{b}\beta,\tilde{w}\beta\big)\overset{M}{\longmapsto }  \big(-\tfrac{k}{w}\beta-b\alpha,w\alpha\big)
\end{align}
for any $\alpha,\beta$ (that is, by simultaneously mapping $(w,b)\mapsto (\tilde{w},\tilde{b})$ and $(\alpha,\beta)\mapsto (\beta,-\alpha)$).
This result immediately establishes the equality
\begin{align}
  Z_{\tfrac{k}{\tilde{w}}\alpha-\tilde{b}\beta,\tilde{w}\beta}\bigg(\frac{\frac{k}{\tilde{w}}\tilde{t}+\tilde{b}}{\tilde{w}}\bigg)=  Z_{-\tfrac{k}{w}\beta-b\alpha,w\alpha}\bigg( \frac{\frac{k}{w}t+b}{w}\bigg) \label{eq:StransfDef}
\end{align}
between defect torus correlators upon enacting the modular S-transformation $t\mapsto \tilde{t}=-1/t$ on the modular parameter $t$.

\subsubsection{Partition function of defect-ending fields}

It is then not difficult to convince oneself that for a fixed orbit $\xi\in\mathcal{O}(g,h)$, interchanging the roles of the two commuting group elements $g,h\in S_N$ in the definition of the parameters $\lambda_\xi$, $\kappa_\xi$ and $\mu_\xi$ is precisely implemented by the transformation \eqref{eq:abtransf} upon identifying $k=\lambda_\xi\mu_\xi$, $d=\lambda_\xi$ and $b=\kappa_\xi$. Hence, performing the modular S-transformation $t\mapsto -1/t$ (using the result \eqref{eq:StransfDef} specialized to the case $\alpha=1$, $\beta=0$), the partition function of the defect-ending fields reads
\begin{align}
    (Z_N)_{0,1}(t) = \frac{1}{N!}\sum_{\substack{g,h\in S_N\\ gh=hg}}\chi_\rho([h])\prod_{\xi\in \mathcal{O}(g,h)} {Z}_{-\kappa_\xi,\lambda_\xi }(t_\xi)\,.\label{eq:ZSNdef}
\end{align}
Comparing with \eqref{eq:ZNt} and \eqref{eq:TrZ}, we observe that: 1.\ there is an additional factor of $\chi_\rho([h])$ in the sum over $h$ and 2.\ the seed CFT torus correlators are decorated by the winding numbers of the defect which depend on $g$ and $h$. As for the appearance of the character $\chi_\rho([h])$, this means that instead of projecting onto $S_N$-invariant states, the sum over $h$ is now projecting the twisted sectors onto states which transform in the representation $\rho$ of $S_N$. On the other hand, the effect of summing over the winding numbers of the seed theory defect at the level of \eqref{eq:ZSNdef} is less obvious. We will see below that for generic defect lines $\mathcal{L}$, this sum indeed imposes a projection on the tensor product of Hilbert spaces of defect ending fields in the seed CFT, so that \eqref{eq:ZSNdef} is indeed a consistent partition function, as required by the defect Cardy condition. We will make this claim very manifest in the cases where $\mathcal{L}$ enacts an non-anomalous invertible symmetry on the seed CFT Hilbert space.

\subsubsection{Grandcanonical ensemble}

From this point on, let us specialize on the two invertible representations of $S_N$ which appear universally for all $N$ (the trivial representation and the sign representation) so that it makes sense to consider the partition functions of a grandcanonical ensemble of theories. In these cases, it is not difficult to establish that a generic torus correlator where the symmetric product orbifold defect $\boldsymbol{\mathcal{L}}_\rho$ winds $\alpha$ times around the spatial cycle and $\beta$ times around the temporal cycle can be expressed as
\begin{align}
(Z_N)_{\alpha,\beta}(t)=\sum_{\substack{\{m_l\}\\\sum_l m_l =N}}\prod_{l} \frac{1}{m_l!}\, T^{(\alpha,\beta)}_{m_l} {Z}(t)  \,,  
\end{align}
where we have introduced the \emph{defect Hecke operators}
\begin{align}
    T^{(\alpha,\beta)}_{k} {Z}(t) = \frac{1}{k}\epsilon^{k(\alpha+\beta)}\sum_{w|k}\epsilon^{\frac{k}{w}\alpha+w\beta}\sum_{b=0}^{w-1} \epsilon^{b(w+1)\beta} {Z}_{\frac{k}{w}\alpha - b\beta,w\beta}\bigg(\frac{\frac{k}{w}t+b}{w}\bigg)\,.
\end{align}
The parameter $\epsilon$ is set to $+1$ when $\rho$ is the trivial representation and to $-1$ when $\rho$ is the sign representation. As in the case of the pure bulk torus partition function, we therefore obtain that the grandcanonical ensemble
\begin{align}
  {\mathfrak{Z}}_{\alpha,\beta}(p;t)  =1+\sum_{N=1}^\infty p^N (Z_N)_{\alpha,\beta}(t) 
\end{align}
of the defect torus correlators $(Z_N)_{\alpha,\beta}$ can be exponentiated as
\begin{subequations}
\label{eq:GrandCanDef}
\begin{align}
{\mathfrak{Z}}_{\alpha,\beta}(p;t) &= \exp\bigg(\sum_{k=1}^\infty p^k {{T}}^{(\alpha,\beta)}_k Z(t)\bigg)\\
&=\exp\bigg(\sum_{a,w=1}^\infty   \frac{[\epsilon^{(\alpha+\beta)}p]^{aw}}{aw}\epsilon^{a\alpha+w\beta}\sum_{b=0}^{w-1} \epsilon^{b(w+1)\beta} {Z}_{a\alpha - b\beta,w\beta}\bigg(\frac{at+b}{w}\bigg)\bigg)\,.
\end{align}
\end{subequations}
We also note that if the seed theory $X$ is equipped with a spin structure and an extra $\text{U}(1)$ current with fugacity $y$ (for example in \cite{Eberhardt:2020bgq}), then the grandcanonical defect torus correlator will instead read\footnote{Setting, for the sake of clarity, $\epsilon=1$, as it can be reintroduced straightforwardly based on the form of \eqref{eq:GrandCanDef}.}
\begin{align}
\mathfrak{Z}_{\alpha,\beta}\begin{bmatrix}m\\n\end{bmatrix}(p,y;t)&=\exp\left(\sum_{a,w= 1}^{\infty}\frac{p^{aw}}{aw} \sum_{b=0}^{w-1}Z_{a\alpha-b\beta,w\beta}\begin{bmatrix}am+bn\\wn\end{bmatrix}\left(y^a;\frac{at+b}{w}\right)\right)\,.\label{eq:ZabGrand}
\end{align}
Here, the doublet $\begin{bmatrix}m \\ n\end{bmatrix}$ with $m,n\in\{0,\frac{1}{2}\}$ encodes the boundary conditions
\begin{equation}
\psi(z+\tau)=(-1)^{2m}\psi(z)\,,\quad\psi(z+1)=(-1)^{2n}\psi(z)
\end{equation}
for any spinor $\psi$. These are the same as for the local bulk partition function, meaning that in the seed theory, the defect $\mathcal{L}$ is assumed to act trivially on the fermionic part of $X$.

\subsubsection{Cyclic projection}
\label{subsec:cyclic}

Let us now specialize to the case $\alpha=0$ and $\beta=1$, so that \eqref{eq:GrandCanDef} computes the grandcanonical partition function of the defect-ending fields for the maximally-fractional defect $\boldsymbol{\mathcal{L}}_\rho$ (recall that we are taking $\rho$ to be either the trivial or the sign representation of $S_N$). In order to proceed with manipulating \eqref{eq:GrandCanDef} into a DMVV-like form, we conjecture that for \emph{any} topological defect $\mathcal{L}$ in the seed theory $X$, we can recast
\begin{align}
    \frac{1}{w}\sum_{b=0}^{w-1} \epsilon^{b(w+1)}  Z_{-b,w}\bigg(\frac{at+b}{w}\bigg) &= \mathrm{Tr}_{\mathcal{H}(\mathcal{L}^w)} \Pi_w q^{\frac{a}{w}(L_0-\frac{c}{24})}\bar{q}^{\frac{a}{w}(\bar{L}_0-\frac{c}{24})}
    \equiv  Z_{0,w}^\prime\bigg(\frac{at}{w}\bigg)\,,\label{eq:spproj}
\end{align}
where $\Pi_w$ is a projector on the Hilbert space $\mathcal{H}(\mathcal{L}^w)$ of fields on which the defect $\mathcal{L}^w$ can end.

First, in the trivial case when $\mathcal{L}$ is the identity defect and $\epsilon=1$, we have already remarked above that \eqref{eq:spproj} holds with $\Pi_w$ being the projector onto states with spin $h-\bar{h}\in w\mathbb{Z}$.
Allowing for $\epsilon=-1$ (the sign-defect), this projection instead selects spins such that 
\begin{align}
    h-\bar{h}+\frac{1}{2}w(w+1) \in w\mathbb{Z}\,.\label{eq:epsProj}
\end{align}
We observe that \eqref{eq:epsProj} gives the spin $\frac{h-\bar{h}}{w}$ of the single-particle defect ending fields of the sign defect $\boldsymbol{\mathcal{L}}_\mathrm{sign}$ to be integer for $w\in 2\mathbb{Z}+1$ and half-integer for $w\in 2\mathbb{Z}$.

For a general seed-CFT topological defect line $\mathcal{L}$, the form of the projector $\Pi_w$ strongly depends on the nature of $\mathcal{L}$. In the cases when $\mathcal{L}$ is invertible and non-anomalous, it is possible to unambiguously extend \cite{Chang:2018iay} the action of the operator $\mathcal{L}$ also to the Hilbert spaces $\mathcal{H}(\mathcal{L}^{n})$ so that one may express
\begin{equation}
Z_{m,n}(t)=\text{Tr}_{\mathcal{H}(\mathcal{L}^{n})}[\mathcal{L}^{m}q^{L_0-\frac{c}{24}}\bar{q}^{\bar{L}_0-\frac{c}{24}}]\,.
\end{equation}
As a consequence, the relation \eqref{eq:spproj} can be seen to hold with the projector\footnote{Indeed, we can prove that the operator $\Pi_w$ given by \eqref{eq:PiInv} satisfies $\Pi_w^2=\Pi_w$ under the assumption that ${\mathcal{L}}^{-w\alpha}e^{2\pi i(L_0-\overline{L}_0)}=\text{id}$ on $\mathcal{H}({\mathcal{L}}^{\alpha w})$. This, in turn, follows from the modular properties \eqref{eq:modular} of the correlators $Z_{m,n}(t)$.}
\begin{equation}
\Pi_w=\frac{1}{w}\sum_{b=0}^{w-1}\mathcal{L}^{-b}e^{2\pi i\frac{b}{w}(L_0-\bar{L}_0)}\,.\label{eq:PiInv}
\end{equation}
Finally, instead of attempting to prove the validity of \eqref{eq:spproj} for a general (possibly anomalous and non-invertible) topological defect, we demonstrate that it holds in the cases of various explicit realizations of TDLs: see Appendix \ref{subsec:su2} for the TDLs $\mathcal{L}^{(\frac{k}{2})}$ in $\mathfrak{su}(2)_k$ WZW models which, for odd $k$, are invertible but anomalous, Appendix \ref{subsec:ising} for the invertible defect $\mathcal{L}^{(\varepsilon)}$ and non-invertible duality defect $\mathcal{L}^{(\sigma)}$ in the Ising model, and, Appendix \ref{subsec:G21} for the non-invertible and non-duality defect of the $(G_2)_1$ WZW model.

Applying the result \eqref{eq:spproj} to the grand canonical partition function \eqref{eq:GrandCanDef} for $\alpha=0$, $\beta=1$, we can first rewrite
\begin{align}
  {\mathfrak{Z}}_{0,1}(p;t) =\exp\bigg(\sum_{a,w=1}^\infty   \frac{p^{aw}}{a}\epsilon^{(a+1)w} {Z}_{0,w}^\prime\bigg(\frac{at}{w}\bigg)\bigg)\,.\label{eq:Z01first}
\end{align}
Notice that the sign $\epsilon^{(a+1)w}$ which appears in \eqref{eq:Z01first} in the case $\boldsymbol{\mathcal{L}}_\text{sign}$ furnishes a natural interpretation that the single-particle states have bosonic statistics for even $w$ and fermionic statistics for odd $w$. Then, since we are cycling through $a$ particles, one picks up an additional sign $(-1)^{a+1}$ if these are fermions. Comparing this result with the spin of the single-particle states, as dictated by the projection \eqref{eq:epsProj}, we conclude that the defect-ending fields of the sign-defect $\boldsymbol{\mathcal{L}}_\text{sign}$ violate spin-statistics. This is not so surprising because these are non-local states. Indeed, as illustrated by many examples in Appendices \ref{defects} and \ref{defectsFree} (see also Section \ref{subsec:SymNT4} below),
for a generic non-trivial topological defect $\mathcal{L}$ in a bosonic CFT $X$, the spins of the defect-ending fields are not restricted to be integral while the fields themselves are treated as bosonic.

\subsubsection{The single-particle spectrum}
\label{subsec:single_particle}

Coming back to \eqref{eq:Z01first}, one can straightforwardly manipulate it further into
\begin{align}
    \mathfrak{Z}_{0,1}(p;t)=\prod_{w=1}^{\infty}\exp\left(-\epsilon^w\,\text{Tr}_{\mathcal{H}(\mathcal{L}^{w})}\left[\Pi_w\log\left(1-\epsilon^w p^w q^{\frac{L_0}{w}}\bar{q}^{\frac{\bar{L}_0}{w}}\right)\right]\right)\,.\label{eq:Z01intermediate}
\end{align}
Denoting by $\mathcal{H}'({\mathcal{L}}^{w})$ the subspace of $\mathcal{H}({\mathcal{L}}^{w})$ surviving the projection by $\Pi_w$, we can recast \eqref{eq:Z01intermediate} into a DMVV-like form
\begin{equation}
\mathfrak{Z}_{0,1}(p;t)=\prod_{w=1}^{\infty}\text{det}_{\mathcal{H}'({\mathcal{L}}^{w})}\left[1-\epsilon^w p^w q^{\frac{L_0}{w}}\overline{q}^{\frac{\overline{L}_0}{w}}\right]^{-\epsilon^w}\,.
\end{equation}
In line with the above discussion, we notice that in the case of the sign defect ($\epsilon=-1$), this can be interpreted as a partition function on a Fock space where even values of $w$ contribute with \emph{bosonic} creation operators while odd values of $w$ contribute with \emph{fermionic} creation operators.\footnote{See \cite{Dijkgraaf:1999za} for a different construction (based on discrete torsion) endowing twisted sectors with definite statistics depending on their cycle lengths.} On the other hand, when $\rho$ is the trivial representation of $S_N$, all particles are bosonic, as it was the case for the local bulk fields. Overall, one can conclude that analogously to the Hilbert space of local bulk fields, the Hilbert space of defect ending fields for  generic maximally-fractional defects $\boldsymbol{\mathcal{L}}_0$ and $\boldsymbol{\mathcal{L}}_\text{sign}$ furnishes an interpretation in terms of multiparticle states, with the single-particle partition function
\begin{subequations}
\label{eq:ZspDef}
\begin{align}
Z^{\text{s.p.}}_{0,1}(t)&=\sum_{w=1}^{\infty}\text{Tr}_{\mathcal{H}'({\mathcal{L}}^{w})}\left[q^{\frac{L_0}{w}}\overline{q}^{\frac{\bar{L}_0}{w}}\right]\\
&=\sum_{w=1}^{\infty}Z'_{0,w}\left(\frac{t}{w}\right)
\end{align}
\end{subequations}
where $Z'_{0,w}$ is the seed theory defect partition function restricted to the subspace $\mathcal{H}'({\mathcal{L}}^{w})$. 

Notice that a twist-$w$ defect-ending field for the maximally fractional defect $\boldsymbol{\mathcal{L}}$ is taken from the seed-theory Hilbert space $\mathcal{H}(\mathcal{L}^w)$ of defect-ending fields for $w$ copies of the seed-theory defect $\mathcal{L}$: this was to be expected because locally, the preimage of the neighbourhood of a twist-$w$ puncture in the covering space contains $w$ copies of the defect line. Also note that the dependence on $\epsilon$ has \emph{not} completely dropped out from \eqref{eq:ZspDef}, as the nature of the projection $'$ of the Hilbert space $\mathcal{H}({\mathcal{L}}^{w})$ depends on whether we are taking $\epsilon=+1$ or $\epsilon=-1$: as we have already remarked, in the case of the sign-defect ($\epsilon=-1$) the defect-ending fields with odd $w$ have integer spin and behave as fermions, while the states with even $w$ have half-integer spin and behave as bosons. Moreover, when all defects are trivial, we recover the usual single-particle spectrum of the symmetric orbifold. 

Finally, notice that in the cases when the action of $\mathcal{L}$ can be consistently extended to the spaces $\mathcal{H}(\mathcal{L}^{w})$ (such as when the defect line $\mathcal{L}$ corresponds to an invertible and non-anomalous symmetry of the theory $X$), one can manipulate the grandcanonical ensemble $\mathfrak{Z}_{\alpha,\beta}$ of the maximally-fractional defect torus correlator for generic $\alpha,\beta$ as
\begin{subequations}
\label{eq:Zalphabeta}
\begin{align}
\mathfrak{Z}_{\alpha,\beta}(p;t)
&=\prod_{w=1}^{\infty}\exp\left(-\epsilon^{w\beta}\text{Tr}_{\mathcal{H}(\mathcal{L}^{w\beta})}\left[\Pi_w\log\left(1-\epsilon^\alpha\epsilon^{w(\alpha+\beta)} p^w\mathcal{L}^\alpha q^{\frac{L_0}{w}}\bar{q}^{\frac{\bar{L}_0}{w}}\right)\right]\right)\\
&=\prod_{w=1}^{\infty}\text{det}_{\mathcal{H}'({\mathcal{L}}^{w\beta})}\left[1-\epsilon^\alpha\epsilon^{w(\alpha+\beta)} \mathcal{L}^\alpha p^wq^{\frac{L_0}{w}}\overline{q}^{\frac{\overline{L}_0}{w}}\right]^{-\epsilon^{w\beta}}\,,
\end{align}
\end{subequations}
where now $\Pi_w$ is explicitly given by \eqref{eq:PiInv}. This yields the single-particle defect torus correlators 
\begin{subequations}
\label{eq:spGen}
\begin{align}
Z^{\text{s.p.}}_{\alpha,\beta}(t)&=\sum_{w=1}^{\infty}\text{Tr}_{\mathcal{H}'({\mathcal{L}}^{w\beta})}\left[\epsilon^{\alpha(w+1)}{\mathcal{L}}^{\alpha}q^{\frac{L_0}{w}}\overline{q}^{\frac{\overline{L}_0}{w}}\right]\\
&=\sum_{w=1}^{\infty} \epsilon^{\alpha(w+1)} Z'_{\alpha,w\beta}\left(\frac{t}{w}\right)\,.
\end{align}
\end{subequations}
In the case when the seed theory $X$ contains a fermionic component, the results \eqref{eq:ZspDef} and \eqref{eq:spGen} have to be modified as (focusing on the NS-NS sector twisted by $(-1)^{F+\bar{F}}$, for the sake of concreteness)
\begin{align}
 Z^{\text{s.p.}}_{\alpha,\beta}\begin{bmatrix}0 \\\frac{1}{2}\end{bmatrix}(y;t)=\sum_{w=1}^{\infty} \epsilon^{\alpha(w+1)} Z'_{\alpha,w\beta}\begin{bmatrix}0\\\frac{w}{2}\end{bmatrix}\left(y;\frac{t}{w}\right)\,.\label{eq:spSpin}
\end{align}
That is, for odd $w$, the seed theory correlator contributes to the single-particle correlator with the NS-NS sector while for $w$ even, it contributes with the R-R sector.

\subsubsection{Geometric construction}

\begin{figure}[!htpb]
\centering
\begin{tikzpicture}
\begin{scope}
\draw[thick, blue] (1.5,0) -- (1.5,3);
\node[right] at (1.5,1.5) {$\boldsymbol{\mathcal{L}}_0$};
\draw[thick] (0,0) -- (0,3) -- (3,3) -- (3,0) -- (0,0);
\node[below] at (1.5,0) {$g$};
\node[left] at (0,1.5) {$h$};
\end{scope}
\node at (4.5,1.5) {$=$};
\begin{scope}[xshift = 6cm]
\draw[thick, blue] (1,0) -- (0,1);
\draw[thick, blue] (3,1) -- (1,3);
\draw[thick] (0,0) -- (0,3) -- (3,3) -- (3,0) -- (0,0);
\node[above] at (0.6,0.6) {$\mathcal{L}$};
\node[above] at (2.1,2.1) {$\mathcal{L}$};
\end{scope}
\end{tikzpicture}
\caption{The $g$-twisted sector contribution to the defect symmetric orbifold partition function. The defect $\boldsymbol{\mathcal{L}}_0$ is lifted to the covering space imposed by the monodromy of the symmetric orbifold.}
\label{fig:defect-lift-torus}
\end{figure}
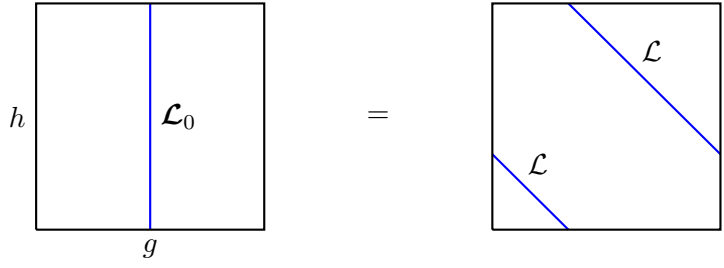

We will now make a couple of comments about the geometrical interpretation of the above construction of the defect torus correlators in the symmetric product orbifold theory. We will focus on the simple case of the maximally-fractional defect $\boldsymbol{\mathcal{L}}_0$ for which $\rho$ is the trivial representation.
Let $\Sigma$ be the torus we formulate the $\text{Sym}^N(X)$ theory on. For every pair $g,h$ in $S_N$, we can associate a covering space $\widetilde{\Sigma}$ of the torus by `stitching together' the $N$ copies of the fields in $X$. That is, we define $\widetilde{\Sigma}$ such that the multi-valued fields on $\Sigma$ lift to single-valued fields on $\widetilde{\Sigma}$. The degree of the covering space $\widetilde{\Sigma}$ will be $N$. The defect $\boldsymbol{\mathcal{L}}_0$ will then lift to the seed theory defect $\mathcal{L}$ on some cycle of $\widetilde{\Sigma}$, see Figure \ref{fig:defect-lift-torus}. This cycle is determined by the preimage of the location of $\boldsymbol{\mathcal{L}}_0$ under the map $\widetilde{\Sigma}\to\Sigma$.
As usual, we would like to compute the grandcanonical ensemble
\begin{equation}
\mathfrak{Z}_{\boldsymbol{\mathcal{L}}_0}(p;t)=\sum_{N=0}^{\infty}p^N(Z_{N})_{\boldsymbol{\mathcal{L}}_0}(t)
\end{equation}
of the torus correlation functions $(Z_{N})_{\boldsymbol{\mathcal{L}}_0}(t)$ where the defect $\boldsymbol{\mathcal{L}}_0$ wraps some cycle in $\Sigma$.
Since  $(Z_{N})_{\boldsymbol{\mathcal{L}}_0}(t)$ is found by summing over all covering spaces of degree $N$, then $\log\mathfrak{Z}_{\boldsymbol{\mathcal{L}}_0}(p;t)$ will be expressed in terms of \textit{connected} covering spaces with degree weighted by $p$, weighted by automorphism. That is, we can write
\begin{equation}
\mathfrak{Z}_{\boldsymbol{\mathcal{L}}_0}(p;t)=\exp\left(\sum_{\widetilde{\Sigma}\text{ connected}}\frac{p^{\text{deg}(\widetilde{\Sigma}\to\Sigma)}}{|\text{Aut}(\widetilde{\Sigma}\to\Sigma)|}Z_{\mathcal{L}}(\widetilde{\Sigma})\right)\,,\label{eq:GrandCanGeom}
\end{equation}
where $Z_{\mathcal{L}}$ is the torus correlation function of the seed theory $X$ with the insertion of the preimage $\mathcal{L}$ of the original defect $\boldsymbol{\mathcal{L}}_0$ on $\widetilde{\Sigma}$.

For the sake of concreteness, let us parameterize the base torus as
\begin{equation}
\Sigma=\mathbb{C}^2/\mathbb{Z}\oplus t\mathbb{Z}\,,
\end{equation}
with $\text{Im}(t)>0$ being the modulus of the torus. Now, let the $A$-cycle be $z\to z+1$ and the $B$-cycle $z\to z+t$. We can consider a maximally-fractional defect $\boldsymbol{\mathcal{L}}_0$ placed on the homology cycle $B$. As we have done in the preceding sections, let us denote the grand canonical partition function in this case by $\mathfrak{Z}_{0,1}(p;t)$, where the subscript tells us that $\boldsymbol{\mathcal{L}}_0$ wraps the $A$-cycle zero times and the $B$-cycle once. As discussed above, $\mathfrak{Z}_{0,1}$ is computed by summing over all connected covering spaces of the base torus. These covering spaces are also tori (by the Riemann-Hurwitz formula) and are labeled by positive integers $a,b,w$ with $b=0,1,\ldots,w-1$. The degree of such a covering map is $aw$, and the degree of its automorphism group is also $aw$. The covering map $\Gamma:\widetilde{\Sigma}\to\Sigma$ acts on the homology cycles $\tilde{A},\tilde{B}$ of $\widetilde{\Sigma}$ as
\begin{equation}
\Gamma(\tilde{A})=wA\,,\quad\Gamma(\tilde{B})=aB+bA\,.
\end{equation}
The pre-image of the defect $\boldsymbol{\mathcal{L}}_0$ is thus $N$ times the pre-image of $B$ under this map, i.e. $w\tilde{B}-b\tilde{A}$. Furthermore, the modular parameter of the covering space is $(at+b)/w$. Altogether, the grandcanonical torus correlator \eqref{eq:GrandCanDef} therefore reduces to \eqref{eq:GrandCanDef} with $\epsilon=+1$.

\subsection{Defects in sphere correlation functions}

Aside from the torus partition function, we can also consider the case of a sphere correlation function of local bulk twist fields in the symmetric product orbifold $\text{Sym}^N(X)$ in the presence of a nontrivial defect which wraps around one or more of the local bulk fields. These are important objects to consider as they should be holographically computed by on-shell closed-string amplitudes on a suitable $\mathrm{AdS}_3$ background. For simplicity (and for connection to holographic calculations), let us focus purely on the case of single-cycle twist fields. These are states $\sigma_w$ in the twisted sector corresponding to permutations of cycle types $(1\cdots w)$. In addition, these $w$-twist vacua can be decorated by non-identity states in the theory $X$, as per the single-particle content which is dictated by the partition function \eqref{eq:ZspDef}.
The correlation function
\begin{equation}
\Braket{\sigma_{w_1}(x_1)\cdots\sigma_{w_n}(x_n)}
\end{equation}
on the sphere is calculated by passing to a covering space $\Gamma:\Sigma\to\mathbb{CP}^1$ such that
\begin{equation}
\Gamma(z)\sim x_i+\mathcal{O}((z-z_i)^{w_i})
\end{equation}
near $n$ marked points $z_i$ on $\Sigma$ \cite{Arutyunov:1997gt,Lunin:2000yv,Pakman:2009zz,Roumpedakis:2018tdb,Dei:2019iym}. The complex structure of the covering space is found by demanding that $\Gamma$ is a holomorphic map. In principle, there are many such distinct choices of covering spaces, and one needs to sum over them, weighted by the appropriate path integral on the covering space (see \cite{Lunin:2000yv,Knighton:2024qxd} for details).

Calculation of sphere correlators involving defect loops will strongly depend on whether we are inserting an invertible or non-invertible topological defect. For the non-invertible ones, we will focus on the \emph{duality defects} in the $\text{Sym}^N(X)$ theory which carry the trivial representation of $S_N$, that is the defects $\boldsymbol{\mathcal{L}}_0$ such that the fusion $\boldsymbol{\mathcal{L}}_0\boldsymbol{\mathcal{L}}^\dagger_0$ results in a superposition of invertible group-like defects.

\subsubsection{Invertible defects wrapping local bulk fields}
\label{subsec:sphereInv}

Let us consider wrapping an invertible maximally-fractional defect loop $\boldsymbol{\mathcal{L}}$ around a subset of punctures on a sphere. Deforming the loop and making use of the invertibility property, this can be directly related to calculating the sphere correlator where defect loops $\boldsymbol{\mathcal{L}}$ wrap the individual punctures. Provided that \emph{local} bulk fields of the $\text{Sym}^N(X)$ theory were inserted at the punctures, these get mapped to (generally different) local bulk fields by the action of $\boldsymbol{\mathcal{L}}$. Hence, the calculation of the defect sphere correlator reduces to the calculation of a sphere correlator of local bulk fields, where (a subset of) the insertions were acted upon by $\boldsymbol{\mathcal{L}}$. 

While this does not provide any qualitatively new structures in terms of computing the correlators by means of the covering-map technology, acting with invertible defect loops inside correlators generally enables neat derivation of conservation laws (Ward identities) associated with the global invertible symmetries enacted by $\boldsymbol{\mathcal{L}}$. For $\boldsymbol{\mathcal{L}}_0$ (fixing the trivial representation of $S_N$), these arise as natural uplift of whatever conservation laws were present in the seed theory $X$ to the individual $w_i$-twisted sectors of the symmetric-product orbifold. Given that correlators in $\text{Sym}^N(X)$ compute amplitudes of closed strings on $\mathrm{AdS}_3\times \mathrm{S}^3 \times X$ at $k=1$ unit of pure NS-NS flux (where we typically take $X=\mathbb{T}^4$), the invertible symmetries associated with $\boldsymbol{\mathcal{L}}_0$ become global symmetries of the space-time in which the string propagates. Locally on the worldsheet, these are enacted by the topological defect which acts trivially on the $\mathfrak{psu}(1,1|2)_1$ factor and is equal to $\mathcal{L}$ in the $X$ factor of the worldsheet theory.

On the other hand, as discussed in Section \ref{subsec:SymFusion}, wrapping a defect loop $\boldsymbol{\mathcal{L}}_\mathrm{sign}$ around local bulk punctures in a sphere correlator, one can derive the selection rule \eqref{eq:wSelection} on the cycle-lengths $w_i$. Incarnation of this invertible defect in the worldsheet CFT can be used to derive the corresponding selection rule on the spectral flow parameters of vertex-operator insertions in order for worldsheet correlators (and hence amplitudes) to be non-zero. 

\subsubsection{Non-invertible duality defects and disorder correlators}
\label{subsec:sphereDual}

The computation becomes more interesting when the defect loop $\boldsymbol{\mathcal{L}}$ wrapping a subset of punctures in a sphere correlator is a non-invertible duality defect. Attempting to deform such a loop around individual local bulk insertions, one does not only end up with a sphere correlator of local bulk fields acted upon by $\boldsymbol{\mathcal{L}}$ (as was the case for the invertible defects) but generally also picks up contributions from sphere correlators where \emph{invertible} topological defect lines run between punctures with insertions of \emph{disorder fields} -- the fields on which invertible defects may terminate. Hence, inserting duality defect loops into sphere correlators may generally provide potentially interesting \emph{order-disorder dualities} between correlators involving only local bulk fields and correlators involving disorder fields, such as the Kramers-Wannier duality of the Ising model \cite{Frohlich:2004ef}. In order to calculate contributions involving networks of non-trivial maximally-fractional invertible defects joining punctures with non-local insertions of disorder fields fields, one has to carefully follow the covering map prescription.

\begin{figure}[!htpb]
\centering
\begin{tikzpicture}[scale = 0.9]
\begin{scope}
\draw[thick] (0,0) circle (2);
\draw[thick] (0,0) [partial ellipse = 0:-180:2 and 0.5];
\draw[thick, dashed] (0,0) [partial ellipse = 0:180:2 and 0.4];
\fill (-0.8,-1.15) circle (0.05);
\fill (-0.8,1.15) circle (0.05);
\fill (0.8,-1.15) circle (0.05);
\fill (0.8,1.15) circle (0.05);
\draw[thick, blue] (0,1.15) [partial ellipse = 0:360:1.1 and 0.3];
\end{scope}
\draw[thick, latex-] (3,0) -- (6.5,0);
\node[above] at (4.75,0) {$\Gamma$};
\begin{scope}[xshift = 9.5cm]
\draw[thick, purple] (-1.15,0) [partial ellipse = 0:180:0.85 and 0.2];
\draw[thick, purple, dashed] (-1.15,0) [partial ellipse = 0:-180:0.85 and 0.2];
\draw[thick,, dashed, blue, fill = white] (-1.5,-0.18) circle (0.2);
\draw[thick, blue, fill = white] (-0.75,0.18) circle (0.2);
\draw[thick] (0,0) [partial ellipse = 0:360:2 and 3];
\draw[thick] (0.2,0) [partial ellipse = 100:260:0.5 and 1];
\draw[thick] (-0.2,0) [partial ellipse = -70:70:0.4 and 0.9];
\fill (-1.5,-0.18) circle (0.05);
\fill (-0.75,0.18) circle (0.05);
\fill (1.5,0.18) circle (0.05);
\fill (0.75,-0.18) circle (0.05);
\end{scope}
\end{tikzpicture}
\caption{The pre-image of a defect $\mathcal{L}$ in the symmetric orbifold under a covering map $\Gamma$. Here, the four local operators are taken to lie in the twisted sector of the permutation $(1\,2)$. The pre-image of the defect wraps the pre-images of the twist fields, and the fusion $\mathcal{L}^{\dagger}\mathcal{L}$ wraps a handle of the torus.}
\label{fig:DualityPreimage}
\end{figure}
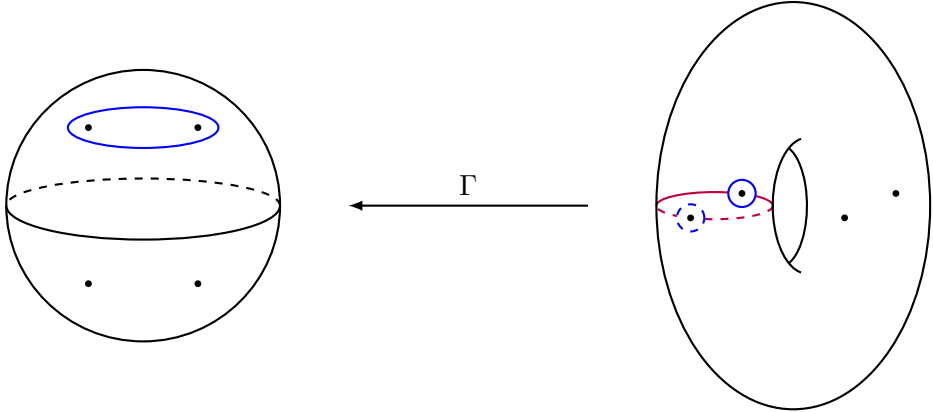

As in the case of the torus correlators considered above, given a maximally-fractional topological defect on the sphere, then in the computation of the correlation functions, we need to `lift' the defect to the covering space. Concretely, let us assume that we have a network of maximally-fractional defects which is supported on a curve $\gamma$ in the base sphere on which the $\text{Sym}^N(X)$ theory lives. The defects terminate on punctures with the insertions of non-local single-particle twist-$w_i$ defect-ending fields whose spectra are encoded by the partition functions \eqref{eq:ZspDef}. Then, given a covering map $\Gamma:\Sigma\to\mathbb{CP}^1$, the corresponding defect network one has to consider in the covering space is taken to live on the preimage $\tilde{\gamma}$ of $\gamma$ under the covering map $\Gamma$. This defect network $\tilde{\gamma}$ then joins insertions of non-local defect-ending fields which are taken from the seed-theory Hilbert spaces $\mathcal{H}({\mathcal{L}}^{w_i})$. In Figure \ref{fig:DualityPreimage}, we provide a concrete illustation of such a contribution to a sphere correlator in the case when a non-invertible defect loop wraps local twist-2 insertions in a 4-point sphere correlator.

While we will stop short of performing any explicit calculations of $\mathrm{Sym}^N(X)$ correlators involving insertions of disorder-fields of maximally-fractional defects, we note that the above discussion provides useful lessons as to the nature of the worldsheet backgrounds whose amplitudes would give rise to such correlators. In particular, it seems to be impossible to realize such holographic duality purely within the framework of the conventional string perturbation theory on $\mathrm{AdS}_3\times \mathrm{S}^3 \times \mathbb{T}^4$ with $k=1$ units of pure NS-NS flux since the spectrum of closed strings (even off-shell) on this background does not allow for non-integer eigenvalues of the operator $J_0^3-\bar{J}_0^3$ and is therefore unable to accommodate vertex operators corresponding to the non-local disorder fields of the spacetime CFT. In Section \ref{sec:worldsheet} we will indeed explicitly construct modular-invariant worldsheet partition functions which \emph{differ} from the one describing $\mathrm{AdS}_3\times \mathrm{S}^3 \times \mathbb{T}^4$ at $k=1$ and whose on-shell states are in one-to-one correspondence with single-particle disorder fields of maximally-fractional defects in the $\mathrm{Sym}^N(X)$ CFT.

Finally, also note that while we have focused on the case of $\boldsymbol{\mathcal{L}}_0$ (trivial representation of $S_N$) the discussion in this section would have equally well applied to the duality defects $\boldsymbol{\mathcal{L}}_\rho$ where, of course, the representation $\rho$ has to fuse with its conjugate to produce a direct sum containing only the trivial and the sign representation. One would then generally have to worry about the statistics of the $\boldsymbol{\mathcal{L}}_\mathrm{sign}$ defect-ending fields depending on their cycle-length, as discussed in Sections \ref{subsec:cyclic} and \ref{subsec:single_particle}.

\subsection[Topological defects in \texorpdfstring{$\mathrm{Sym}^{N}(\mathbb{T}^4)$}{SymNT4}]{\boldmath Topological defects in \texorpdfstring{$\mathrm{Sym}^{N}(\mathbb{T}^4)$}{SymNT4}}
\label{subsec:SymNT4}

In light of the applications for the tensionless holographic duality to be considered in Section \ref{sec:worldsheet}, let us make some of the above results very explicit in the case when the seed theory $X$ is taken to be the CFT of four free bosons and four free fermions on a $\mathbb{T}^4$. We will focus on considering examples of defects in the seed theory, which satisfy definite simple gluing conditions on the free boson and fermion currents.\footnote{See \cite{Fuchs:2007tx} for a more complete discussion of topolgical defects for a single free boson.} Structurally clearer incarnations of these defects, which arise in the theory involving only one free boson, are reviewed in Appendices \ref{subapp:noncompact} and \ref{subapp:compact}.
For the sake of simplicity, we will focus on considering non-trivial topological defect lines only for the bosons and let all defect operators act as the identity on the fermions.

\subsubsection{The theory without defects}

Let us start by reviewing the bulk spectrum of the seed theory. As the modular-invariant partition function factorizes into the theory of the four free bosons and the fermionic partition function, these can be treated separately.

Beginning with the fermionic part of the seed CFT, the bulk partition function will be simply that of a tensor product of four free-fermion CFTs. Introducing chemical potentials $\zeta_1,\zeta_2$ for the Cartan elements of the $\mathfrak{so}(4)\cong \mathfrak{su}(2)\oplus \mathfrak{su}(2)$ algebra generated by fermion bilinears, the partition function reads
\begin{align}
    Z^\mathrm{f}\begin{bmatrix}\mu \\ \nu \end{bmatrix}(\zeta_1,\zeta_2;t) = \frac{1}{2}\frac{1}{|\eta(t)|^4}
    \left|\,\vartheta\begin{bmatrix}\mu \\ \nu \end{bmatrix}\bigg(\frac{\zeta_1+\zeta_2}{2};t\bigg)\,\vartheta\begin{bmatrix}\mu \\ \nu \end{bmatrix}\bigg(\frac{\zeta_1-\zeta_2}{2};t\bigg)\right|^2\,,\label{eq:4fermion_part}
\end{align}
where $\mu,\nu\in \{0,\frac{1}{2}\}$ determine the boundary conditions on the fermions: NS-NS sector for $\mu=\nu=\frac{1}{2}$, the NS-NS sector twisted by $(-1)^{F+\bar{F}}$ for $\mu=0$, $\nu=\frac{1}{2}$, the R-R sector for $\mu=\frac{1}{2}$, $\nu=0$ and the R-R sector twisted by $(-1)^{F+\bar{F}}$ for $\mu=\nu=0$.

On the other hand, the theory of the four free bosons compactified on a $\mathbb{T}^4$ has a comparably richer structure to offer. Its (Narain) moduli can be encoded in terms of the metric $g$ and the anti-symmetric tensor $B$. In terms of these, it is customary to define $E=g+B$ (so that $E^T=g-B$). Here and in the following, we are mostly suppressing the (spacetime) $\mathbb{T}^4$ indices. The modular-invariant partition function of the four free bosons can then be written as
\begin{align}
    Z^\mathrm{b}(t) = \frac{\Theta(t)}{|\eta(t)|^8}
\,,\label{eq:narain_part}
\end{align}
where $\Theta(t)$ denotes the Narain lattice theta function
\begin{align}
      \Theta(t)=\sum_{n,u\in \mathbb{Z}^{\otimes 4}}q^{h_\mathrm{L}(n,u)}\,\bar{q}^{h_\mathrm{R}(n,u)}\,.\label{eq:NarainTheta}
\end{align}
Here the zero-mode parts $h_\mathrm{L}$ and $h_\mathrm{R}$ of the left and right conformal dimensions $\Delta_\mathrm{L}=h_\mathrm{L}+M$ and $\Delta_\mathrm{R}=h_\mathrm{R}+\bar{M}$ are functions of the quantized momenta $n$, winding numbers $u$ (both running over 4-tuples of integers), as well as the Narain moduli. Also, $M,\bar{M}$ denote the oscillator contributions to the left and right conformal weights. In particular, $h_\mathrm{L}$ and $h_\mathrm{R}$ can be expressed as
\begin{subequations}
\label{eq:hLhR}
\begin{align}
    h_\mathrm{L}&= \frac{1}{4}(k_\mathrm{L})^T g^{-1}k_\mathrm{L}\,,\\
    h_\mathrm{R}&= \frac{1}{4}(k_\mathrm{R})^T g^{-1}k_\mathrm{R}\,,
\end{align}
\end{subequations}
where we have introduced the left and right momenta $k_\mathrm{L}$ and $k_\mathrm{R}$. This in turn can be given in terms of the momenta $n$, the winding numbers $u$ and the Narain moduli as
\begin{subequations}
\label{eq:kLkR}
    \begin{align}
        k_\mathrm{L} &= n+Eu\,,\\
        k_\mathrm{R} &= n-E^T u\,.
    \end{align}
\end{subequations}
The vertex operators corresponding to the momenta $n$ and winding numbers $u$ can then be represented as (up to cocycle prefactors)
\begin{align}
    V_{n,u}(z,\bar{z})=e^{ik_\mathrm{L}^T X_\mathrm{L}}(z)\,e^{ik_\mathrm{R}^T X_\mathrm{R}}(\bar{z}) = e^{in^TX+ i\tilde{X}^Tu }(z,\bar{z})\,,
\end{align}
where we define
\begin{subequations}
    \begin{align}
        X(z,\bar{z})&=X_\mathrm{L}(z)+X_\mathrm{R}(\bar{z})\,,\\
        \tilde{X}(z,\bar{z}) &=E^T X_\mathrm{L}(z)- E X_\mathrm{R}(\bar{z})\,.
    \end{align}
\end{subequations}
Altogether, the local bulk partition function of the superconformal $\mathbb{T}^4$ sigma model reads
\begin{subequations}
\label{eq:ZT4munu}
\begin{align}
      Z_{\mathbb{T}^4}\begin{bmatrix}\mu \\ \nu \end{bmatrix}(\zeta_1,\zeta_2;t) &= Z^\mathrm{b}(t)\,  Z^\mathrm{f}\begin{bmatrix}\mu \\ \nu \end{bmatrix}(\zeta_1,\zeta_2;t)\\
      &=\frac{1}{2}\frac{1}{|\eta(t)|^{12}}\,\Theta(t)\,
    \left|\,\vartheta\begin{bmatrix}\mu \\ \nu \end{bmatrix}\bigg(\frac{\zeta_1+\zeta_2}{2};t\bigg)\,\vartheta\begin{bmatrix}\mu \\ \nu \end{bmatrix}\bigg(\frac{\zeta_1-\zeta_2}{2};t\bigg)\right|^2\,.
\end{align}
\end{subequations}
Under the action of $\mathrm{PSL}(2;\mathbb{Z})$, this partition function transforms as
    \begin{align}
        \Big|e^{-\frac{\pi i}{4(c\tau+d)}(\zeta_1^2+\zeta_2^2)c}\Big|^2 Z_{\mathbb{T}^4}\begin{bmatrix}\mu \\ \nu \end{bmatrix}\left(\frac{\zeta_1}{c\tau+d},\frac{\zeta_2}{c\tau+d};\frac{a\tau+b}{c\tau+d}\right) &= Z_{\mathbb{T}^4}\begin{bmatrix}a\mu+b\nu \\ c\mu+d\nu \end{bmatrix}(\zeta_1,\zeta_2;\tau)\,.\label{eq:ZT4modular}
    \end{align}
Uplifting this to the symmetric-product orbifold theory $\mathrm{Sym}^N(\mathbb{T}^4)$ and focusing on the NS-NS sector twisted by $(-1)^{F+\bar{F}}$, one would obtain the single-particle partition function
\begin{align}
    Z_{\mathbb{T}^4}^{\text{s.p.}}\begin{bmatrix} 0 \\ \frac{1}{2} \end{bmatrix}(\zeta_1,\zeta_2;t)=  \sum_{w=1}^\infty Z_{\mathbb{T}^4}^\prime\begin{bmatrix} 0 \\ \frac{w}{2} \end{bmatrix}\bigg(\zeta_1,\zeta_2;\frac{t}{w}\bigg)\,,\label{eq:ZspT4}
\end{align}
where $^\prime$ imposes the projection onto $\mathbb{T}^4$ which satisfy
\begin{align}
    h_\mathrm{L}-h_\mathrm{R}+M-\bar{M}\in w\mathbb{Z}\,.\label{eq:ProjspT4}
\end{align}
Finally, recall that it is \emph{not} true that each four-by-four matrix $E$ is to be associated with a distinct point in the Narain moduli space: parameters $E$ related by the action of the \emph{T-duality group} \cite{Buscher:1987qj,Buscher:1987sk,Giveon:1994fu} (which, for a $\mathbb{T}^4$, is isomorphic to $O(4,4;\mathbb{Z})$) give rise to the same partition function \eqref{eq:narain_part}. In the simple case of a square torus with radii $R_1=\ldots=R_4\equiv R$ and $B=0$, part of this T-duality group acts by simply inverting the radius as $R \leftrightarrow \alpha^\prime/R$.

\subsubsection{Shift-symmetry defects}
\label{subsec:ShiftT4}

Let us first construct a family of defects for the trivial gluing conditions
\begin{subequations}
\label{eq:trivial1}
    \begin{align}
        \p X^{(1)}(z) &=\p X^{(2)}(z)\,,\\
        \bar{\p} X^{(1)}(\bar{z}) &=\bar{\p} X^{(2)}(\bar{z})\,.
    \end{align}
\end{subequations}
These will be given by the defect operators
\begin{align} 
    {\mathcal{L}}^{(U,N)} = \sum_{n,u\in\mathbb{Z}^{\otimes 4}} e^{2\pi i n^T U}e^{2\pi i N^T u}\,\sum_{M,\bar{M}}|n,u,M,\bar{M}\rangle\,\langle n,u,M,\bar{M}|\,,\label{eq:Dbtr}
\end{align}
where the parameters $U,N$ are real four-component vectors which are defined modulo 1. They implement the translations $X\longrightarrow X+2\pi U$, $\tilde{X}\longrightarrow \tilde{X}+2\pi N$ along the $\mathbb{T}^4$ and its dual. These are clearly invertible symmetries of the free-boson CFT as 
\begin{align}
    {\mathcal{L}}^{(U_1,N_1)}{\mathcal{L}}^{(U_2,N_2)}={\mathcal{L}}^{(U_1+U_2,N_1+N_2)}\,,
\end{align}
so that, in particular, ${\mathcal{L}}^{(U,N)}{\mathcal{L}}^{(-U,-N)}=\mathrm{id}$. As we have already remarked above, we are taking the defect to act trivially on the free-fermion factor of the CFT. Inserting the defect loops ${\mathcal{L}}^{(U,N)}$ into sphere correlators, they make manifest the corresponding conservation laws for the momenta $n_i$ and winding numbers $u_i$ of the vertex operators inserted at punctures. 

On the other hand, inserting the defect operators \eqref{eq:Dbtr} into a torus correlator (first along the spatial cycle and then performing suitable modular transformations to achieve general winding numbers $r,s$), one can find 
\begin{align}
     (Z_{\mathbb{T}^4})_{r,s}^{(U,N)}\begin{bmatrix}\mu \\ \nu \end{bmatrix}(\zeta_1,\zeta_2;t) &=\frac{\Theta_{r,s}^{(U,N)}(t)}{|\eta(t)|^8}\, Z^\mathrm{f}\begin{bmatrix}\mu \\ \nu \end{bmatrix}(\zeta_1,\zeta_2;t) \,,\label{eq:ZWNrs}
\end{align}
where we have introduced the defect Narain theta function $\Theta_{r,s}^{(U,N)}(t)$. This is parameterized in terms of the defect kernels at fixed momentum $n$ and winding $u$
\begin{align}
   K_{r,s}^{(U,N)}(n,u;t)= e^{2\pi irs N^TU}e^{2\pi i r(n^TU + N^T u) }q^{h_\mathrm{L}(n+sN,u+sU)}\bar{q}^{h_\mathrm{R}(n+sN,u+sU)}\label{eq:DefectKernel}
\end{align}
as
\begin{align}
  \Theta_{r,s}^{(U,N)}(t)&= \sum_{n,u\in\mathbb{Z}^{\otimes 4}} K_{r,s}^{(U,N)}(n,u;t)\,.
\end{align}
We have also noted that the four free fermions simply contribute with \eqref{eq:4fermion_part}. As a consistency check, one readily verifies that for $U=N=0$, this reduces to the modular-invariant bulk partition function. 

Under the action of the $\mathrm{PSL}(2;\mathbb{Z})$, the correlators \eqref{eq:ZWNrs} share the transformation property \eqref{eq:ZT4modular} of the local bulk partition function \eqref{eq:ZT4munu} except now also the defect winding numbers $(r,s)$ transform as a modular doublet. For $r=0$ and $s=1$, the correlator \eqref{eq:ZWNrs} turns into the parition function of fields on which the defect $\mathcal{L}^{(U,N)}$ can terminate. This encodes the spectrum
\begin{align}
    \mathcal{H}(\mathcal{L}^{(U,N)}) = \bigoplus_{n,u\in\mathbb{Z}^{\otimes 4}} \mathcal{V}_{n+N, u+U}\,,
\end{align}
where the states are build upon vacua with momentum $n$ and winding $u$ which have conformal dimension $h_\mathrm{L}(n+N,u+U)$ in the left and $h_\mathrm{R}(n+N,u+U)$ in the right. Hence, for generic values of $N$ and $U$, the spins of these defect-ending fields are non-integer, as we may express
\begin{align}
    h_\mathrm{L}\big(n+N,u+U\big) -h_\mathrm{R}\big(n+N,u+U\big) = (n+N)^T(u+U)\notin \mathbb{Z}\,.\label{eq:spinT4}
\end{align}
Also, we notice that changing the summation indices from $(n,u)$ to $(-n,-u)$, one can readily verify that
\begin{align}
    Z_{r,s}^{(U,N)}(t)=\smash{Z_{-r,-s}^{(U,N)}}(t)=Z_{r,s}^{(-U,-N)}(t)\,,
\end{align}
namely that the defect ${\mathcal{L}}^{(U,N)}$ and its conjugate $({\mathcal{L}}^{(U,N)})^\dagger=\mathcal{L}^{(-U,-N)}$ give rise to the same (Virasoro-specialized) defect torus correlators.

Uplifting the defect \eqref{eq:Dbtr} to the symmetric product orbifold theory $\mathrm{Sym}^N(\mathbb{T}^4)$, one obtains the maximally fractional defect $\boldsymbol{\mathcal{L}}_\rho^{(U,N)}$. In order for the partition function of the defect-ending fields to factorize into DMVV-like Fock-space partition function
we will assume $\rho$ to be either the trivial representation $\rho=0$ or the sign representation. For the sake of simplicity, we will also assume that we are fixing the NS-NS boundary conditions for the free fermions in the symmetric-product orbifold theory, twisted by the insertion of $\smash{(-1)^{F+\bar{F}}}$ into the trace. It is then useful to note that the equation \eqref{eq:spproj} holds, where, for $\rho=0$, the projector $\smash{\Pi_w}$ on $\smash{\mathcal{H}[(\mathcal{L}^{(U,N)})^w]}$ can be explicitly evaluated as\footnote{In addition, we should be projecting on integer spins $\frac{L_0-\bar{L}_0}{w}$ for even number of $\mathbb{T}^4$ fermions and half-integer spins for odd number of $\mathbb{T}^4$ fermions. This feature is unchanged by the presence of the defect since it acts trivially on the $\mathbb{T}^4$ fermions.}
\begin{align}
    \Pi_w 
    &= \delta_{\mathbb{Z}_w}\bigg[L_0-\bar{L}_0-(n+wN)^T(u+wU)+n^T u\bigg]\,.\label{eq:T4Pd}
\end{align}
Here we have introduced the indicator function $\delta_{\mathbb{Z}_w}(m)$ on the additive identity in $\mathbb{Z}_w$ (that is, $\delta_{\mathbb{Z}_w}(m)$ is equal to $1$ whenever $m\equiv 0\,\mathrm{mod}\,w$ and 0 otherwise). Note that \eqref{eq:T4Pd} indeed defines a good projector on $\mathcal{H}[(\mathcal{L}^{(U,N)})^w]$ because owing to \eqref{eq:spinT4}, it simply imposes the solvable constraint 
\begin{align}
    n^T u+M-\bar{M}\in w\mathbb{Z}\label{eq:SpinContraintT4}
\end{align}
on the momenta and windings $n,u\in\mathbb{Z}^{\otimes 4}$. The actual spins $\frac{L_0-\bar{L}_0}{w}$ of the single-particle defect-ending fields in the $w$-cycle twisted sector are for generic values of $U$ and $N$ non-integral, which underlines their non-local nature. Also note that if we were to consider the sign defect (that is $\epsilon=-1$ in \eqref{eq:spproj}), an additional term $\frac{1}{2}(w+1)w$ would appear in \eqref{eq:T4Pd} and \eqref{eq:SpinContraintT4}. This would contribute into the single-particle spin with a half-integer for $w\in 2\mathbb{Z}$ even though the $w$-twisted states with even $w$ were observed to obey bosonic statistics. Again, this violation of spin-statistics should not be surprising due to the inherent non-locality of the defect-ending fields.
Finally, substituting into \eqref{eq:spSpin}, this would yield the corresponding single-particle torus correlators
\begin{align}
(Z_{\mathbb{T}^4}^{\text{s.p.}})_{\alpha,\beta}^{(U,N)} \begin{bmatrix}0 \\\frac{1}{2}\end{bmatrix}(\zeta_1,\zeta_2;t)=\sum_{w=1}^{\infty} \epsilon^{\alpha(w+1)} \,
  (Z_{\mathbb{T}^4}')_{\alpha,w\beta}^{(U,N)}
\begin{bmatrix}0\\\frac{w}{2}\end{bmatrix}\left(\zeta_1,\zeta_2;\frac{t}{w}\right)\,,\label{eq:spSpinT4}
\end{align}
where $'$ imposes the projection onto the image of $\Pi_w$ inside $\mathcal{H}[(\mathcal{L}^{(U,N)})^w]$.

More complicated gluing conditions may yield other invertible defects, depending on the symmetries of the Narain lattice at a particular point in the moduli space \cite{Dulat:2000xj}. For instance, the gluing conditions $ \p X^{(1)}(z) =-\p X^{(2)}(z)$, $ \bar{\p} X^{(1)}(\bar{z}) =-\bar{\p} X^{(2)}(\bar{z})$ would have given rise to an invertible $\mathbb{Z}_2$ defect for any value of $E$. Gauging such defect would then yield the orbifold theory $\mathbb{T}^4/\mathbb{Z}_2$ (see Appendices \ref{subsubapp:reflection} and \ref{subsubapp:reflectionCompact} for more details about this defect in the case of the CFT of a single free boson).

\subsubsection{T-duality defects}

One of the main reasons why we are predominantly interested into the simple shift-symmetry defects \eqref{eq:Dbtr} is that at rational points in the Narain moduli space, they may arise in the fusion channel $\mathcal{L}^\mathrm{T}(\mathcal{L}^\mathrm{T})^\dagger$ of a non-invertible duality defect $\mathcal{L}^\mathrm{T}$. 

This is demonstrated in detail in Appendix \ref{subsubapp:Tduality} in the simpler case of a single free boson, where the defect $\mathcal{L}^\mathrm{T}$ interchanges the momentum vector $n/R$ with the winding vector $wR$ when acting on the vertex operators. Since this is an exact symmetry of the Hilbert space only when $R$ is equal to the self-dual radius, the defect $\mathcal{L}^\mathrm{T}$ has to act non-invertibly by killing the states which would have landed outside of the local bulk CFT Hilbert space. In order for at least some of the states to survive the action of $\mathcal{L}^\mathrm{T}$, one has to fix $R^2$ to be a rational number. 

In the case of $\mathbb{T}^4$, reflecting the rich structure of the T-duality group $O(4,4;\mathbb{Z})$, we expect there to be a large number of duality operators $\mathcal{L}^\mathrm{T}$ implementing similar non-invertible symmetries at rational points in the Narain moduli space \cite{Gukov:2002nw}. For the sake of clarity, we will not attempt a complete discussion of these. Instead, we simply specialize on the case of a square torus with $R_1=\ldots=R_4\equiv R = \sqrt{p/q}$, $B=0$ and take $\mathcal{L}^\mathrm{T}$ to act as in \eqref{eq:LTbos} on each of the four constituent free-boson CFTs.\footnote{See \cite{Damia:2024xju,Cordova:2023qei} for more complicated examples of defects in 2d sigma models.} It then follows from \eqref{eq:LTLT} that $\mathcal{L}^\mathrm{T}$ obeys the fusion rule
\begin{align}
    {\mathcal{L}}^{\mathrm{T}}({\mathcal{L}}^{\mathrm{T}})^\dagger=\sum_{\alpha\in(\mathbb{Z}_p)^{\otimes 4}}\sum_{\beta\in(\mathbb{Z}_q)^{\otimes 4}}\mathcal{L}^{(\frac{\alpha}{p},\frac{\beta}{q})}\,,\label{eq:TdualityFusion}
\end{align}
where the defects appearing in the fusion channel are precisely the shift-symmetry defects \eqref{eq:Dbtr} with $U = \alpha/p$ and $N=\beta/q$.

It is then straightforward to uplift the whole setup to the symmetric-product orbifold CFT, where the fusion rule \eqref{eq:TdualityFusion} holds for the maximally-fractional defects $\boldsymbol{\mathcal{L}}_0^\mathrm{T}$ and $\boldsymbol{\mathcal{L}}_0^{(\alpha/p,\beta/q)}$.
Blowing up a defect loop $\boldsymbol{\mathcal{L}}_0^\mathrm{T}$ inside a correlator involving local bulk field insertions, one can derive order-disorder (Kramers-Wannier-type) dualities relating local bulk field correlators with correlators of non-local disorder fields which live in the Hilbert spaces of defect-ending fields of the shift-symmetry defects $\boldsymbol{\mathcal{L}}_0^{(\alpha/p,\beta/q)}$. As we have already commented above, in order to discuss manifestations of such order-disorder dualities at the level of amplitudes of closed strings propagating in an $\mathrm{AdS}_3$ background, one first has to answer the question of how to represent the non-local defect-ending fields in the $\mathrm{Sym}^N(\mathbb{T}^4)$ defect spacetime CFT using local vertex operators representing on-shell closed strings at the level of the worldsheet CFT.

\section{\boldmath \texorpdfstring{$\mathrm{AdS}_3$}{AdS3} worldsheet realization of topological defects }\label{sec:worldsheet}

In this section, we make concrete proposals on the tensionless-string duals of the maximally-fractional defects in the $\mathrm{Sym}^N(\mathbb{T}^4)$ CFT which we discussed in the previous section.

\subsection{The tensionless worldsheet}

Before we discuss the holographic duals of topological defects, let us review the computation of the worldsheet partition function describing the pure closed-string $\mathrm{AdS}_3$ background at $k=1$ \cite{Eberhardt:2018ouy,Eberhardt:2020bgq}.
The worldsheet theory dual to the symmetric orbifold $\text{Sym}^N(\mathbb{T}^4)$ is described by type IIB string theory on $\text{AdS}_3\times\text{S}^3\times\mathbb{T}^4$ with $k=1$ unit of pure NS-NS flux \cite{Eberhardt:2018ouy}. In the hybrid formalism of Berkovits, Vafa, and Witten \cite{Berkovits:1999im}, the worldsheet sigma model is
\begin{equation}
\mathfrak{psu}(1,1|2)_1\oplus(\text{topologically twisted }\mathbb{T}^4)\oplus(\text{ghosts})\,.\label{eq:worldsheet-theory}
\end{equation}
The first factor is a super-WZW model on the supergroup $\text{PSU}(1,1|2)$ at level $k=1$ (determined by the amount of NS-NS flux). This factor describes the $\text{AdS}_3\times\text{S}^3$ geometry in a Green-Schwarz-like description making the target space supersymmetry manifest. The topologically-twisted $\mathbb{T}^4$ is the $\mathbb{T}^4$ of the target space geometry with the conformal weights of the free fermions twisted to yield a sigma model with vanishing central charge. For all intents and purposes, the corresponding spectrum can be achieved by considering R-R boundary conditions for the $\mathbb{T}^4$ fermions twisted by the insertion of $(-1)^F$ in the worldsheet torus partition function.
Finally, the ghost system contains the usual $b,c$ conformal ghost system, as well as a scalar $\rho$ of central charge $c(\rho)=28$.

\subsubsection[Free-field realization of \texorpdfstring{$\mathfrak{psu}(1,1|2)_1$}{psu(1,1|2)}]{\boldmath Free-field realization of \texorpdfstring{$\mathfrak{psu}(1,1|2)_1$}{psu(1,1|2)}}

The special feature of this worldsheet theory is that the $\text{PSU}(1,1|2)$ WZW model at $k=1$ admits a free-field realization \cite{Dei:2020zui}. Specifically, it was shown in \cite{Beem:2023dub,Dei:2023ivl} that the $\mathfrak{psu}(1,1|2)_1$ algebra can be realized in terms of a single bosonic $\beta\gamma$ system and two fermionic first order systems $(p_a,\theta^a)$ $(a=1,2)$ with action
\begin{equation}
S=\frac{1}{2\pi}\int(\beta\overline{\partial}\gamma+p_a\overline{\partial}\theta^a)\,.
\end{equation}
The free fields $\beta,p_a$ have conformal weight $\Delta=1$, while the fields $\gamma,\theta^a$ have weight $\Delta=0$. Since we are concerned with closed strings, the full worldsheet theory also includes right-moving analogues of these fields.

The above free field realization is morally similar to the Wakimoto representation of $\mathfrak{sl}(2,\mathbb{R})_k$. Specifically, the generators of the subalgebra $\mathfrak{sl}(2,\mathbb{R})_1\oplus\mathfrak{su}(2)_1\subset\mathfrak{psu}(1,1|2)_1$ are given by
\begin{equation}
\begin{gathered}
J^+=\beta\,,\quad J^3=\beta\gamma+\frac{1}{2}(p_a\theta^a)\,,\quad J^-=(\beta\gamma)\gamma+(p_a\theta_a)\gamma\,,\\
K^+=p_2\theta^1\,,\quad K^3=-\frac{1}{2}(p_1\theta^1)+\frac{1}{2}(p_2\theta^2)\,,\quad K^-=p_1\theta^2\,.
\end{gathered}
\end{equation}
Just as in the Wakimoto representation, we should think of the field $\gamma$ as a complex holographic coordinate on the $\text{AdS}_3$ boundary.\footnote{Here we are working in Euclidean signature where $\gamma$ is complex and $\bar{\gamma}$ is its complex conjugate. In Lorenzian signature, $\gamma$ and $\bar{\gamma}$ would be real and independent.} This can be seen by noting that under $\text{SL}(2,\mathbb{R})$ transformations $\gamma$ transforms as
\begin{equation}
\gamma\to\frac{a\gamma+b}{c\gamma+d}\,.
\end{equation}
In fact, every field in the $\mathfrak{psu}(1,1|2)_1$ can be given interpretations as natural objects living on the boundary of $\text{AdS}_3$. We can think of $\text{PSU}(1,1|2)$ as the global group of $\mathcal{N}=(4,0)$ superconformal transformations in two dimensions. Analysing the transformation laws of the various free fields under the generators of $\mathfrak{psu}(1,1|2)$ shows that $\gamma$ transforms like a complex coordinate $x$ on the boundary, while the fermionic partners $\theta^a$ transform as holomorphic $\mathcal{N}=(2,0)$ superspace coordinates $\vartheta^a$. This suggests a bulk-boundary dictionary
\begin{equation}
\begin{split}
\gamma\longleftrightarrow x\,,&\quad\beta\longleftrightarrow\partial_x\,,\\
\theta^a\longleftrightarrow\vartheta^a\,,&\quad p_a\longleftrightarrow\frac{\partial}{\partial\vartheta_a}\,.
\end{split}
\end{equation}
relating worldsheet fields to geometric objects on the conformal boundary of $\text{AdS}_3$. The right-moving fields on the worldsheet similarly map to antiholomorphic quantities on the boundary. 

\subsubsection{The worldsheet representations}

The highest-weight representations of the $\mathfrak{psu}(1,1|2)_1$ model are spanned by states $\ket{m}$ satisfying
\begin{equation}
\beta_0\ket{m}=m\ket{m+1}\,,\quad\gamma_0\ket{m}=\ket{m-1}\,,\quad (p_a)_0\ket{m}=0\,,
\end{equation}
as well as all of their descendants. Each such representation is labeled uniquely by the fractional part $\lambda$ of the quantum number $m$. Let us denote this representation by $\mathcal{F}_\lambda$. The (super-)character of the representation $\mathcal{F}_\lambda$ is defined to be
\begin{equation}
\chi_{\lambda}(t,z;\tau)=\text{Tr}_{\mathcal{F}_\lambda}\left[ (-1)^{F} q^{L_0}x^{J^3_0}y^{K^3_0}\right]\,,
\end{equation}
where $q=e^{2\pi i\tau}$, $x=e^{2\pi it}$, and $y=e^{2\pi iz}$ are fugacities associated with the worldsheet conformal dimension, the boundary conformal dimension, and the boundary R-symmetry charges respectively. This character is readily computed by noting that the $\beta,\gamma$ system decouples from the $p_a,\theta^a$ system: the character of the $\beta\gamma$ system is
\begin{equation}
\sum_{m\in\mathbb{Z}+\lambda}\frac{x^m}{\eta(\tau)^2}\,,
\end{equation}
while the free fields $p_a,\theta^a$ contribute a factor of
\begin{equation}
\frac{\vartheta_1(\tfrac{t+z}{2};\tau)\vartheta_1(\tfrac{t-z}{2};\tau)}{\eta(\tau)^2}\,.
\end{equation}
Thus, we have
\begin{equation}
\chi_\lambda(t,z;\tau)=\sum_{m\in\mathbb{Z}+\lambda}x^m\frac{\vartheta_1(\tfrac{t+z}{2};\tau)\vartheta_1(\tfrac{t-z}{2};\tau)}{\eta(\tau)^4}\,.
\end{equation}
In addition to the highest-weight representations defined above, the $\text{PSU}(1,1|2)$ WZW model contains an infinite tower of non-highest weight representations related by the so-called spectral flow automorphism. This acts on the modes of the $\beta\gamma$ system and the $p_a,\theta^a$ system as
\begin{equation}
\begin{split}
\sigma^w(\gamma_n)=\gamma_{n+w}\,,&\quad\sigma^w(\beta_n)=\beta_{n-w}\,,\\
\sigma^w(\theta^1_n)=\theta^1_{n+w}\,,&\quad\sigma^w((p_1)_n)=(p_1)_{n-w}\,,
\end{split}
\end{equation}
and acts trivially on $p_2$ and $\theta^2$. This can be translated into the corresponding action on the generators $L_0$, $J^3_0$, $K^3_0$ as
\begin{equation}
\sigma^w(L_0)=L_0+w(K^3_0-J^3_0)\,,\quad\sigma^w(J^3_0)=J^3_0+\frac{w}{2}\,,\quad\sigma^w(K^3_0)=K^3_0+\frac{w}{2}\,.
\end{equation}
Thus, since $\sigma^w$ is an automorphism, we have
\begin{equation}
\begin{split}
\text{Tr}_{\sigma^w(\mathcal{F}_\lambda)}[q^{L_0}x^{J^3_0}y^{K^3_0}]&=\text{Tr}_{\mathcal{F}_\lambda}[(-1)^F q^{\sigma^{w}(L_0)}x^{\sigma^{w}(J^3_0)}y^{\sigma^{w}(K^3_0)}]\\
&=x^{\frac{w}{2}}y^{\frac{w}{2}}\text{Tr}_{\mathcal{F}_{\lambda}}[ (-1)^F q^{L_0}(q^{-w}x)^{J^3_0}(q^{w}x)^{K^3_0}]\,.
\end{split}
\end{equation}
This enables us to express the character of the representation $\sigma^w(\mathcal{F}_\lambda)$ as \cite{Creutzig:2012sd}
\begin{equation}
\begin{split}
\chi^{w}_{\lambda}(t,z;\tau)&=x^{\frac{w}{2}}y^{\frac{w}{2}}\chi_{\lambda}(t-w\tau,z+w\tau;\tau)\\[2mm]
&=x^{\frac{w}{2}}y^{\frac{w}{2}}\sum_{m\in\mathbb{Z}+\lambda}e^{2\pi im(t-w\tau)}\frac{\vartheta_1(\tfrac{t+z}{2};\tau)\vartheta_1(\tfrac{t-z}{2}-w\tau;\tau)}{\eta(\tau)^4}\\
&=x^{w}q^{-\frac{w^2}{2}}\sum_{m\in\mathbb{Z}+\lambda}x^mq^{-wm}\frac{\vartheta_1(\tfrac{t+z}{2};\tau)\vartheta_1(\tfrac{t-z}{2};\tau)}{\eta(\tau)^4}\,,\label{eq:psu_characters}
\end{split}
\end{equation}
where in the last line, we have used the $\vartheta$-function identity
\begin{equation}
\vartheta_1(\tfrac{t-z}{2}-w\tau;\tau)=q^{-\frac{w^2}{2}}x^{\frac{w}{2}}y^{-\frac{w}{2}}\vartheta_1(\tfrac{t-z}{2};\tau)\,.
\end{equation}
Looking into the behaviour of the supercharacters under the modular transformations, one finds that the S-transformation is given as
\begin{align}
e^{\frac{\pi i}{2\tau}(t^2-z^2)}\chi^{w}_{\lambda}(\tfrac{t}{\tau},\tfrac{z}{\tau};-\tfrac{1}{\tau})&=\sum_{w'\in\mathbb{Z}}\int_0^1 d\lambda'\, S_{\lambda,\lambda'}^{w,w'}\,\chi^{w'}_{\lambda'}(t,z;\tau)\,,\label{eq:SMod}
\end{align}
where we introduce the modular S-matrix
\begin{align}
    S_{\lambda,\lambda'}^{w,w'}=i\,\frac{|\tau|}{\tau}\,e^{2\pi i [w'(\lambda-\frac{1}{2})+w(\lambda'-\frac{1}{2})]}\,.\label{eq:SMat}
\end{align}
On the other hand, the modular T-transformation acts as
\begin{align}
    \chi^{w}_{\lambda}(t,z;\tau+1)&=e^{2\pi i (\frac{w^2}{2}-mw+\frac{1}{12})}\chi^{w}_{\lambda}(t,z;\tau)\,.
\end{align}

\subsubsection[The global \texorpdfstring{$\mathrm{AdS}_3$}{AdS3} partition function]{\boldmath The global \texorpdfstring{$\mathrm{AdS}_3$}{AdS3} partition function}

The spectrum of the $\mathfrak{psu}(1,1|2)_1$ WZW model which is relevant for the pure closed-string global $\text{AdS}_3\times\text{S}^3\times\mathbb{T}^4$ background with $k=1$ unit of NS-NS flux takes the block diagonal form
\begin{equation}
\mathcal{H}_{\text{PSU}}=\bigoplus_{w\in\mathbb{Z}}\int_{0}^{1}\mathrm{d}\lambda\left[\sigma^w(\mathcal{F}_\lambda)\oplus\sigma^w(\overline{\mathcal{F}_{\lambda}})\right]\,.\label{eq:HPSU}
\end{equation}
Thus, the corresponding modular-invariant partition function of the $\mathfrak{psu}(1,1|2)_1$ WZW model can be written as
\begin{equation}
Z_{\text{PSU}}(t,z;\tau)=\frac{1}{2}\sum_{w\in\mathbb{Z}}\int_{0}^{1}\mathrm{d}\lambda\,|\chi_{\lambda}^{w}(t,z;\tau)|^2\,.
\end{equation}
Using Poisson resummation on the characters $\chi_{\lambda}^{w}$, the partition function can be formally recast in terms of a sum over delta functions as
\begin{equation}
Z_{\text{PSU}}(t,z;\tau)=\frac{1}{2}\sum_{r,w\in\mathbb{Z}}|q|^{w^2}\left|\frac{\vartheta_1(\tfrac{t+z}{2};\tau)\vartheta_1(\tfrac{t-z}{2};\tau)}{\eta(\tau)^4}\right|^2\delta^{(2)}(t-w\tau-r)\,.
\end{equation}
In order to obtain a full worldsheet partition function, we also need to include the partition function of the topologically twisted $\mathbb{T}^4$ CFT and $(\rho,\sigma)$ ghost systems, as specified in \eqref{eq:worldsheet-theory}. We simply quote the results
\begin{subequations}
    \begin{align}
        Z_{\mathbb{T}^4}(\tau)&=\left|\frac{\vartheta_1(0;\tau)^2}{\eta(\tau)^6}\right|^2\Theta(\tau)\,,\\
        Z_{\rho\sigma}(\tau)&=\left|\frac{\eta(\tau)^4}{\vartheta_1(0;\tau)^2}\right|^2\,.\label{eq:Zghost}
    \end{align}
\end{subequations}
Here, $\Theta(\tau)$ is the Narain theta function defined in \eqref{eq:NarainTheta} which keeps track of the winding and momentum sectors in the $\mathbb{T}^4$. Putting everything together a noticing that the fermionic oscillators cancel between the ghost and the $\mathbb{T}^4$ part, the full worldsheet partition function therefore reads
\begin{subequations}
\label{eq:WSPartitionFunction}
\begin{align}
Z(t,z;\tau)&=\frac{\Theta(\tau)}{|\eta(\tau)|^4}\sum_{w\in\mathbb{Z}}\,\int_{0}^{1}\mathrm{d}\lambda\,|\chi_{\lambda}^{w}(t,z;\tau)|^2 \label{eq:WSPartitionFunction1} \\
&=\frac{1}{2}\sum_{r,w\in\mathbb{Z}}|q|^{w^2}\left|\frac{\vartheta_1(\tfrac{t+z}{2};\tau)\vartheta_1(\tfrac{t-z}{2};\tau)}{\eta(\tau)^6}\right|^2\Theta(\tau)\,\delta^{(2)}(t-w\tau-r)\,.\label{eq:WSPartitionFunction2}
\end{align}
\end{subequations}
Substituting the constraint $t=w\tau+r$ into the $\vartheta$-functions in \eqref{eq:WSPartitionFunction2}, this can be further recast as \cite{Eberhardt:2020bgq}
\begin{align}
Z(t,z;\tau)&=\sum_{w,r\in \mathbb{Z}}Z_{r,w}(t;\tau)\,Z_{\mathbb{T}^4}\begin{bmatrix}\frac{r}{2} \\ \frac{w}{2} \end{bmatrix}(z;\tau)\,,\label{eq:Zblocks}
\end{align}
where we have introduced the blocks
\begin{align}
    Z_{r,w}(t;\tau) = \frac{1}{2}\delta^{(2)}(t-w\tau-r)\, e^{-\pi w \mathrm{Im}\,t}    \,.
\end{align}
It is then not hard to verify that the pair $(r,w)$ transforms as a doublet under $\mathrm{PSL}(2;\mathbb{Z})$, namely
    \begin{align}
        \Big|e^{\frac{\pi ic}{2(c\tau+d)}t^2}\Big|^2 Z_{r,w}\left(\frac{t}{c\tau+d};\frac{a\tau+b}{c\tau+d}\right) &= |c\tau+d|^2 Z_{ar+bw,cr+dw}(t;\tau)
    \end{align}
and we also recall the transformation properties \eqref{eq:ZT4modular} of $Z_{\mathbb{T}^4}$.
Summing over all $r,w\in\mathbb{Z}$ in \eqref{eq:Zblocks} therefore guarantees modular invariance of $Z(t,z;\tau)$ (up to the elliptic prefactors due to chemical potentials $t,z$). The appearance of the decomposition \eqref{eq:Zblocks} of $Z(t,z;\tau)$ into the blocks $Z_{w,r}(t;\tau)$ is also a telltale sign of the origin of the global $\mathrm{AdS}_3$ worldsheet theory as a $\mathbb{Z}$-orbifold whose twisted sectors give rise to the spectral flow parameter $w$. One can readily identify the modular-invariant partition function of the theory which is being orbifolded as
\begin{subequations}
\label{eq:psu_product}
\begin{align}
    Z_0(t,z;\tau)&=\frac{1}{2}\left|\frac{\vartheta_1(\tfrac{t+z}{2};\tau)\vartheta_1(\tfrac{t-z}{2};\tau)}{\eta(\tau)^6}\right|^2\Theta(\tau)\,\delta^{(2)}(t)\\
    &=\frac{\Theta(\tau)}{|\eta(\tau)|^4}\int_{0}^{1}\mathrm{d}\lambda\int_{0}^{1}\mathrm{d}\bar{\lambda}\,\chi_{\lambda}^{0}(t,z;\tau)\,\chi_{\bar{\lambda}}^{0}(t,z;\tau)^\ast\,.
\end{align}
\end{subequations}
This appears to encode a spectrum whose $\mathfrak{psu}(1,1|2)_1$ part fully factorizes into the holomorphic and anti-holomorphic sector.
We will come back to this point in Section \ref{subsec:path_integral}, where we will identify the topological defects in the $\mathfrak{psu}(1,1|2)_1$ WZW model, which can be gauged to give rise to the $\mathbb{Z}$-orbifold structure \eqref{eq:Zblocks} of the global $\mathrm{AdS}_3$ partition function $Z(t,z;\tau)$.

Finally, let us consider projecting the worldsheet partition function \eqref{eq:WSPartitionFunction} onto the states which satisfy the mass-shell constraints $h=\bar{h}=0$, where $h,\bar{h}$ denote the total worldsheet conformal weights. These can be directly read off from \eqref{eq:WSPartitionFunction1} as coefficients multiplying $\tau$ and $\bar{\tau}$. One obtains
\begin{subequations}
    \begin{align}
        h&=\frac{w^2}{2}-mw +h_\mathrm{L} +M\,,\\
        \bar{h}&=\frac{w^2}{2}-\bar{m}w +h_\mathrm{R} +\bar{M}\,,
    \end{align}
\end{subequations}
where we recall that $m,\bar{m}\in\mathbb{Z}+\lambda$ and that $h_\mathrm{L}+M$ with $h_\mathrm{R}+M$ denote the dimensions of states in the $\mathbb{T}^4$ CFT (see \eqref{eq:hLhR} and \eqref{eq:kLkR}). In particular, the level-matching condition $h-\bar{h}=0$ for the closed strings implies the constraint
\begin{align}
    h_\mathrm{L}-h_\mathrm{R} +M -\bar{M}\in w\mathbb{Z}
\end{align}
on the spin of the $\mathbb{T}^4$ states. We note that this is identical to the constraint \eqref{eq:ProjspT4} provided that we identify the spectral flow parameter with the cycle-length $w$ labelling the twisted sectors of the $\mathrm{Sym}^N(\mathbb{T}^4)$ theory. Moreover, solving the mass-shell constraints for $m,\bar{m}$ and substituting back into \eqref{eq:WSPartitionFunction1}, one can find the on-shell worldsheet partition function as (see \cite{Eberhardt:2018ouy} for more details) 
\begin{align}
    Z^\text{on-shell}(z;t) = \sum_{w=1}^\infty Z_{\mathbb{T}^4}^\prime\begin{bmatrix} 0 \\ \frac{w}{2} \end{bmatrix}\bigg(z;\frac{t}{w}\bigg)=   Z_{\mathbb{T}^4}^{\text{s.p.}}\begin{bmatrix} 0 \\ \frac{1}{2} \end{bmatrix}(z;t)\,.
\end{align}
This concludes the identification of the on-shell closed-string states in the $w$ spectrally-flowed sector propagating on global $\mathrm{AdS}_3\times \mathrm{S}^3\times \mathbb{T}^4$ at $k=1$ unit of NS-NS flux with the single-cycle $w$-twisted states of the $\mathrm{Sym}^N(\mathbb{T}^4)$ CFT.

\subsection{Spacetime defects from the worldsheet}

We are now in a position to study the bulk duals to topological defects in the symmetric orbifold CFT. We will show that topological defects in $\mathrm{Sym}^N(\mathbb{T}^4)$ give rise to modified $\mathrm{AdS}_3$ closed-string backgrounds which describe the extensions of the spacetime CFT defects into the bulk. Worldsheet partition functions of these backgrounds will be seen to encode spectra of on-shell vertex operators which are exactly dual to the single-cycle spectra of defect-ending fields for maximally-fractional defects in the symmetric-product orbifold.

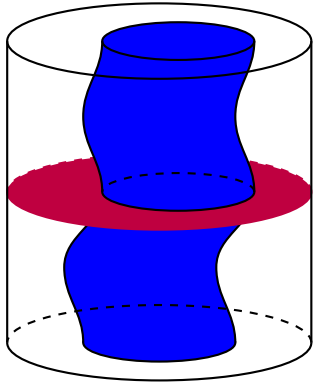
\begin{figure}[!htpb]
\centering
\begin{tikzpicture}
\draw[thick, dashed] (0,0) [partial ellipse = 0:180:1 and 0.25];
\draw[thick, fill = blue, fill opacity = 0.15] (-0.75,2) to[out = -90, in = 90] (-1.25,1) to[out = -90, in = 90] (-1,0) arc (180:360:1 and 0.25) (1,0) to[out = 90, in = -90] (0.75,1) to[out = 90, in = -90] (1.25,2) arc (0:-180:1 and 0.25);
\draw[thick, dashed, purple] (0,2) [partial ellipse = 0:180:2 and 0.5];
\fill[opacity = 0.15, purple] (0,2) [partial ellipse = 0:360:2 and 0.5];
\draw[thick, purple] (0,2) [partial ellipse = 180:360:2 and 0.5];
\draw[thick, fill = blue, fill opacity = 0.15] (1.25,2) to[out = 90, in = -90] (1,3) to[out = 90, in = -90] (1.25,4) arc (0:180:1 and 0.25) to[out = -90, in = 90] (-1,3) to[out = -90, in = 90] (-0.75,2) arc(180:360:1 and 0.25);
\draw[thick, dashed] (0.25,2) [partial ellipse = 0:180:1 and 0.25];
\draw[thick] (0.25,4) [partial ellipse = 0:-180:1 and 0.25];
\draw[thick] (-2,0) -- (-2,4);
\draw[thick] (2,0) -- (2,4);
\draw[thick] (0,4) [partial ellipse = 0:360:2 and 0.5];
\draw[thick] (0,0) [partial ellipse = 180:360:2 and 0.5];
\draw[thick, dashed] (0,0) [partial ellipse = 0:180:2 and 0.5];
\end{tikzpicture}
\caption{A cartoon of a bulk topological defect `brane' in $\text{AdS}_3$ which intersects the boundary on a certain cycle.}
\label{fig:brane-cartoon}
\end{figure}

\subsubsection{Geometric picture}

In terms of bulk geometry, it is natural to suspect that the duals of topological defect lines in the spacetime CFT would look like tensionless brane-like extended objects, which intersect the $\text{AdS}_3$ boundary at the location of the dual topological defects \cite{GarciaEtxebarria:2022vzq,Heckman:2022muc,Heckman:2022xgu,Dierigl:2023jdp,Cvetic:2023plv}. From the point of view of the worldsheet CFT in $\text{AdS}_3$, we can think of this `brane' as inducing a defect line on the intersection of the worldsheet and the location of the brane in the bulk as shown in Figure \ref{fig:brane-cartoon}. This is somewhat analogous to the interpretation of spherical D-branes in $\mathrm{AdS}_3$ as boundary states in the worldsheet CFT \cite{Gaberdiel:2021kkp} except now, the brane-like duals of spacetime topological defects seem to give rise to consistent \emph{closed-string} backgrounds in the $\mathrm{AdS}_3$ bulk.

Whenever an invertible spacetime topological defect loop wraps around local spacetime CFT insertions, it can be shrunk around individual punctures to yield a correlator of local fields. Correspondingly, as we have already pointed out in Section \ref{subsec:sphereInv}, such a configuration would not give rise to a new background for the closed strings propagating in the bulk: indeed, this can be understood in terms of a tensionless brane-like object collapsing around the points on the worldsheet where the vertex operators are inserted and which are pinned to the boundary of $\mathrm{AdS}_3$. These insertions are then mapped to (generally different) vertex operators which, in turn, are dual to the local spacetime fields arising from the action of the corresponding spacetime defect. The calculation of spacetime CFT correlators involving invertible topological defect loops can therefore be reduced to the calculation of amplitudes of closed strings propagating on the `standard' global $\mathrm{AdS}_3\times \mathrm{S}^3\times \mathbb{T}^4$ background at $k=1$ unit of NS-NS flux.

On the other hand, as soon as we have to consider spacetime CFT correlators involving topological defect lines terminating at punctures with non-local defect-ending field insertions, the calculation cannot generally be reduced to computing amplitudes involving closed-string excitations in the usual $k=1$ global $\mathrm{AdS}_3\times \mathrm{S}^3\times \mathbb{T}^4$ background, as these can only give rise to \emph{local} spacetime single-particle states. As discussed in Section \ref{subsec:sphereDual}, such a situation may for instance arise when placing non-invertible maximally-fractional defect loops into correlators of local spacetime fields and attempting to shrink these around punctures. The corresponding tensionless brane-like duals in the bulk then become supported at the points on the conformal boundary where the spacetime defect-ending fields are being inserted, which prevents them from collapsing. They give rise to modified spectra of closed strings whose partition functions we will now determine.

\subsubsection{Worldsheet partition functions for spacetime defects}

For the sake of illustration, let us first write down explicit proposals for the modular-invariant `global' worldsheet CFT partition functions describing closed strings in the presence of two specific classes of spacetime topological defects. When restricted to the on-shell states, these will be seen to give rise to the expected single-particle spectra of defect-ending fields in the spacetime CFT.

\subsubsection*{\boldmath Shift-symmetry defects in \texorpdfstring{$\mathbb{T}^4$}{T4}}

Let us start by exhibiting the modular-invariant worldsheet partition function which we claim to encode the worldsheet counterparts of the defect-ending fields for the maximally-fractional defect $\boldsymbol{\mathcal{L}}_0^{(U,N)}$ in the $\mathrm{Sym}^N(\mathbb{T}^4)$ (see Section \ref{subsec:ShiftT4} for the notation). It reads
\begin{align}
    &Z^{(U,N)}(t,z;\tau)=\nonumber\\
    &\hspace{0.4cm}=\frac{1}{|\eta(\tau)|^4}\sum_{w\in\mathbb{Z}}\,\int_{0}^{1}\mathrm{d}\lambda\,\sum_{n,u\in \mathbb{Z}^{\otimes 4}}\,\chi_{\lambda+\Delta\lambda(n,u;w) }^{w}(t,z;\tau)\,\chi_{\lambda}^{w}(t,z;\tau)^\ast\, K_{0,w}^{(U,N)}(n,u;\tau)\,,\label{eq:ZUN}
\end{align}
where we have denoted
\begin{align}
 \Delta\lambda(n,u;w)=   w N^T U+n^T U+N^Tu\,.\label{eq:DeltaLambda}
\end{align}
Also recall the definition \eqref{eq:DefectKernel} of the defect kernel $K_{r,s}^{(U,N)}(n,u;\tau)$. As a consistency check, one can set $U=N=0$ to recover the diagonal partition function \eqref{eq:WSPartitionFunction}.

Notice that while the partition function \eqref{eq:ZUN} consistently decomposes as a positive linear combination of the holomorphic and anti-holomorphic characters of the worldsheet CFT chiral algebra \eqref{eq:worldsheet-theory}, it can no longer be factorized into the $\mathfrak{psu}(1,1|2)_1$ part and the topologically-twisted $\mathbb{T}^4$ part, as was the case for the global $\mathrm{AdS}_3$ partition function \eqref{eq:WSPartitionFunction}. Also note that while the claimed modular invariance of the partition function $Z^{(U,N)}(t,z;\tau)$ is far from manifest at the level of the expression \eqref{eq:ZUN}, it can be made crystal clear by recasting \eqref{eq:ZUN} as
\begin{align}
    Z^{(U,N)}(t,z;\tau) = \sum_{w,r\in \mathbb{Z}}Z_{r,w}(t;\tau)\, (Z_{\mathbb{T}^4})_{r,w}^{(U,N)}\begin{bmatrix}\frac{r}{2} \\ \frac{w}{2} \end{bmatrix}(z;\tau)\,.\label{eq:ZUNblocks}
\end{align}
This is very similar to the expression \eqref{eq:Zblocks} except now the $\mathbb{T}^4$ torus correlator contains insertion of the shift-symmetry defect $\mathcal{L}^{(U,N)}$ which winds $r$ times around the spatial cycle and $w$ times around the temporal cycle. This does not spoil its modular properties since the winding numbers $(r,w)$ transform as a doublet under $\mathrm{PSL}(2;\mathbb{Z})$. Indeed, in Section \ref{subsec:path_integral} we will observe how the mechanism giving rise to \eqref{eq:ZUNblocks} can be interpreted in terms of gauging a tensor product of a topological defect in the $\mathfrak{psu}(1,1|2)_1$ WZW model with the defect $\mathcal{L}^{(U,N)}$ in the $\mathbb{T}^4$ worldsheet CFT.

In order to extract the spectrum of on-shell closed strings out of \eqref{eq:ZUN}, one can first read off the left and the right-moving worldsheet conformal  weighths as
\begin{subequations}
    \begin{align}
        h&=\frac{w^2}{2}-\big[m+\Delta\lambda(n,u;w)\big]w +h_\mathrm{L}\big(n+wN,u+wU\big) +M\,,\\
        \bar{h}&=\frac{w^2}{2}-\bar{m}w +h_\mathrm{R}\big(n+wN,u+wU\big) +\bar{M}\,,
    \end{align}
\end{subequations}
where $m,\bar{m}\in\mathbb{Z}+\lambda$. Substituting from \eqref{eq:DeltaLambda} and recalling the result \eqref{eq:spinT4}, one finds that the level-matching constraint $h-\bar{h}=0$ implies
\begin{align}
    n^Tu +M-\bar{M}\in w \mathbb{Z}\,,
\end{align}
which reproduces the projection \eqref{eq:SpinContraintT4} in the spacetime torus partition function for the $\smash{\boldsymbol{\mathcal{L}}_0^{(U,N)}}$ defect-ending fields. Furthermore, solving the mass-shell constraints $h=\bar{h}=0$ for $m$ and $\bar{m}$ and substituting back into \eqref{eq:ZUN}, we obtain the partition function of on-shell closed strings
\begin{align}
      Z^{(U,N),\text{on-shell}}(z;t)= \sum_{w=1}^{\infty}
  (Z_{\mathbb{T}^4}')_{0,w}^{(U,N)}
\begin{bmatrix}0\\\frac{w}{2}\end{bmatrix}\left(z;\frac{t}{w}\right)\,,
\end{align}
which exactly matches the single-particle partition function for the $\smash{\boldsymbol{\mathcal{L}}_0^{(U,N)}}$ defect-ending fields in the twisted NS-NS sector (cf.\ \eqref{eq:spSpinT4}).

\subsubsection*{Sign-defect}

Let us continue with writing down the modular-invariant partition function which is supposed to encode the closed-string vertex operators which are dual to the defect-ending fields for the sign defect $\boldsymbol{\mathcal{L}}_\mathrm{sign}$. This should be possible by making modifications to the global $\mathrm{AdS}_3$ partition function \eqref{eq:WSPartitionFunction} purely within the $\mathfrak{psu}(1,1|2)_1$ sector, as the defect acts trivially on the $\mathbb{T}^4$ states. Our proposal reads
\begin{align}
    Z_\mathrm{sign}(t,z;\tau)&=\frac{\Theta(\tau)}{|\eta(\tau)|^4}\sum_{w\in\mathbb{Z}}\,\int_{0}^{1}\mathrm{d}\lambda\,(-1)^w\,\chi_{\lambda+\frac{1}{2}(w+1)}^{w}(t,z;\tau)\,\chi_{\lambda}^{w}(t,z;\tau)^\ast\,,\label{eq:ZsignWS}
\end{align}
where the signs $(-1)^w$ dictate that we should treat the states with odd spectral flow parameter $w$ as spacetime fermions -- in exact agreement with the expected statistics of the defect-ending fields of $\boldsymbol{\mathcal{L}}_\mathrm{sign}$ as discussed in Sections \ref{subsec:cyclic} and \ref{subsec:single_particle}. This time, the modular invariance of \eqref{eq:ZsignWS} can be readily checked by directly using the transformation property \eqref{eq:SMod} of the $\mathfrak{psu}(1,1|2)_1$ characters. Nevertheless, it is still beneficial to rewrite the partition function in the form 
\begin{align}
Z_\mathrm{sign}(t,z;\tau)&=\sum_{w,r\in \mathbb{Z}}e(r,w) \, Z_{r,w}(t;\tau)\,Z_{\mathbb{T}^4}\begin{bmatrix}\frac{r}{2} \\ \frac{w}{2} \end{bmatrix}(z;\tau)\,.\label{eq:ZsignBlocks}
\end{align}
This differs from the expression \eqref{eq:Zblocks} for the standard global $\mathrm{AdS}_3$ worldsheet partition function by the appearance of the sign factors
\begin{align}
    e(r,w)=(-1)^{rw+r+w}\,.\label{eq:eSign}
\end{align}
It is not hard to check that these satisfy the modular properties
\begin{subequations}
\label{eq:e_modular}
    \begin{align}
        e(-w,r)&=e(r,w)\,,\\
        e(r+w,w)&=e(r,w)\,,
    \end{align}
\end{subequations}
so that \eqref{eq:ZsignBlocks} is indeed modular invariant, guaranteeing one-loop consistency of the worldsheet partition function. It is interesting to note that $e(r,w)$ fails to satisfy the properties of discrete torsion, which would have guaranteed consistency at all loops \cite{Vafa:1986wx,Vafa:1994rv}. This should not worry us too much as it will turn out that the worldsheet calculation reproduces the entire spacetime partition function already at one loop (so that contributions from higher-loop worldsheet amplitudes vanish).

The total left- and right-moving worldsheet conformal dimensions can be extracted from \eqref{eq:ZsignWS} as
\begin{subequations}
    \begin{align}
        h&=\frac{w^2}{2}-\bigg[m+\frac{1}{2}(w+1)\bigg]w +h_\mathrm{L} +M\,,\\
        \bar{h}&=\frac{w^2}{2}-\bar{m}w +h_\mathrm{R} +\bar{M}\,.
    \end{align}
\end{subequations}
The level-matching condition therefore yields the requirement 
\begin{align}
    h_\mathrm{L}-h_\mathrm{R} +M-\bar{M}-\frac{1}{2}(w+1)w \in w\mathbb{Z}\,,\label{eq:sign_proj}
\end{align}
which agrees with the constraint \eqref{eq:epsProj} following from the cyclic projection \eqref{eq:spproj} in the defect CFT of the maximally-fractional defect $\boldsymbol{\mathcal{L}}_\mathrm{sign}$. Indeed, repeating once more the by-now-familiar procedure of putting the closed strings on shell, one ends up with the partition function
\begin{align}
     Z_\mathrm{sign}^\text{on-shell}(z;t) = \sum_{w=1}^{\infty} Z'_{\mathbb{T}^4}\begin{bmatrix}0\\\frac{w}{2}\end{bmatrix}\left(z;\frac{t}{w}\right)\label{eq:ZsignOnShell}
\end{align}
where we have to remember that here $'$ imposes the projection \eqref{eq:sign_proj} on the local seed $\mathbb{T}^4$ Hilbert space. The partition function \eqref{eq:ZsignOnShell} of the on-shell closed strings propagating on the background given by \eqref{eq:ZsignWS} therefore exactly matches the twisted NS-NS single-cycle partition function of the $\boldsymbol{\mathcal{L}}_\mathrm{sign}$ defect-ending fields.

\subsubsection{Thermal backgrounds and generalized orbifolds}
\label{subsec:path_integral}

Having identified the spectra of closed-string states which are dual to the defect-ending fields for some concrete realizations of maximally-fractional defects in the $\mathrm{Sym}^N(\mathbb{T}^4)$ theory, let us proceed with the calculation of the complete vacuum torus amplitude for closed strings propagating on a thermal background in the spirit of \cite{Eberhardt:2020bgq}. We will see that defining this thermal background as an orbifold which arises by gauging suitable defects in the $\mathfrak{psu}(1,1|2)_1$ CFT tensored with a general defect $\mathcal{L}$ in the $\mathbb{T}^4$ worldsheet CFT, the torus amplitudes evaluate precisely to the grandcanonical torus correlators of the corresponding maximally-fractional topological defect $\boldsymbol{\mathcal{L}}_0$ in the spacetime CFT.

In the $k=1$ theory on $\text{AdS}_3\times\text{S}^3\times\mathbb{T}^4$, the worldsheet path integral is dominated entirely by strings which live arbitrarily close to the conformal boundary of $\text{AdS}_3$ \cite{Eberhardt:2018ouy,Eberhardt:2019ywk,Dei:2020zui}. For these string configurations, the worldsheet field $\gamma$ defined above provides a map from the worldsheet $\Sigma$ to the boundary of $\text{AdS}_3$. By the equations of motion for $\beta,\bar\beta$, this map is holomorphic, and so the worldsheet holomorphically `wraps' the $\text{AdS}_3$ boundary a certain number of times. In other words, the moduli-space integrals defining worldsheet amplitudes localize precisely at those points where these holomorphic covering maps from the worldsheet to the $\text{AdS}_3$ boundary exist. 

We will focus on the case of thermal $\text{AdS}_3$, where the conformal boundary is a torus, as the vacuum amplitude calculation should then coincide with the spacetime CFT calculations of torus correlators we considered in Section \ref{sec:sym-defects}. If we take the worldsheet to be a torus as well, then the worldsheet theory is a theory of maps $\gamma$ from the worldsheet torus to the boundary torus. We shall see that this worldsheet topology already provides the complete answer expected from the $\mathrm{Sym}^N(\mathbb{T}^4)$ calculation, thus showing that the worldsheet genus expansion in this theory truncates at genus one \cite{Eberhardt:2020bgq,Eberhardt:2021jvj}. 

Now, let $\mathcal{L}$ be a topological defect in the $\mathbb{T}^4$ CFT, and consider a configuration such that, at the asymptotic boundary of $\text{AdS}_3$, the defect $\mathcal{L}$ wraps the A-cycle (or spatial cycle) $\alpha$ times and the B-cycle (or thermal cycle) $\beta$ times. In the bulk theory, since the strings are glued to the asymptotic boundary of $\text{AdS}_3$, the worldsheet CFT will include the topological defect $\mathcal{L}$ at the preimage of its location on the boundary under the map $\gamma$. Since the map $\gamma$ can have different winding numbers around the different cycles of the $\text{AdS}_3$ boundary torus, the exact defect configuration on the worldsheet depends sensitively on the topological details of the map $\gamma$.

\begin{figure}
\centering
\begin{tikzpicture}[scale = 0.9]
\begin{scope}[xshift = -4cm]
\draw[thick] (0,0) [partial ellipse = 0:360:1.7 and 2.7];
\draw[thick] (0.25,0) [partial ellipse = 100:260:0.8 and 1.3];
\draw[thick] (-0.15,0) [partial ellipse = -83:83:0.7 and 1.2];
\draw[thick, purple] (-1.125,0) [partial ellipse = 180:360:0.575 and 0.125];
\draw[thick, dashed, purple] (-1.125,0) [partial ellipse = 180:0:0.575 and 0.125];
\draw[thick, blue] (0,0) [partial ellipse = 0:360:1.1 and 2.1];
\end{scope}
\begin{scope}[xshift = 4cm]
\draw[very thick, latex-latex] (-3.2,0) -- (3.2,0);
\draw[very thick, latex-latex] (0,-3) -- (0,3);
\draw[thick, purple] (0,0) circle (0.125);
\draw[thick, purple] (0,0) circle (0.25);
\draw[thick, purple] (0,0) circle (0.5);
\draw[thick, purple] (0,0) circle (1);
\draw[thick, purple] (0,0) circle (2);
\draw[thick, blue] (0,0) -- (2.9,2.9);
\fill[white] (2.25,2.25) -- (3,2.25) -- (3,3) -- (2.25,3) -- (2.25,2.25);
\draw[thick] (3,2.25) -- (2.25,2.25) -- (2.25,3);
\node at (2.625,2.625) {$\gamma$};
\end{scope}
\end{tikzpicture}
\caption{Left: Topological defects \textcolor{purple}{$\mathcal{L}^\alpha$} and \textcolor{blue}{$\mathcal{L}^\beta$} wrapping the spatial and thermal cycle, respectively, near the boundary of thermal $\text{AdS}_3$. Right: The support of the defects on the boundary of global $\text{AdS}_3$ before orbifolding.}
\label{fig:unfolding}
\end{figure}
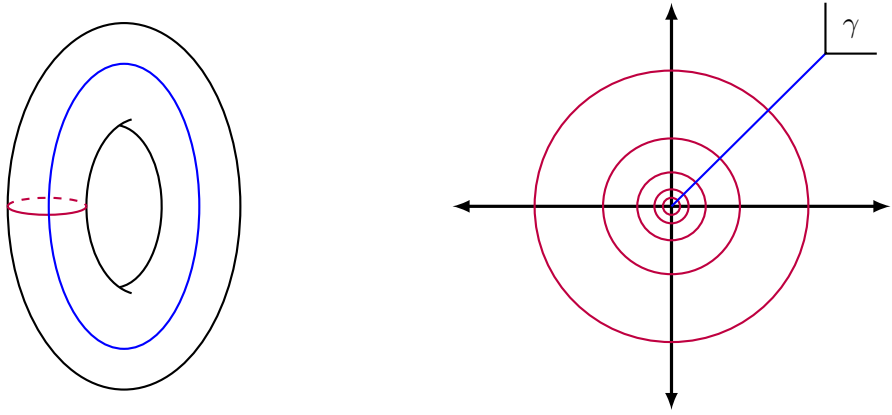

\subsubsection*{\boldmath Thermal $\mathrm{AdS}_3$ as a double orbifold}

One way to implement the various winding sectors is to write the target space as an orbifold. Fundamentally, we are interested in computing a worldsheet partition function on thermal $\text{AdS}_3$. Without the defect, this is a smooth bulk geometry obtained from global $\text{AdS}_3$ by orbifolding the subgroup $\mathbb{Z}\subset\text{SL}(2,\mathbb{C})$ generated by the element
\begin{equation}
e^{2\pi iJ^3_0}=\begin{pmatrix}
x^{1/2} & 0\\ 0 & x^{-1/2}
\end{pmatrix}\in\text{SL}(2,\mathbb{C})
\end{equation}
with $x=e^{2\pi it}$ \cite{Maloney:2007ud}. The worldsheet theory on thermal $\text{AdS}_3$ can then be considered as a $\mathbb{Z}$ orbifold of the worldsheet theory on global $\text{AdS}_3$.

In the case of the superstring on $\text{AdS}_3\times\text{S}^3\times\mathbb{T}^4$, it is actually necessary to orbifold by the $\mathrm{PSU}(1,1|2)$ group element
\begin{equation}
g=e^{2\pi itJ^3_0}e^{-2\pi i\bar{t}\bar{J}^3_0}e^{2\pi izK^3_0}e^{-2\pi i\bar{z}\bar{K}^3_0}\,,
\end{equation}
in order to account for the (spacetime) $\text{U}(1)$ R-symmetry chemical potential. This group element acts on the various worldsheet free fields as
\begin{equation}
\begin{split}
g\cdot\gamma=x^{-1}\gamma\,,&\quad g\cdot\beta=x\beta\,,\\
g\cdot p_1=y^{1/2}p_1\,,&\quad g\cdot\theta^1=y^{-1/2}\theta^1\,,\\
g\cdot p_2=y^{-1/2}p_2\,,&\quad g\cdot\theta^2=y^{1/2}\theta^2\,.
\end{split}
\end{equation}
Here we have defined the fugacity $y=e^{2\pi iz}$. 

Computing the thermal $\mathrm{AdS}_3$ partition function is now done in two steps, which we summarize here. First, we realize that we have to compute the traces
\begin{equation}
\text{Tr}_{\mathcal{H}_\mathrm{PSU}(g^c)}\big[g^{a}q^{L_0}\bar{q}^{L_0}\big]\,,
\end{equation}
where $\mathcal{H}_\mathrm{PSU}(g^c)$ denotes the worldsheet Hilbert space \eqref{eq:HPSU} of the theory on global $\mathrm{AdS}_3$ twisted by $g^c$.
This is simply expressed in terms of $\mathfrak{psu}(1,1|2)_1$ characters \eqref{eq:psu_characters}. 
Now, the Hilbert space of the $\mathfrak{psu}(1,1|2)_1$ WZW model describing the global $\mathrm{AdS}_3$ is composed of unflowed and spectrally-flowed sectors. In the $w$ spectrally-flowed sector, the string winds around the spatial cycle of the  $\text{AdS}_3$ space $w$ times. In terms of boundary conditions for the worldsheet fields, the $w$ spectrally-flowed sector can be defined by the boundary conditions
\begin{equation}
\begin{split}
\gamma(z+1)=e^{2\pi iw}\gamma(z)\,,&\quad \beta(z+1)=e^{-2\pi iw}\beta(z)\,,\\
\theta^1(z+1)=e^{2\pi iw}\theta^1(z)\,,&\quad p_1(z+1)=e^{2\pi iw}p_1(z)\,,
\end{split}
\end{equation}
where by $e^{2\pi iw}\gamma$, we mean the coordinate $\gamma$ transported continuously around the point $\gamma=0$ $w$ times. In terms of the mode expansion for the fields, this set of boundary conditions is equivalent to the standard definition of spectral flow in $\text{AdS}_3$ \cite{Maldacena:2000hw}. Spectral flow can thus be thought of as the sector of the worldsheet theory which is twisted by $w$ copies of the $\text{PSU}(1,1|2)$ group element\footnote{An analogous observation was made in \cite{Knighton:2023mhq} in the context of bosonic strings in $\text{AdS}_3$.}
\begin{equation}
h=e^{-2\pi iJ^3_0}e^{+2\pi i\bar{J}^3_0}e^{-2\pi iK^3_0}e^{2\pi i\bar{K}^3_0}\,.
\end{equation}
This identification suggests that we can think of spectrally-flowed sectors in the $\mathfrak{psu}(1,1|2)_1$ WZW model as twisted sectors in a second $\mathbb{Z}$ orbifold implemented by the element generated by $h$. As we will explicitly see below, this is indeed the case.

For the case of topological defects, however, a slight modification is needed. Let us consider the case for which there is a defect $\mathcal{L}$ of the $\mathbb{T}^4$ theory which wraps $\alpha$ times around the thermal cycle and $\beta$ times around the spatial cycle of thermal $\text{AdS}_3$. Since we make the identification $\gamma\sim x^{-1}\gamma$, the pre-orbifolded theory must also include defect branes at circles of radii $|x|^{m}$ for all $m\in\mathbb{Z}$, as shown in Figure \ref{fig:unfolding}. The defect wrapping the thermal cycle is now represented as a ray extending from the origin in the $\gamma$-plane.

Now, any operator $\mathcal{O}$ in the worldsheet theory which is acted on by the $\text{PSU}(1,1|2)$ matrix $g$ must also pass through one of these defect branes in the $\gamma$ plane. Thus, the `natural' orbifold is that which is generated not by $g$, but by the combination
\begin{equation}
\mathcal{O}\to (g\otimes\mathcal{L}^{\beta})\,\mathcal{O}\,.
\end{equation}
Similarly, if an operator is transported around the origin in the $\gamma$ plane without crossing one of the red defect lines, it will cross the blue one. Thus, the $w$ spectrally-flowed sector is now no longer twisted by the element $h$, but rather satisfies the twisted boundary conditions
\begin{equation}
\mathcal{O}(z+1)=(h\otimes \mathcal{L}^{\alpha})^w\,\mathcal{O}(z)\,.
\end{equation}
That is, it is not just twisted by the operator $h$, but also by the defect operator $\mathcal{L}$.

\subsubsection*{Spherical defects}

The natural setting with which to describe this setup is that of (generalized) orbifolds implemented by gauging topological defects \cite{Frohlich:2009gb,Bhardwaj:2017xup,Chang:2018iay,Thorngren:2019iar,Thorngren:2021yso} in the worldsheet theory. Specifically, given a defect $\mathcal{L}$ in the $\mathbb{T}^4$ CFT, let us define two worldsheet defect operators $\mathcal{L}_1,\mathcal{L}_2$ by taking
\begin{equation}
\mathcal{L}_1=\mathcal{L}_1^{\mathfrak{psu}}\otimes\mathcal{L}^\alpha\,,\quad\mathcal{L}_2=\mathcal{L}_2^{\mathfrak{psu}}\otimes\mathcal{L}^\beta\,,
\end{equation}
where the defects $\mathcal{L}_1^{\mathfrak{psu}}$ and $\mathcal{L}_2^{\mathfrak{psu}}$ in the $\mathfrak{psu}(1,1|2)_1$ WZW model can be realized in terms of explicit operators acting on the states in the $\mathfrak{psu}(1,1|2)_1$ WZW model, namely
\begin{subequations}
\begin{align}
\mathcal{L}_1^{\mathfrak{psu}} &\equiv g=
e^{+2\pi i t J_0^3}e^{-2\pi i \bar{t} \bar{J}_0^3}e^{+2\pi i z K_0^3}e^{-2\pi i\bar{z} \bar{K}_0^3}\,,\label{eq:Lpsu1}\\
\mathcal{L}_2^{\mathfrak{psu}} &\equiv h={{ e^{-2\pi i J_0^3} 
e^{+2\pi i \bar{J}_0^3}e^{-2\pi i K_0^3}  e^{+2\pi i \bar{K}_0^3}}}\,.\label{eq:Lpsu2}
\end{align}
\end{subequations}
Here, $t$ should be thought of as the size of the thermal cycle, that is, the modular parameter of the \emph{spacetime} torus. On the other hand, $z$ will give rise to the chemical potential of the spacetime $\mathbb{T}^4$ fermions. In the case of the defect $\mathcal{L}_1^{\mathfrak{psu}}$, the $\mathfrak{psu}(1,1|2)_1$ chiral currents satisfy the twisted gluing conditions
\begin{subequations}
\label{eq:tr_psu_tw}
    \begin{align}
        J^{3,(1)}(z) &={J}^{3,(2)}({z})\,,\\
        J^{\pm,(1)}(z) &= e^{\pm 2\pi i t} {J}^{\pm,(2)}({z})\,,\\
        K^{3,(1)}(z) &={K}^{3,(2)}({z})\,,\\
        K^{\pm,(1)}(z) &=e^{\pm 2\pi i z} {K}^{\pm,(2)}({z})\,,\\
         S^{\lambda\mu\nu,(1)}(z) &=e^{ 2\pi i  \lambda t} e^{ 2\pi i  \mu z}{S}^{\lambda\mu\nu,(2)}({z})\,,
    \end{align}
\end{subequations}
across the codimension one interface given by the defect, while for the defect $\mathcal{L}_2^{\mathfrak{psu}}$, the corresponding gluing conditions read
\begin{subequations}
\label{eq:tr_psu}
    \begin{align}
        J^{a,(1)}(z) &={J}^{a,(2)}({z})\,,\\
        K^{a,(1)}(z) &={K}^{a,(2)}({z})\,,\\
         S^{\lambda\mu\nu,(1)}(z) &={S}^{\lambda\mu\nu,(2)}({z})\,.
    \end{align}
\end{subequations}
Drawing an analogy with the case of the conformal boundary conditions in the $\mathfrak{psu}(1,1|2)_1$ model, which were analyzed in detail in \cite{Gaberdiel:2021kkp}, the defects $\mathcal{L}_1^{\mathfrak{psu}}$ and $\mathcal{L}_2^{\mathfrak{psu}}$ may therefore be called \emph{(twisted) spherical defects}. Checking that $\mathcal{L}_1^{\mathfrak{psu}}$ and $\mathcal{L}_2^{\mathfrak{psu}}$, as given by the operators \eqref{eq:Lpsu1} and \eqref{eq:Lpsu2}, satisfy Cardy consistency at one loop proceeds rather uneventfully.

\subsubsection*{The torus amplitude}

We claim that the worldsheet dual to the symmetric orbifold partition function $\mathfrak{Z}_{\alpha,\beta}$ calculated in \eqref{eq:ZabGrand} is found by taking the defects $\mathcal{L}_1,\mathcal{L}_2$ to define a (generalized) $\mathbb{Z}\oplus\mathbb{Z}$ orbifold. To this end, we need to sum over all possible `twisted sectors' of the worldsheet theory. Let us take
\begin{equation}
\begin{tikzpicture}[baseline={([yshift=0.25cm]current bounding box.center)}]
\draw[thick] (0,0) -- (0,1) -- (1,1) -- (1,0) -- (0,0);
\node[below] at (0.5,0) {\footnotesize $(c,d)$};
\node[left] at (0,0.5) {\footnotesize $(a,b)$};
\end{tikzpicture}
\end{equation}
to denote the string worldsheet torus correlator where the defect $\mathcal{L}_1$ wraps $a$ times the spatial cycle and $c$ times the temporal cycle, while the defect $\mathcal{L}_2$ wraps $b$ times the spatial cycle and $d$ times the temporal cycle.

Our goal is to compute the amplitude
\begin{equation}
\int_{\mathcal{F}}\frac{\mathrm{d}^2\tau}{\mathrm{Im}\,\tau}\,Z^\text{thermal}_{\alpha,\beta}(t,z;\tau)\,,\label{eq:ModularIntegral}
\end{equation}
where we have introduced the modular-invariant thermal worldsheet partition function
\begin{align}
  Z^\text{thermal}_{\alpha,\beta}=  \sum_{a,b,c,d\in\mathbb{Z}}
e^{\pi (ad-bc)\mathrm{Im}\,t}p^N
\begin{tikzpicture}[baseline={([yshift=0.25cm]current bounding box.center)}]
\draw[thick] (0,0) -- (0,1) -- (1,1) -- (1,0) -- (0,0);
\node[below] at (0.5,0) {\footnotesize $(c,d)$};
\node[left] at (0,0.5) {\footnotesize $(a,b)$};
\end{tikzpicture}\ \,.\label{eq:ZthermalAB}
\end{align}
Note that in \eqref{eq:ZthermalAB}, we have also accounted for the contribution from the sphere partition function and introduced fugacity $p$ for the number of fundamental strings $N=ad-bc$ as in \cite{Eberhardt:2020bgq}.
This describes closed strings propagating on a thermal background which is modified by the presence of the spacetime defect $\mathcal{L}$ winding $(\alpha,\beta)$ times around the conformal boundary. 
In the case that $\mathcal{L}$ is non-invertible, we define $\mathcal{L}^{-1}\equiv \mathcal{L}^{\dagger}$.

Now it is our task to compute the partition function $Z^\text{thermal}_{\alpha,\beta}$.
Let us first focus on a fixed twisted sector labelled by integers $a,b,c,d$. For the $\mathfrak{psu}(1,1|2)_1$ fermions, 
the partition function is easily expressed in terms of $\vartheta$-functions as
\begin{equation}
\left|\frac{1}{\eta(\tau)^2}\,\vartheta\begin{bmatrix}a(t+z)/2\\ c(t+z)/2\end{bmatrix}\left(0;\tau\right)\,\vartheta\begin{bmatrix}a(t-z)/2\\ c(t-z)/2\end{bmatrix}\left(0;\tau\right)\right|^2\,,
\end{equation}
The entries of the theta functions determine the twisted boundary conditions of the worldsheet fermions $p_a,\theta^a$. Specifically, the first theta function is the expression for the $p_1,\theta^1$ system, while the second is for the $p_2,\theta^2$ system. Here we use the conventions for $\vartheta$-functions defined below \eqref{eq:4fermion_part}. We also recall the expression \eqref{eq:Zghost} for the partition function of the $\rho\sigma$ ghost system, which we write as
\begin{equation}
Z_{\rho\sigma}(\tau)=\left|\frac{\eta(\tau)^4}{\vartheta\SmallMatrix{ 0\\ 0}(0;\tau)}\right|^2\,.
\end{equation}
At the same time, the defect torus correlation function of the defect $\mathcal{L}$ in the $(a,b,c,d)$-twisted sector of the topologically-twisted $\mathbb{T}^4$ CFT reads
\begin{equation}
(Z_{\mathbb{T}^4})_{a\alpha+b\beta,c\alpha+d\beta}\begin{bmatrix}0 \\ 0\end{bmatrix}(0;\tau)\,.
\end{equation}
Finally, noting the dependence of the spherical defects $\mathcal{L}_1^{\mathfrak{psu}}$ and $\mathcal{L}_2^{\mathfrak{psu}}$ on the $J_0^3$ current, the bosonic part of the $\mathfrak{psu}(1,1|2)_1$ model contributes with (cf.\ also eq.\ (3.23) of \cite{Eberhardt:2020bgq})
    \begin{align}
&\frac{1}{2}\frac{\mathrm{Im}\,t}{|\eta(\tau)^2|^2}
\big|e^{{i\pi}t(ad-bc)}\big|^2
\delta^{(2)}\left(at-b+\tau(ct-d)\right)=\nonumber\\
&\hspace{4cm}=\frac{1}{2}\frac{1}{|ct-d|^2}\frac{\mathrm{Im}\,t}{|\eta(\tau)^2|^2}e^{-2\pi(ad-bc)\mathrm{Im}\,t}\,\delta^{(2)}\left(\tau-\frac{at-b}{-ct+d}\right)\,.
\end{align}
It is now manifest that the torus amplitude $\delta$-function localizes precisely at those values of $\tau$ where the worldsheet holomorphically wraps the boundary (spacetime) torus. We have also remembered to insert the effective inverse volume $\mathrm{Im}\,t$ or the orbifold group generated by the worldsheet defect $\mathcal{L}_1$.

Putting all ingredients together, we obtain that the twisted sector labelled by $a,b,c,d$ contributes to the thermal partition function $Z_{\alpha,\beta}^\mathrm{thermal}$ as
\begin{align}
&
\begin{tikzpicture}[baseline={([yshift=0.25cm]current bounding box.center)}]
\draw[thick] (0,0) -- (0,1) -- (1,1) -- (1,0) -- (0,0);
\node[below] at (0.5,0) {\footnotesize $(c,d)$};
\node[left] at (0,0.5) {\footnotesize $(a,b)$};
\end{tikzpicture}\ (t,z;\tau)\,=\nonumber\\
&\hspace{1cm}=\frac{\mathrm{Im}\,t}{|ct-d|^2|\vartheta\SmallMatrix{0\\0}(0;\tau)|^2}(Z_{\mathbb{T}^4})_{a\alpha+b\beta,c\alpha+d\beta}\begin{bmatrix}0\\0\end{bmatrix}(0;\tau)\,\,\delta^{(2)}\left(\tau-\frac{at-b}{-ct+d}\right)\times\nonumber\\
&\hspace{1.4cm}\times e^{-2\pi(ad-bc)\mathrm{Im}\,t}\left|\,\vartheta\begin{bmatrix}a(t+z)/2\\ c(t+z)/2\end{bmatrix}\left(0;\tau\right)\,\vartheta\begin{bmatrix}a(t-z)/2\\ c(t-z)/2\end{bmatrix}\left(0;\tau\right)\right|^2\,.\label{eq:TorusIntegrand}
\end{align}
One can readily check that the r.h.s.\ of \eqref{eq:TorusIntegrand} transforms under $\text{PSL}(2;\mathbb{Z})$ as
\begin{align}
    \begin{tikzpicture}[baseline={([yshift=0.25cm]current bounding box.center)}]
\draw[thick] (0,0) -- (0,1) -- (1,1) -- (1,0) -- (0,0);
\node[below] at (0.5,0) {\footnotesize $(c,d)$};
\node[left] at (0,0.5) {\footnotesize $(a,b)$};
\end{tikzpicture}\ \left( t,z;\frac{A\tau+B}{C\tau+D}\right) = |C\tau +D|^2\ \begin{tikzpicture}[baseline={([yshift=0.25cm]current bounding box.center)}]
\draw[thick] (0,0) -- (0,1) -- (1,1) -- (1,0) -- (0,0);
\node[below] at (0.5,0) {\footnotesize $(Ca+Dc,Cb+Dd)$};
\node[left] at (0,0.5) {\footnotesize $(Aa+Bc,Ab+Bd)$};
\end{tikzpicture} ( t,z;\tau)\,.
\end{align}
Hence, upon summing \eqref{eq:TorusIntegrand} over $a,b,c,d\in\mathbb{Z}$, we verify that the thermal partition function $Z_{\alpha,\beta}^\mathrm{thermal}$ is modular invariant (as usual, up to a factor which cancels with the transformation of the integration measure in \eqref{eq:ModularIntegral}) and can therefore be meaningfully integrated over the fundamental domain. 

A standard trick is to trade the sum over $a,b,c,d$ and the integral over the fundamental domain $\mathcal{F}$ for a sum over $a,d\in\mathbb{Z}$, $b=0,\ldots,d-1$ with $c=0$ and an integral over the full upper-half-plane $\mathbb{H}$. We can also use the symmetry $(a,b,c,d)\to (-a,-b,-c,-d)$ to restrict the sum over $d$ to start from $d=1$ (excluding the case $c=d=0$ as it does not give rise to a holomorphic covering).
Setting $c=0$ simplifies the above expression considerably, since we can write
\begin{equation}
\begin{split}
&\vartheta\begin{bmatrix}a(t+z)/2\\ c(t+z)/2\end{bmatrix}\left(0;\tau\right)\,\vartheta\begin{bmatrix}a(t-z)/2\\ c(t-z)/2\end{bmatrix}\left(0;\tau\right)=\\
&\hspace{3cm}\stackrel{c\to 0}{=}\vartheta\begin{bmatrix}0 \\ 0\end{bmatrix}\left(\frac{a(t+z)}{2};\tau\right)\vartheta\begin{bmatrix}0 \\ 0\end{bmatrix}\left(\frac{a(t-z)}{2};\tau\right)\,.
\end{split}
\end{equation}
Furthermore, the delta function in the string partition function allows us to set $t=(d\tau+b)/a$, so that the theta function combination becomes
\begin{equation}
\begin{split}
&\vartheta\begin{bmatrix}0 \\ 0\end{bmatrix}\left(\frac{az}{2}+\frac{d\tau}{2}+\frac{b}{2};\tau\right)\vartheta\begin{bmatrix}0 \\ 0\end{bmatrix}\left(-\frac{az}{2}+\frac{d\tau}{2}+\frac{b}{2};\tau\right)=\\
&\hspace{4.5cm}=|q|^{-\frac{d^2}{2}}\vartheta\begin{bmatrix}\frac{b}{2}\\ \frac{d}{2}\end{bmatrix}\left(\frac{az}{2};\tau\right)\vartheta\begin{bmatrix}\frac{b}{2}\\ \frac{d}{2}\end{bmatrix}\left(-\frac{az}{2};\tau\right)\,,
\end{split}
\end{equation}
where in the second line we have used standard theta function identities relating half-lattice shifts in the chemical potential to change in spin structure. Using this result, we can write the $(a,b,c,d)$ sector partition function \eqref{eq:TorusIntegrand} as
\begin{equation}\label{eq:string-torus-nearly-final}
\begin{tikzpicture}[baseline={([yshift=0.25cm]current bounding box.center)}]
\draw[thick] (0,0) -- (0,1) -- (1,1) -- (1,0) -- (0,0);
\node[below] at (0.5,0) {\footnotesize $(0,d)$};
\node[left] at (0,0.5) {\footnotesize $(a,b)$};
\end{tikzpicture}\ (t,z;\tau)=\frac{\text{Im}t}{d^2}e^{-\pi ad\,\text{Im}t}(Z_{\mathbb{T}^4})_{a\alpha+b\beta,d\beta}\begin{bmatrix}\frac{b}{2}\\\frac{d}{2}\end{bmatrix}(az;\tau)\delta^{(2)}\left(\tau-\frac{at-b}{d}\right)\,.
\end{equation}
Here, we have used the fact that $|q|^{-\frac{d^2}{2}}=e^{\pi ad\,\text{Im}\,t}$ in the presence of the delta function. We have also used the cancellation between the factor of $|\vartheta\SmallMatrix{0\\0}(\tau)|^2$ from the $\mathbb{T}^4$ partition function with the same factor in the denominator of the ghost partition function to rewrite.
\begin{equation}
\frac{(Z_{\mathbb{T}^4})_{a\alpha+b\beta,d\beta}\begin{bmatrix}0\\0\end{bmatrix}(0;\tau)}{\left|\vartheta\SmallMatrix{0\\0}(0;\tau)\right|^2}\left|\,\vartheta\begin{bmatrix}\frac{b}{2}\\ \frac{d}{2}\end{bmatrix}\left(\frac{az}{2};\tau\right)\,\vartheta\begin{bmatrix}\frac{b}{2}\\ \frac{b}{2}\end{bmatrix}\left(-\frac{az}{2};\tau\right)\right|^2=(Z_{\mathbb{T}^4})_{a\alpha+b\beta,d\beta}\begin{bmatrix}\frac{b}{2}\\\frac{d}{2}\end{bmatrix}\left(az;\tau\right)\,.
\end{equation}

The end result of the worldsheet partition function is now given as a sum of the partition function \eqref{eq:string-torus-nearly-final} over $a,d>0$ and $b=0,\ldots,d-1$ and integrating over $\mathbb{H}$. After relabelling $b\to -b$ and noting that the contribution due to sphere topologies exactly cancels with the elliptic prefactor $e^{-\pi ad\,\mathrm{Im}\,t} $, this gives the sum over defect Hecke operators appearing in the exponent of \eqref{eq:ZabGrand}, that is 
\begin{align}
 &\int_{\mathcal{F}}\frac{\mathrm{d}^2\tau}{\tau_2}  \sum_{a,d=1}^\infty \sum_{b=0}^{d-1} \,e^{\pi ad\,\mathrm{Im}\,t}\,p^{ad}\begin{tikzpicture}[baseline={([yshift=0.25cm]current bounding box.center)}]
\draw[thick] (0,0) -- (0,1) -- (1,1) -- (1,0) -- (0,0);
\node[below] at (0.5,0) {\footnotesize $(0,d)$};
\node[left] at (0,0.5) {\footnotesize $(a,b)$};
\end{tikzpicture}=\nonumber\\
&\hspace{4cm}=\sum_{a,d=1}^\infty\sum_{b=0}^{d-1}\frac{p^{ad}}{ad}(Z_{\mathbb{T}^4})_{a\alpha-b\beta,d\beta}\begin{bmatrix}\frac{b}{2}\\\frac{d}{2}\end{bmatrix}\left(az;\frac{at+b}{d}\right)\,,
\end{align}
which in turn can be exactly identified as the connected contribution to the spacetime defect torus correlator $\mathfrak{Z}_{\alpha,\beta}$ for the maximally-fractional defect $\boldsymbol{\mathcal{L}}_0$ with winding numbers $(\alpha,\beta)$ with the choice of the twisted NS spin structure along the boundary torus.\footnote{The untwisted spacetime NS sector can be obtained by sending $z\to z+1$, while the R sector would have been obtained from a BTZ geometry in the bulk \cite{Eberhardt:2020bgq}.} Exponentiating this result to account also for the disconnected worldsheets, one readily recovers the full expression for the grandcanonical defect torus correlator $\mathfrak{Z}_{\alpha,\beta}(p,z;t)$, as we have calculated it in Section \ref{sec:sym-defects}. We can therefore conclude that maximally-fractional defects $\boldsymbol{\mathcal{L}}_0$ winding $(\alpha,\beta)$-times around the spacetime torus
are holographically dual to closed strings propagating on thermal backgrounds described by the worldsheet partition function \eqref{eq:ZthermalAB}.

A couple of comments are now in order. First, note that while our calculation only recovered the R-sector spin structure in the symmetric orbifold, we can recover the expressions for all other spin structures by performing (spacetime) modular transformations on the integrated result. Second, the above-derived holographic correspondence can be readily extended for the sign-defect $\boldsymbol{\mathcal{L}}_\mathrm{sign}$ by noting that replacing
\begin{align}
    \begin{tikzpicture}[baseline={([yshift=0.25cm]current bounding box.center)}]
\draw[thick] (0,0) -- (0,1) -- (1,1) -- (1,0) -- (0,0);
\node[below] at (0.5,0) {\footnotesize $(c,d)$};
\node[left] at (0,0.5) {\footnotesize $(a,b)$};
\end{tikzpicture}\quad\longrightarrow\quad (-1)^{(ad-bc)(\alpha+\beta)}\, e(a\alpha,c\alpha) \,e(b\beta,d\beta) \hspace{2mm}  \begin{tikzpicture}[baseline={([yshift=0.25cm]current bounding box.center)}]
\draw[thick] (0,0) -- (0,1) -- (1,1) -- (1,0) -- (0,0);
\node[below] at (0.5,0) {\footnotesize $(c,d)$};
\node[left] at (0,0.5) {\footnotesize $(a,b)$};
\end{tikzpicture} 
\end{align}
in the definition \eqref{eq:ZthermalAB} of $Z_{\alpha,\beta}^\mathrm{thermal}$ does not spoil its modular invariance. The sign factors $e(\alpha,\beta)$ were defined in \eqref{eq:eSign}, while the overall factor $(-1)^{(ad-bc)(\alpha+\beta)}$ can be absorbed into a redefinition of the fugacity $p$. Upon integrating over the fundamental domain, we recover the exact sign structure observed in \eqref{eq:GrandCanDef} for $\epsilon=-1$. 

Also, instead of performing the full $\mathbb{Z}\oplus \mathbb{Z}$ double orbifold by summing over all $a,b,c,d\in\mathbb{Z}$, one could have stopped after gauging just the defect $\mathcal{L}_2$ (which is associated with the sum over $b,d$) while leaving behind just a single insertion of $\mathcal{L}_1$ around the spatial cycle of the worldsheet torus (i.e.\ setting $a=1,c=0$). This has the effect of implementing the identification $\gamma\sim e^{-2\pi it}\gamma$ and, at the same time, reinterpreting the spacetime modular parameter $t$ as the worldsheet chemical potential for the $\mathfrak{sl}(2,\mathbb{R})$ current $J_0^3$. Doing this, one obtains the global defect partition functions of the type \eqref{eq:ZUNblocks} and \eqref{eq:ZsignBlocks}, noting that the twisted sectors of this orbifold exactly give rise to the spectrally-flowed sectors upon identifying $d=w$. Given an explicit expression for the seed $\mathbb{T}^4$ defect $\mathcal{L}$, such partition functions can, in turn, be recast as manifest linear combinations of characters of the worldsheet chiral algebra \eqref{eq:worldsheet-theory}. This enables one to read off the representations appearing in the closed-string vertex operators for such backgrounds (as in \eqref{eq:ZUN} and \eqref{eq:ZsignWS}).

\section{Discussion}\label{sec:discussion}

In this paper we have discussed construction of topological defects in general symmetric-product orbifold CFTs. While this is a worthwhile endeavour in its own right, one has to bear in mind that these are CFTs which (or their deformations) often enter holographic dualities \cite{Belin:2014fna,Haehl:2014yla}.
Indeed, in the case of a particular choice for the seed theory, we have exploited this to investigate implications of the presence of topological defects in two-dimensional spacetime on the physics of tensionless closed strings propagating in three-dimensional bulk.

In the symmetric-product orbifold we have concentrated on considering the \emph{maximally-fractional defects} whose action on local operators preserves the whole of the $S_N$ which permutes the individual copies of the seed theory. These defects carry representation labels of both the seed-theory chiral algebra, as well as of the orbifold group $S_N$. They satisfy fusion algebra which comes from tensoring the $S_N$ fusion rules with those of the seed theory. Thus, they not only provide for a natural uplift of whatever symmetry was present in the seed theory, but are also capable of encoding the symmetry structure of the $S_N$ representations.

Checking the Cardy's consistency condition, we have obtained the spectra of non-local states on which such defects may terminate, finding that they transform in the $S_N$ representation determined by the defect label. Narrowing our scope further to consider defects carrying either the trivial or the sign representation of the symmetric group\footnote{As we have discussed, these defects may nevertheless still enter order-disorder dualities involving more complicated non-invertible duality-like $S_N$ representations} while keeping the seed CFT defect general, we have shown that the grandcanonical ensemble of the disorder-field spectra exponentiates and furnishes a multiparticle interpretation in terms of states with definite statistics: bosonic for the trivial representation and mixed bosonic / fermionic for the sign representation. Knowledge of the disorder fields is crucial whenever attempting to derive conservation laws for CFT correlators associated with non-invertible symmetries: moving a non-invertible defect line past a puncture leaves behind a trailing defect line which terminates at a non-local disorder field inserted at the said puncture. Such a procedure typically yields interesting relations (order-disorder dualities) between CFT correlators of local bulk fields a correlators of non-local disorder fields.

To quantify the effects which the topological defects wrapping the spacetime have on calculating string amplitudes in the bulk, it was crucial to realize that, in the case of the tensionless $\mathrm{AdS}_3/\mathrm{CFT}_2$ holography at $k=1$ units of pure NS-NS flux, the worldsheet which is to contribute to a given spacetime correlator  holomorphically covers the conformal boundary. A maximally-fractional spacetime defect therefore leaves its imprint on the worldsheet in the form the preimage under the covering map. First, we have noted that \emph{invertible} maximally-fractional defect loops wrapping local fields are simply uplifted to invertible worldsheet defect loops wrapping local insertions of closed-string vertex operators, on which they implement a global worldsheet symmetry. On the other hand, starting with a \emph{non-invertible} maximally-fractional defect loop in the spacetime, this can be first reduced to calculating correlators of the non-local disorder fields connected by maximally-fractional topological defect lines. We have then shown that reinterpreting these correlators in terms of worldsheet amplitudes necessitates modifying the closed-string background to accommodate vertex operators dual to disorder fields. Our discussion in this paper yields the worldsheet partition functions of these modified backgrounds for \emph{any} maximally-fractional spacetime defect carrying the trivial or the sign representation of $S_N$. Algebraically, these partition functions arise upon \emph{gauging} a suitable worldsheet defect in the $\mathfrak{psu}(1,1|2)_1$ WZW model tensored with the seed theory defect which was used to define the maximally-fractional spacetime defect in question, in agreement with the ideas of \cite{Harlow:2018tng}.
Geometrically, such backgrounds should describe brane-like extensions of the spacetime defects into the bulk.

\vspace{1cm}

\noindent Finally, we close with a discussion of possible extensions of this work and potential future directions.

\paragraph{Non-local correlators} In this work we mostly considered the computation of torus partition functions of CFTs with topological defects winding various cycles. Another quantity of interest in field theories with topological defects/non-invertible symmetries are non-local correlation functions of disorder fields. While we briefly discussed how such correlation functions look schematically in the symmetric product orbifold in Section \ref{sec:sym-defects}, we did not discuss them in detail. It would in particular be interesting to check if we can reproduce correlation functions of \emph{non-local} disorder fields in the symmetric orbifold from a worldsheet computation of amplitudes of closed strings on the backgrounds modified by the presence of the defects. Such amplitudes would involve insertions of \emph{local} vertex operators whose spectra we have identified in this work. In particular, the discussion in Section \ref{sec:worldsheet} suggests that a non-local maximally-fractional disorder field in the $w$ single-cycle twisted sector of the symmetric orbifold should be holographically dual to a local worldsheet operator in the $w$ spectrally-flowed sector which is twisted by $w$ copies of the seed defect $\mathcal{L}$ in the $\mathbb{T}^4$ CFT. It would be nice to see this duality more concretely by explicitly calculating correlation functions of such operators in the worldsheet theory and comparing them to 
correlators of disorder fields in the symmetric orbifold. In particular, in this way one would study worldsheet string manifestations of spacetime order-disorder (Kramers-Wannier-like) dualities \cite{Frohlich:2004ef,Kaidi:2021xfk} and other consequences of whatever non-invertible symmetries are present in the theory which lives on the conformal boundary.

\paragraph{Higher tension/supergravity}
The analysis of this paper, from the bulk side, relied heavily on the 
special properties of the tensionless ($k=1$) worldsheet theory on $\text{AdS}_3\times\text{S}^3\times\mathbb{T}^4$. While interesting from a technical point of view, this regime of string theory is far removed from the usual supergravity limit which is dual to strongly-coupled field theory. In order to understand this more conventional AdS/CFT regime, it is necessary to study worldsheet theories with non-minimal tension. Fortunately, worldsheet string theory on $\text{AdS}_3$ backgrounds supported by pure, generic NS-NS flux are still in principle solvable, being described by an $\text{SL}(2,\mathbb{R})$ WZW model. On the CFT side, these backgrounds have recently been reconsidered, and have been argued to be dual to non-compact symmetric orbifold CFTs deformed by a non-normalizable marginal operator \cite{Balthazar:2021xeh,Eberhardt:2021vsx,Dei:2022pkr,Sriprachyakul:2024gyl} (see \cite{Hikida:2023jyc,Knighton:2023xzg,Knighton:2024qxd,Sriprachyakul:2024gyl} for precision checks of this duality).
Given our general discussion of topological defects in symmetric orbifold theories, it would then be interesting to investigate the worldsheet dual of topological defects in this non-tensionless holographic setting. Such an endeavour could shed more light on the details of the geometry of the brane-like bulk duals, as our construction in this paper was largely algebraic and confined to the stringy regime at $k=1$.

\paragraph{Turning on the R-R flux}

In \cite{Fiset:2022erp}, the worldsheet operator corresponding to turning on infinitesimal R-R flux was found, and it was shown in conformal perturbation theory that this deformation is holographically dual to a particular twisted marginal operator in the boundary symmetric orbifold. It would be interesting to study the fate of the worldsheet dual to topological defects that we have identified in this paper and see how it reacts once the R-R flux is switched on. One could also try approaching this problem by discussing whether various maximally-fractional defects survive the exactly marginal deformation in the twisted sector which deforms away from the tensionless point \cite{David:2002wn,Burrington:2012yq,Gaberdiel:2015uca,Guo:2020gxm,Apolo:2022fya}, possibly following the flow all the way into the semiclassical supergravity regime \cite{Seiberg:1999xz,Benjamin:2022jin}.

\acknowledgments
We thank Matthias Gaberdiel, Kiarash Naderi, Beat Nairz, Sara Kalisnik Hintz, Ondra Hulik, Joris Raeymaekers, Marco Meineri and Carlo Maccaferri for useful discussions. We would like to specially thank Matthias Gaberdiel for collaboration at the initial stage. 
BK and JV are also grateful to the organizers and participants of the workshop Speakable and Unspeakable in Quantum Gravity for providing a stimulating environment which enabled completion of this work.
BK is supported by STFC consolidated grants ST/T000694/1 and ST/X000664/1. The work of VS is supported by a grant from the Swiss National Science Foundation. VS acknowledges the support from NCCR SwissMAP which is also funded by the Swiss National Science Foundation. JV is supported by the ERC Starting Grant 853507.

\appendix

\section{Topological defects in 2d CFT}\label{section2}

The usual setup for a 2d CFT $X$ is to consider correlation functions on a generic Riemann surface with handles, conformal boundaries and punctures (where the states of the Hilbert space of $X$ are to be inserted). However, it is of physical and mathematical importance to also consider gluing 2 conformal field theories $X_1$ and $X_2$ along a 1d interface called a defect line \cite{Oshikawa:1996dj,Petkova:2000ip,Bachas:2001vj} (see \cite{Chang:2018iay} for a modern review). A particularly interesting and simple case is when the 2 CFTs are taken to have matching central charge and the gluing conditions on the chiral stress-energy currents are such that the precise location of the defect line does not matter as long as the line stays in the same (first) homotopy class of the Riemann surface. 
This is achieved by requiring that the conditions 
\begin{subequations}
    \begin{align}
        T^{(1)}(z)&=T^{(2)}(z)\,,\\
        \bar{T}^{(1)}(\bar{z})&=\bar{T}^{(2)}(\bar{z})\,,
    \end{align}
\end{subequations}
on the holomorphic and anti-holomorphic components of the stress-energy tensors of the two CFTs 
are satisfied along the interface.
Such defect lines are known as \emph{topological defects}. From now on, the defects we are considering are all topological. 

\subsection{Defect operators}

\begin{figure} [htpb!]
\centering
\begin{subfigure}[b]{0.4\textwidth}
         \centering
         \begin{tikzpicture}
\begin{scope}
\draw[thick, purple] (0,-1) [partial ellipse = 0:-180:1.72 and 0.25];
\draw[thick, dashed, purple] (0,-1) [partial ellipse = 0:180:1.72 and 0.25];
\draw[thick] (0,0) circle (2);
\draw[thick] (0,0) [partial ellipse = 0:-180:2 and 0.4];
\draw[thick, dashed] (0,0) [partial ellipse = 0:180:2 and 0.4];
\node at (1,1.2) {$\boldsymbol{\times}$};
\node at (-1,1) {$\boldsymbol{\times}$};
\node at (0,-2) {$\boldsymbol{\times}$};
\node at (0,-2.4) {$V$};
\node at (0,-1) {\textcolor{purple}{$\mathcal{L}$}};
\end{scope}
\end{tikzpicture}
         \caption{Sphere correlator with a defect line wrapping one of the punctures.}
     \end{subfigure}
     \hspace{2cm}
     \begin{subfigure}[b]{0.4\textwidth}
         \centering
     \begin{tikzpicture}
\begin{scope}[xshift=8cm]
\draw[thick] (0,0) circle (2);
\draw[thick] (0,0) [partial ellipse = 0:-180:2 and 0.4];
\draw[thick, dashed] (0,0) [partial ellipse = 0:180:2 and 0.4];
\node at (1,1.2) {$\boldsymbol{\times}$};
\node at (-1,1) {$\boldsymbol{\times}$};
\node at (0,-2) {\textcolor{purple}{$\boldsymbol{\times}$}};
\node at (0,-2.4) {\textcolor{purple}{${\mathcal{L}}(V)$}};
\end{scope}
\end{tikzpicture}
     \caption{Equivalent correlator with a new insertion at the south pole.}
     \end{subfigure}
     \caption{On the definition of the defect map ${\mathcal{L}}$.}
\label{defect}
\end{figure}
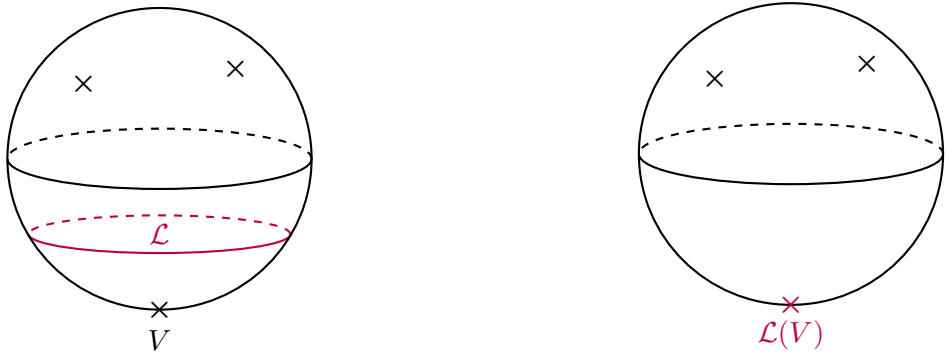

Let us first consider calculating a correlation function on a sphere where part of the insertions are taken to live in a CFT $X_1$ and the part in a CFT $X_2$. The two CFTs are glued by means of a closed topological defect loop. By considering the situation (see fig.~\ref{defect}) where the defect winds around a single puncture (with an insertion, say, in CFT $X_1$), one can deduce that the defect loop can be associated with a map between the Hilbert spaces of the two CFTs, namely
\begin{align}
{\mathcal{L}}  : \mathcal{H}_1 \longrightarrow \mathcal{H}_2 \,. 
\end{align}
Note that in general, this map may not be invertible. That is, denoting by ${\mathcal{L}}^\dagger$ the corresponding conjugate map from $\mathcal{H}_2$ to $\mathcal{H}_1$, it may not be true that ${\mathcal{L}}{\mathcal{L}}^\dagger = 1_{\mathcal{H}_2}$ and $  {\mathcal{L}}^\dagger {\mathcal{L}}= 1_{\mathcal{H}_1}$. 

\subsubsection{Decomposition into irreps and fusion}

For the sake of simplicity, let us fix the two CFTs to be the same from now on, that is $X_1=X_2=X$.
The requirement of invariance of gluing conditions under a continuous deformation of a defect line implies that the action of the defect commutes with the action of Virasoro generators of $X$, that is
\begin{subequations}
\begin{align}
[L_n,{\mathcal{L}}]&=0\,,\\
[\bar L_n,{\mathcal{L}}]&=0\,.\label{eq:gluing}
\end{align}
\end{subequations}
If the CFT $X$ has a chiral algebra with currents $\{W^{(i)}\}$ that is larger than the Virasoro algebra, one may instead impose the conditions ${\mathcal{L}}W^{(i)}_n=\Omega(W^{(i)}_n) {\mathcal{L}}$, ${\mathcal{L}}\bar{W}^{(i)}_n=\bar{\Omega}(\bar{W}^{(i)}_n) {\mathcal{L}}$, where $\Omega,\bar{\Omega}$ are automorphisms of the chiral algebra $\{W^{(i)}\}$. Assuming that the full spectrum $\mathcal{H}$ of the CFT $X$ can be decomposed diagonally into irreducible representations $\mathcal{V}_j$ as
\begin{align}
    \mathcal{H} = \bigoplus_{j}  \mathcal{V}_j \otimes \overline{\mathcal{V}_j}\,,\label{eq:spec}
\end{align}
the gluing conditions \eqref{eq:gluing} imply that the defect operator can be expanded as
\begin{align}
   {\mathcal{L}}= \sum_{j} {\mathcal{L}}_{j}\Pi_{j}\,,
\end{align}
where $\Pi_{j}$ is intertwiner between $\mathcal{V}_j \otimes \overline{\mathcal{V}_{{j}}}$ and $\mathcal{V}_{\Omega(j)} \otimes \overline{\mathcal{V}_{\bar{\Omega}(j)}}$, namely
\begin{align} \Pi_{j}=\sum_{N,\bar{N}}|\Omega(j,N)\rangle|\bar{\Omega}(j,\bar{N})\rangle\,\langle j,N|\langle j,\bar{N}|\,,
\end{align}
where $N,\bar{N}$ are some multi-indices labelling the orthonormalized $W$-descendants constituting the modules $\mathcal{V}_j$, $\overline{\mathcal{V}_{{j}}}$. Note that $\Pi_j$ may in general not exist for all $j$ once we choose $\Omega\neq \bar{\Omega}$. 
The defect coefficients ${\mathcal{L}}_{j}$ then need to be chosen subject to various sewing constraints (such as the Cardy condition, which is to be discussed in the next subsection). Note that by putting $\Omega=\bar{\Omega}=\mathrm{id}$ and ${\mathcal{L}}_{j} =1$ for all $j$, one ends up with a \emph{trivial} (or identity) defect. Also note that at the level of the defect operators, the \emph{fusion} of two topological defects is implemented simply by operator multiplication. For instance, in the case when $\Omega = \bar{\Omega}=\mathrm{id}$, we have
\begin{align}
    \Pi_i\Pi_j = \delta_{ij}\Pi_j
\end{align}
and so 
\begin{align}
    {\mathcal{L}}^{(1)}{\mathcal{L}}^{(2)} = 
    \sum_{j} {\mathcal{L}}_{j}^{(1)} {\mathcal{L}}_{j}^{(2)}\Pi_{j}= {\mathcal{L}}^{(2)}{\mathcal{L}}^{(1)}\,.
\end{align}
We will also say that a given defect is \emph{elementary} if the corresponding defect operator cannot be written as a positive-integer linear combination of other defect operators. A defect is called \emph{group-like} if it fuses with an arbitrary elementary defect to another elementary defect.

\subsection{Torus correlation functions}

As in the case of boundary states, the defect operator ${\mathcal{L}}$ has to satisfy a consistency condition similar to the Cardy condition for the boundary states \cite{Cardy:1986gw, Petkova:2000ip}.  

\subsubsection{Bulk channel}

Consider first the vacuum correlation function on a torus with modular parameter $\tilde{\tau}$ and let us insert a closed defect loop such that it winds once around the space-like cycle and zero-times around the $\tilde{\tau}$ cycle. This is the same as computing the trace
\begin{equation}
\begin{aligned}
Z^{\mathcal{L}}_{1,0}(\tilde{\tau})=\text{Tr}_{\mathcal{H}}\left( {\mathcal{L}}\tilde q^{L_0-c/24}\bar{\tilde{q}}^{\bar L_0-c/24} \right)\,,\label{eq:closed}
\end{aligned}
\end{equation}
where $\tilde q\equiv \exp(2\pi i\tilde\tau)$ and $\bar{}$ denotes the usual complex conjugation. Note that although we have taken \eqref{eq:closed} to be Virasoro-specialized, we may in principle also choose to include chemical potentials of other $W$-currents. Focusing from now on the case where $\Omega = \mathrm{id}$, one can evaluate \eqref{eq:closed} as
\begin{align}
    Z^{\mathcal{L}}_{1,0}(\tilde{\tau})=\sum_j {\mathcal{L}}_j \chi_j(\tilde{\tau})\chi_j(\bar{\tilde{\tau}})\,,\label{eq:closed_eval}
\end{align}
where
\begin{align}
    \chi_j(\tau) = \text{Tr}_{\mathcal{V}_j}\left(  q^{L_0-c/24}\right)
\end{align}
denotes the Virasoro-specialized character of $\mathcal{V}_j$. In the case of the trivial defect, the expression \eqref{eq:closed_eval} clearly reduces to the (modular-invariant) partition function encoding the bulk spectrum \eqref{eq:spec} of the CFT $X$.

\subsubsection{General winding}

In general, we will denote by $Z^{\mathcal{L}}_{m,n}(\tau)$ the result of calculating a vacuum correlator on a torus with modular parameter $\tau$ where the defect ${\mathcal{L}}$ wraps $m$ times around the space-like cycle and $n$ times around the $\tau$-cycle. As the pair of labels $(m,n)$ forms a modular doublet, we \emph{define} $Z^{\mathcal{L}}_{m,n}$ to satisfy
    \begin{align}
    Z_{m,n}^{\mathcal{L}}\left(\frac{a\tau+b}{c\tau+d}\right)  &= Z_{am+bn,cm+dn}^{\mathcal{L}}(\tau)\,.\label{eq:modular}
    \end{align}
A generic correlator $Z_{m,n}^{\mathcal{L}}(\tau)$ can then be computed by starting with $Z^{\mathcal{L}}_{\mathrm{gcd}(m,n),0}(\tau)$, which can be directly evaluated by \eqref{eq:closed}, inserting $\mathrm{gcd}(m,n)$ powers of the defect operator ${\mathcal{L}}$ in the trace, and then applying the rules \eqref{eq:modular} suitably many times.\footnote{Here we always take $\mathrm{gcd}(a,b)>0$ and understand $Z^{\mathcal{L}}_{n,0}$ for $n<0$ as a torus correlator with $n$ insertions of ${\mathcal{L}}^\dagger$ along the spacelike cycle of the torus.} In other words, defining $g_{m,n}\equiv \mathrm{gcd}(m,n)$ and introducing an explicit label for the defect in the correlator, we have 
\begin{align}
    Z^{\mathcal{L}}_{m,n}(\tau) =  Z^{{\mathcal{L}}^{g_{m,n}}}_{\mu,\nu}(\tau)\,,\label{eq:gcd}
\end{align}
where ${\mathcal{L}}^{g_{m,n}}$ denotes the defect ${\mathcal{L}}$ fused $g_{m,n}$-times with itself and
we have noted that for $(m,n)\neq (0,0)$, we can always write $(m,n)=(g_{m,n}\mu,g_{m,n}\nu)$ for $\mu,\nu\in\mathbb{Z}$ such that $\mathrm{gcd}(\mu,\nu)=1$. As we will demonstrate on a number of examples in Appendix \ref{defects}, taking suitable sums over $m,n$ (possibly decorated with discrete torsion) will result in modular-invariant quantities which can be interpreted in terms of (generalized) orbifold partition functions. We will also observe that the Virasoro-specialized torus defect correlators $Z^{\mathcal{L}}_{m,n}(\tau)$ are invariant under charge-conjugation of the Hilbert space. Realizing that $C = S^2$, we can therefore write
\begin{align}
Z^{\mathcal{L}}_{m,n}(\tau)=Z^{\mathcal{L}}_{m,n}(S^2\tau)=Z^{\mathcal{L}}_{-m,-n}(\tau)=Z^{{\mathcal{L}}^\dagger}_{m,n}(\tau)\,,
\end{align}
where in the last equality, we have noted that $Z^{{\mathcal{L}}^\dagger}_{m,n}(\tau)=Z^{\mathcal{L}}_{-m,-n}(\tau)$.

\subsubsection{Defect-ending fields}

In particular, after a modular S-transformation $\tilde\tau=-1/\tau$, the spatial and temporal cycles of the torus are interchanged, so that the transformation \eqref{eq:modular} specializes to
\begin{align}
    Z_{1,0}^{{\mathcal{L}}}(\tilde{\tau}) = Z_{0,1}^{{\mathcal{L}}}(\tau)\,.\label{eq:Stransf}
\end{align}
The correlator $Z^{\mathcal{L}}_{0,1}(\tau)$ can then be interpreted as the bulk partition of the CFT $X$ twisted by the defect. Indeed, it encodes the spectrum $\mathcal{H}_{{\mathcal{L}}}$ of \emph{defect-ending fields}. These are the fields which, upon being inserted at a puncture, interpolate between the given defect ${\mathcal{L}}$ and the trivial defect. Or, in other words, fields on which a defect can end. 
In particular, we can write
\begin{align}
    Z^{\mathcal{L}}_{0,1}({\tau}) = \text{Tr}_{\mathcal{H}_{{\mathcal{L}}}}\left( \tilde q^{L_0-c/24}\bar{\tilde{q}}^{\bar L_0-c/24} \right) = \sum_{k,\bar{k}} M_{k,\bar{k}} \chi_k(\tau) \chi_{\bar{k}}(\bar{\tau})\,,\label{eq:disorder}
\end{align}
where the positive integers $M_{k,\bar{k}}$ determine the multiplicity with which the representation $\mathcal{V}_k\otimes \overline{\mathcal{V}_k}$ appears in $\mathcal{H}_{{\mathcal{L}}}$. These may not be diagonal in $k$ even when the spectrum of local bulk fields is diagonal (as in \eqref{eq:spec}).
More generally, inserting along the $\tau$-cycle of the torus a pair of topological defects given by the defect operators ${\mathcal{L}}^{(1)},{\mathcal{L}}^{(2)}$ results in a twisted partition function encoding the spectrum $\mathcal{H}_{{\mathcal{L}}^{(1)}{\mathcal{L}}^{(2)}}$ of defect-changing fields interpolating between the two defects.
Given the expressions \eqref{eq:closed_eval} and \eqref{eq:disorder} and the modular S-duality relation \eqref{eq:Stransf}, the statement of the \emph{defect Cardy condition} can be summarized by requiring that the defect coefficients ${\mathcal{L}}_{j}$ are chosen such that the multiplicities $M_{k,\bar{k}}$ appearing in \eqref{eq:disorder} are positive integers. 

\subsubsection{Verlinde lines in RCFTs}

In particular, for $\Omega=\mathrm{id}$, and, provided that the action of the modular S-transformation on the characters can be implemented as 
\begin{align}
    \chi_j(-\tfrac{1}{\tau}) = \sum_{k}S_{kj} \chi_k(\tau)
\end{align}
(such as for an RCFT), one obtains the relation
\begin{align}
   M_{k,\bar{k}}= \sum_j {\mathcal{L}}_j S_{kj}S_{\bar{k}j}^\ast\,.
\end{align}
Moreover, note that in these cases the Cardy condition admits a solution (the \emph{Verlinde topological defect lines})
\begin{subequations}
    \begin{align}
        {\mathcal{L}}_j^{(\alpha)} &= \frac{S_{\alpha j}}{S_{\mathbb{I}j}}\,,\label{cardycoeff}\\
        M_{k,\bar{l}}^{(\alpha)} &= \tensor{N}{_{\alpha k}^{\bar{k}}}\,.
    \end{align}
\end{subequations}
where $\alpha$ labels any $W$-irrep $\mathcal{V}_\alpha$ ($\mathbb{I}$ labelling the vacuum irrep) and $\tensor{N}{_{\alpha k}^{\bar{k}}}$ are the Verlinde fusion coefficients. It is then straightforward to show that the fusion algebra of such defects is precisely the Verlinde algebra, namely that
\begin{align}
    {\mathcal{L}}^{(\alpha)} {\mathcal{L}}^{(\beta)} = \sum_\gamma\tensor{N}{_{\alpha\beta}}^{\gamma} {\mathcal{L}}^{(\gamma)}\,.
\end{align}
In particular, observe that ${\mathcal{L}}^{(\alpha)}$ is group-like if and only if $\alpha$ is a simple current. The defect operator ${\mathcal{L}}^{(\alpha)}$ then implements the symmetry associated with the simple current $\alpha$ on $\mathcal{H}$.

\section{Examples of topological defects in RCFTs}\label{defects}

In this appendix we will exhibit examples of various topological defects in a number of CFTs, mainly focusing on the Verlinde lines in rational conformal field theories.
These give rise to both invertible and non-invertible symmetries. We will also observe how to use torus correlators involving defect insertions to diagnose if a particular symmetry can be gauged or if it is anomalous and does not give rise to an orbifold CFT. In a number of cases of defects (including anomalous and non-invertible ones), we will also explicitly check the validity of our conjecture \eqref{eq:spproj}.

\subsection{Ising CFT}
\label{subsec:ising}

The Ising CFT (central charge $\frac{1}{2}$) is the simplest of the unitary series of Virasoro minimal models. It is built from three Virasoro irreps $\mathcal{V}_\mathbb{I}$, $\mathcal{V}_\varepsilon$, $\mathcal{V}_\sigma$ with conformal dimensions $h_\mathbb{I} = 0$, $h_\varepsilon = \frac{1}{2}$, $h_\sigma = \frac{1}{16}$. These modules satisfy the Verlinde fusion rules
\begin{subequations}
\begin{align}
    \varepsilon \times \varepsilon &= \mathbb{I}\,,\\
    \varepsilon \times \sigma &= \sigma\,,\\
    \sigma \times \sigma &= \mathbb{I}+\varepsilon
\end{align}
\end{subequations}
(plus the obvious ones with the identity module). The irrep $\varepsilon$ therefore corresponds to a simple current. These can be obtained using the Verlinde with the modular S-matrix
\begin{align}
    S = \frac{1}{2}\begin{pmatrix}
   1 & 1 & \sqrt{2} \\
   1 & 1 & -\sqrt{2}\\
   \sqrt{2} & -\sqrt{2} & 0
    \end{pmatrix}\,.
\end{align}
The three irreps assemble into the diagonal spectrum
\begin{align}
    \mathcal{H}\ =\ \mathcal{V}_\mathbb{I}\otimes \overline{\mathcal{V}_\mathbb{I}} \ \oplus \ \mathcal{V}_\varepsilon \otimes \overline{\mathcal{V}_\varepsilon} 
     \ \oplus \ \mathcal{V}_\sigma \otimes \overline{\mathcal{V}_\sigma} \,,
    \end{align}
which is encoded in the (modular invariant) bulk partition function
\begin{align}
    Z(\tau) = |\chi_\mathbb{I}(\tau)|^2 +|\chi_\varepsilon(\tau)|^2+|\chi_\sigma(\tau)|^2\,.\label{eq:Zising}
\end{align}
The three Verlinde defect operators on $\mathcal{H}$ are then explicitly written as
\begin{subequations}
\begin{align}
    {\mathcal{L}}^{(\mathbb{I})} &= \Pi_{\mathbb{I}\otimes\bar{\mathbb{I}}}+\Pi_{\varepsilon\otimes\bar{\varepsilon}} +\Pi_{\sigma\otimes\bar{\sigma}}\,,\\
    {\mathcal{L}}^{(\varepsilon)} &= \Pi_{\mathbb{I}\otimes\bar{\mathbb{I}}}+\Pi_{\varepsilon\otimes\bar{\varepsilon}} -\Pi_{\sigma\otimes\bar{\sigma}}\,,\\
    {\mathcal{L}}^{(\sigma)} &= \sqrt{2}\Pi_{\mathbb{I}\otimes\bar{\mathbb{I}}}-\sqrt{2}\Pi_{\varepsilon\otimes\bar{\varepsilon}} \,.
\end{align}
\end{subequations}
They inherit the fusion rules of the corresponding Virasoro modules, namely
\begin{align}
\mathcal{L}^{(\varepsilon)}\mathcal{L}^{(\sigma)}=
\mathcal{L}^{(\sigma)}
=\mathcal{L}^{(\sigma)}\mathcal{L}^{(\varepsilon)}\,,\qquad (\mathcal{L}^{(\sigma)})^2 = {\mathcal{L}}^{(\mathbb{I})} +{\mathcal{L}}^{(\varepsilon)}\,.
\end{align}

\subsubsection{The defect \texorpdfstring{${\mathcal{L}}^{(\varepsilon)}$}{Lepsilon}}

The (self-conjugate) defect ${\mathcal{L}}^{(\varepsilon)} $ is group-like and implements the $\mathbb{Z}_2$ symmetry transformation ${\mathcal{L}}^{(\varepsilon)}(\sigma) = -\sigma$. It is closely related to the defect ${\mathcal{L}}^{(-,-)}=(-1)^{F+\bar{F}}$ in the free fermion CFT.
A generic vacuum torus correlator involving ${\mathcal{L}}^{(\varepsilon)}$ can be compactly expressed as
\begin{align}
    Z_{m,n}^{(\varepsilon)}(\tau) = \frac{1}{2}(-1)^{mn}\Big[
    |\chi_\mathbb{I}+\chi_\varepsilon|^2 +(-1)^n|\chi_\mathbb{I}-\chi_\varepsilon|^2 +2(-1)^m |\chi_\sigma|^2
    \Big]
\end{align}
and it is readily verified that it obeys the rules \eqref{eq:modular}. In particular, from $Z_{0,1}^{(\varepsilon)}(\tau)$ we can read off the spectrum of operators on which the defect ${\mathcal{L}}^{(\varepsilon)} $ can end
\begin{align}
    \mathcal{H}({{\mathcal{L}}^{(\varepsilon)}}) = (\mathcal{V}_\mathbb{I}\otimes \overline{\mathcal{V}_\varepsilon})  \ \oplus \ (\mathcal{V}_\varepsilon \otimes \overline{\mathcal{V}_\mathbb{I}} )
     \ \oplus \ (\mathcal{V}_\sigma \otimes \overline{\mathcal{V}_\sigma})\,.
\end{align}
We observe, that it contains operators with non-integer spin. 
By performing a sequence of partial fusion and crossing operations, one can also show that it is possible to rewrite
\begin{align}
    Z_{m,n}^{(\varepsilon)}(\tau) =\mathrm{Tr}_{\mathcal{H}\left[({{\mathcal{L}}^{(\varepsilon)}})^n\right]}\left[({{\mathcal{L}}^{(\varepsilon)}})^m q^{L_0-\frac{1}{48}}\bar{q}^{L_0-\frac{1}{48}}\right] \,,\label{eq:Zepsmn}
\end{align}
where we have implicitly used the (unique) defect $({\mathcal{L}}^{(\varepsilon)})^{m+n}$ to resolve the 4-valent defect junction on the torus in order to define $({{\mathcal{L}}^{(\varepsilon)}})^m$ insertion in the trace over $\mathcal{H}\left[({{\mathcal{L}}^{(\varepsilon)}})^n\right]$. Since there is no additional phase appearing in front of the trace in \eqref{eq:Zepsmn}, the $\mathbb{Z}_2$ symmetry is non-anomalous and can be gauged. The corresponding $\mathbb{Z}_2$-orbifold torus partition function can be computed as the manifestly modular-invariant sum
\begin{align}
    \frac{1}{|\mathbb{Z}_2|}\sum_{m,n\in\mathbb{Z}_2}Z_{m,n}^{(\varepsilon)}(\tau)= |\chi_\mathbb{I}(\tau)|^2 +|\chi_\varepsilon(\tau)|^2+|\chi_\sigma(\tau)|^2\,.
\end{align}
In this simple case, we get back the bulk partition function \eqref{eq:Zising}.

\subsubsection{The defect \texorpdfstring{${\mathcal{L}}^{(\sigma)}$}{Lsigma}}

On the other hand, the defect ${\mathcal{L}}^{(\sigma)}$ is non-invertible (but self-conjugate). Since its fusion can still be expressed as a sum of invertible defects, it is a duality defect.
In particular, it enacts the Kramers-Wannier duality on 2d Ising correlators. The expression for torus correlators $Z_{m,n}^{(\sigma)}(\tau)$ is much more complicated than in the case of ${\mathcal{L}}^{(\varepsilon)}$. 
First, we note that
\begin{align}
    ({\mathcal{L}}^{(\sigma)})^{g} = \left\{\begin{array}{ll}
    2^{\frac{g}{2}-1}\big[
{\mathcal{L}}^{(\mathbb{I})}+{\mathcal{L}}^{(\varepsilon)}
\big]   &  \text{for $g\geqslant 1$ even}\,,\\
 2^{\frac{g-1}{2}}  {\mathcal{L}}^{(\sigma)}      & \text{for $g\geqslant 1$ odd}\,.
    \end{array}\right.
\end{align}
Recalling the property \eqref{eq:gcd}, the form of $Z^{(\sigma)}_{m,n}(\tau)$ therefore depends on the $\mathbb{Z}_2$ parity of $g_{m,n}\equiv \mathrm{gcd}(m,n)$. Let us decompose $(m,n)=(g_{m,n}\mu,g_{m,n}\nu)$ where $\mathrm{gcd}(\mu,\nu)=1$.
For $g_{m,n}\in 2\mathbb{Z}$, we simply have 
\begin{align}
  Z^{(\sigma)}_{m,n}(\tau)=2^{\frac{g_{m,n}}{2}-1}\Big[Z^{(\mathbb{I})}_{\mu,\nu}(\tau)+Z^{(\varepsilon)}_{\mu,\nu}(\tau)\Big]  \,,\label{eq:Zmnsig1}
\end{align}
while for $g_{m,n}\in 2\mathbb{Z}+1$ we obtain 
\begin{align}
    Z^{(\sigma)}_{m,n}(\tau)=\left\{\begin{array}{ll}
        2^{\frac{g_{m,n}-2}{2}}\Big(e^{-\frac{i\pi}{8}(-\mu\nu+\mu^2-1)}(\chi_\mathbb{I}+\chi_\varepsilon)(\bar{\chi}_\mathbb{I}-\bar{\chi}_\varepsilon)+\text{c.c.}\Big) &  \text{$\mu+1,\nu\in 2\mathbb{Z}$}\\
        2^{\frac{g_{m,n}-1}{2}}\Big(e^{-\frac{i\pi}{8}(\mu\nu+\nu^2-1)}(\chi_\mathbb{I}+\chi_\varepsilon)\bar{\chi}_\sigma+\text{c.c.}\Big) & \text{$\mu,\nu+1\in 2\mathbb{Z}$} \\
        2^{\frac{g_{m,n}-1}{2}}\Big(e^{-\frac{i\pi}{8}
        \mu\nu}(\chi_\mathbb{I}-\chi_\varepsilon)\bar{\chi}_\sigma+\text{c.c.}\Big) & \text{$\mu,\nu\in 2\mathbb{Z}+1$}
    \end{array}\right.\,.\label{eq:Zmnsig2}
\end{align}
In particular, the torus correlator $Z^{(\sigma)}_{0,1}(\tau)$ encodes the spectrum of operators on which the defect ${\mathcal{L}}^{(\sigma)}$ can end, namely
\begin{align}
    \mathcal{H}({{\mathcal{L}}^{(\sigma)}}) = (\mathcal{V}_\mathbb{I}\otimes \overline{\mathcal{V}_\sigma})  \ \oplus \ (\mathcal{V}_\varepsilon \otimes \overline{\mathcal{V}_\sigma} )
     \ \oplus \ (\mathcal{V}_\sigma \otimes \overline{\mathcal{V}_\mathbb{I}})\ \oplus \ (\mathcal{V}_\sigma \otimes \overline{\mathcal{V}_\varepsilon})\,.
\end{align}
Again, we note that in contrast with the spectrum of local bulk fields, the spectrum of the ${{\mathcal{L}}^{(\sigma)}}$ defect-ending fields contains fields with non-integral spin.

\subsubsection{Checking the cyclic projection}

Finally, let us explicitly test the validity of our conjecture \eqref{eq:spproj}.

\subsubsection*{Invertible cases}

For the defects ${{\mathcal{L}}^{(\mathbb{I})}}$, ${{\mathcal{L}}^{(\varepsilon)}}$, this follows from the fact that one is able to directly write $Z_{m,n}^{(\varepsilon)}$ as a trace over $\mathcal{H}(\mathcal{L}^n)$ as in \eqref{eq:Zepsmn}, so that \eqref{eq:spproj} holds with
\begin{align}
    \Pi_d = \frac{1}{d}\sum_{b=0}^{d-1}{{\mathcal{L}}}^{-b}e^{\frac{2\pi i b}{d}(L_0-\bar{L}_0)}\,.\label{eq:PidInv}
\end{align}
Indeed, since both ${{\mathcal{L}}^{(\mathbb{I})}}$, ${{\mathcal{L}}^{(\varepsilon)}}$ are group-like, then in order to define the insertion of ${{\mathcal{L}}}^{-b}$ into a trace over $\mathcal{H}(\mathcal{L}^d)$, one can unambiguously resolve the 4-valent defect junction using the defect ${{\mathcal{L}}}^{d-b}$.
The r.h.s.\ of \eqref{eq:PidInv} can then be shown to define a projector on $\mathcal{H}(\mathcal{L}^d)$ on the grounds of covariance of $Z_{m,n}(\tau)$ under the modular T-transformation. Substituting the explicit results for  $Z_{m,n}(\tau)$, one obtains\footnote{Here $\delta_{\mathbb{Z}_d}(m)$ denotes the $\delta$-distribution on $\mathbb{Z}_d$, that is the indicator function on $m\equiv 0\,\mathrm{mod}\,d$.}
\begin{align}
    \Pi_d^{(\mathbb{I})} = \delta_{\mathbb{Z}_d}(L_0-\bar{L}_0)
\end{align}
for the identity defect ${{\mathcal{L}}^{(\mathbb{I})}}$ (empty torus),
while for the defect ${{\mathcal{L}}^{(\varepsilon)}}$ one has
\vspace{-4mm}
        
    \begin{align}
        \Pi_d^{(\varepsilon)} =\bigg\{ \begin{array}{ll} 
        &\\
        d\in 2\mathbb{Z}: & 
        \delta_{\mathbb{Z}_d}(L_0-\bar{L}_0)\Pi_{\mathbb{I}\otimes \bar{\mathbb{I}}}+\delta_{\mathbb{Z}_d}(L_0-\bar{L}_0)\Pi_{\varepsilon\otimes\bar{\varepsilon}}  \\
        & \hspace{5cm}+\delta_{\mathbb{Z}_d}(L_0-\bar{L}_0+\tfrac{d}{2})\Pi_{\sigma\otimes\bar{\sigma}}\,,\\
         d\in 2\mathbb{Z}+1 :   & 
         \delta_{\mathbb{Z}_d}(L_0-\bar{L}_0+\tfrac{d}{2})\Pi_{\mathbb{I}\otimes \bar{\varepsilon}}+\delta_{\mathbb{Z}_d}(L_0-\bar{L}_0+\tfrac{d}{2})\Pi_{\varepsilon\otimes\bar{\mathbb{I}}} \\
         & \hspace{5cm}+\delta_{\mathbb{Z}_d}(L_0-\bar{L}_0)\Pi_{\sigma\otimes\bar{\sigma}}\,.
        \end{array}
    \end{align}
Both $\Pi_d^{(\mathbb{I})}$ and $\Pi_d^{(\varepsilon)}$ are manifest projectors on $\mathcal{H}[(\mathcal{L}^{(\mathbb{I})})^d] $ and $\mathcal{H}[(\mathcal{L}^{(\varepsilon)})^d] $.

\subsubsection*{The non-invertible case}
    
For the non-invertible defect $\mathcal{L}^{(\sigma)}$, one needs to perform a more complicated sequence of partial fusion and crossing moves in order to recast the correlators $Z_{m,n}(\tau)$ in terms of traces over $\mathcal{H}(\mathcal{L}^{d})$. Nevertheless, one can still explicitly substitute the results \eqref{eq:Zmnsig1} and \eqref{eq:Zmnsig2} for $Z_{m,n}^{(\sigma)}(\tau)$ into the l.h.s.\ of \eqref{eq:spproj} to compute the r.h.s.\ of \eqref{eq:spproj} directly. First, note that for $d\in 2\mathbb{Z}+1$, one has the spectrum of $(\mathcal{L}^{(\sigma)})^{d}$ defect-ending fields
\begin{align}
   \mathcal{H}\big[(\mathcal{L}^{(\sigma)})^{d}\big] &=  2^{\frac{d-1}{2}}\Big[(\mathcal{V}_\mathbb{I}\otimes\overline{\mathcal{V}_\sigma})\,\oplus\, (\mathcal{V}_\varepsilon\otimes\overline{\mathcal{V}_\sigma}) \oplus (\mathcal{V}_\sigma\otimes\overline{\mathcal{V}_\mathbb{I}})\,\oplus\, (\mathcal{V}_{\sigma}\otimes\overline{\mathcal{V}_\varepsilon})\Big]\,,\label{eq:Hodd}
\end{align}
while for $d\in 2\mathbb{Z}$, one has
\begin{align}
    \mathcal{H}\big[(\mathcal{L}^{(\sigma)})^{d}\big] &= 2^{\frac{d}{2}-1}\Big[(\mathcal{V}_\mathbb{I}\otimes\overline{\mathcal{V}_\mathbb{I}})\,\oplus\, (\mathcal{V}_\varepsilon\otimes\overline{\mathcal{V}_\varepsilon})\,\oplus\, \nonumber\\
    &\hspace{4cm}\oplus\,(\mathcal{V}_\mathbb{I}\otimes\overline{\mathcal{V}_\varepsilon})\,\oplus\, (\mathcal{V}_\varepsilon\otimes\overline{\mathcal{V}_\mathbb{I}})\,\oplus\, 2(\mathcal{V}_\sigma\otimes\overline{\mathcal{V}_\sigma})\Big]\,.\label{eq:Heven}
\end{align}
That is, as a consequence of the non-trivial $g$-function of $\mathcal{L}^{(\sigma)}$, the irreducible Virasoro modules enter $\mathcal{H}\big[(\mathcal{L}^{(\sigma)})^{d}\big]$ with non-trivial multiplicities. To facilitate the ensuing calculation, it is also convenient to note that one can rewrite the l.h.s.\ of \eqref{eq:spproj} as
\begin{align}
     \frac{1}{d}\sum_{b\in \mathbb{Z}_d} Z^{\mathcal{L}}_{-b,d}(\tfrac{a\tau+b}{d}) =  \frac{1}{d}\sum_{k|d}\sum_{\substack{\beta\in \mathbb{Z}_k\\ {\mathrm{gcd}(\beta,k)=1}}} Z^{\mathcal{L}^\frac{d}{k}}_{-\beta,k}(\tfrac{a\tau}{d}+\tfrac{\beta}{k})\,.
\end{align} 
Then, after a straightforward but rather lengthy manipulation, one finds that for $d\in 2\mathbb{Z}+1$ the conjecture \eqref{eq:spproj} holds with
\begin{align}
\Pi_d^{(\sigma)}&=\sum_{m\in\mathbb{Z}_d}\sum_{i=1}^{M_d^{\mathbb{I}\sigma}(m)}\delta_{\mathbb{Z}_d}(L_0-\bar{L}_0+\tfrac{1}{16}-m)\Pi_{\mathbb{I}\otimes \bar{\sigma}}^{(m,i)}+\nonumber\\
&\hspace{4cm}+\sum_{m\in\mathbb{Z}_d}\sum_{i=1}^{M_d^{\varepsilon\sigma}(m)}\delta_{\mathbb{Z}_d}(L_0-\bar{L}_0-\tfrac{7}{16}-m)\Pi_{\varepsilon\otimes \bar{\sigma}}^{(m,i)}+\mathrm{h.c.}\,,\label{eq:dodd}
\end{align}
where the superscripts $(m,i)$ were introduced to differentiate between projectors on the $2^{\frac{d-1}{2}}$ degenerate copies of the Virasoro irreducible modules which are present in \eqref{eq:Hodd}. Furthermore, $M_d^{\mathbb{I}\sigma}(m)$ with $M_d^{\varepsilon\sigma}(m)$ are numbers which can be expressed as
\begin{subequations}
    \begin{align}
        M_d^{\mathbb{I}\sigma}(m) &= \frac{1}{d}\sum_{k|d}2^\frac{\frac{d}{k}-1}{2}e^{-\frac{i\pi}{8}(k^2-1)} c_k\big[m - \tfrac{(k^2-1)(k-1)}{16}\big]\,,\\
        M_d^{\varepsilon\sigma}(m) &= \frac{1}{d}\sum_{k|d}2^\frac{\frac{d}{k}-1}{2}e^{-\frac{i\pi}{8}(k^2-1)} c_k\big[m - \tfrac{(k^2-1)(k-1)}{16}-\tfrac{k-1}{2}\big]\,,
    \end{align}
\end{subequations}
where by $c_k(m)$, we have denoted the Ramanujan $c$-function, namely
\begin{align}
    c_k(m) = \sum_{\substack{\alpha\in \mathbb{Z}_k\\
    \mathrm{gcd}(\alpha,k)=1}} e^{2\pi i \frac{\alpha}{k}m}
\end{align}
for $m\in \mathbb{Z}_k$. As can be checked with MATHEMATICA, both $M_d^{\mathbb{I}\sigma}(m)$ and $M_d^{\varepsilon\sigma}(m)$ are positive-integer valued functions on $\mathbb{Z}_d$ (that is, $d$-periodic in $m$). They also satisfy
\begin{align}
    \sum_{m\in \mathbb{Z}_d}M_d^{\mathbb{I}\sigma}(m)  =\sum_{m\in \mathbb{Z}_d}M_d^{\varepsilon\sigma}(m) = 2^{\frac{d-1}{2}}\,,
\end{align}
so that the indices $(m,i)$ in the sums in \eqref{eq:dodd} exactly resolve the $2^{\frac{d-1}{2}}$-fold degeneracy of the Virasoro modules in $\mathcal{H}\big[(\mathcal{L}^{(\sigma)})^{d}\big]$ (see \eqref{eq:Hodd}). The expression \eqref{eq:dodd} therefore defines a good projector on $\mathcal{H}\big[(\mathcal{L}^{(\sigma)})^{d}\big]$ for $d\in 2\mathbb{Z}+1$. 
For even $d$, one would instead obtain
\begin{align}
\Pi_d^{(\sigma)}&=\sum_{m\in\mathbb{Z}_d}\sum_{i=1}^{M_d^{\sigma\sigma}(m)}\delta_{\mathbb{Z}_d}(L_0-\bar{L}_0-m)\Pi_{\sigma\otimes \bar{\sigma}}^{(m,i)}+\nonumber\\
&\hspace{2cm}+\sum_{m\in\mathbb{Z}_d}\sum_{i=1}^{M_d^{\mathbb{I}\mathbb{I}}(m)}\delta_{\mathbb{Z}_d}(L_0-\bar{L}_0-m)\Pi_{\mathbb{I}\otimes \bar{\mathbb{I}}}^{(m,i)}+\nonumber\\
&\hspace{2cm}+\sum_{m\in\mathbb{Z}_d}\sum_{i=1}^{M_d^{\varepsilon\varepsilon}(m)}\delta_{\mathbb{Z}_d}(L_0-\bar{L}_0-m)\Pi_{\varepsilon\otimes \bar{\varepsilon}}^{(m,i)}+\nonumber\\
&\hspace{2cm}+\Bigg(\sum_{m\in\mathbb{Z}_d}\sum_{i=1}^{M_d^{\mathbb{I}\varepsilon}(m)}\delta_{\mathbb{Z}_d}(L_0-\bar{L}_0+\tfrac{1}{2}-m)\Pi_{\mathbb{I}\otimes \bar{\varepsilon}}^{(m,i)}+\mathrm{h.c.}\Bigg)\,.
\label{eq:deven}
\end{align}
In particular, the numbers $M_d^{\sigma\sigma}(m)$ can be expressed as
\begin{align}
    M_d^{\sigma\sigma}(m) = \frac{2}{d}\sum_{\substack{k|\frac{d}{2}\\ \text{$k$ odd}}} 2^{\frac{d}{2k}-1} c_k(m)
\end{align}
and can be identified as the lengths of the Varshamov-Tenengolts error-correcting codes $VT_m(\frac{d}{2})$ \cite{Sloane:2002}, thus being positive-integer valued functions on $\mathbb{Z}_d$ for all $d$. Finally, the numbers $M_d^{\mathbb{I}\mathbb{I}}(m)$, $M_d^{\varepsilon\varepsilon}(m)$ and $M_d^{\mathbb{I}\varepsilon}(m)$ can be expressed as
\begin{subequations}
\begin{align}
   M_d^{\mathbb{I}\mathbb{I}}(m) &= \frac{1}{d}\sum_{\substack{k|d\\ \text{$\frac{d}{k}$ even}\\ \text{$k$ even}}} 2^{\frac{d}{2k}} c_k(m)+
   \frac{1}{d}\sum_{\substack{k|d\\ \text{$\frac{d}{k}$ even}\\ \text{$k$ odd}}} 2^{\frac{d}{2k}-1} c_k(m)+\nonumber\\
   &\hspace{2cm}+\frac{1}{d}\sum_{
   \substack{k|d\\ 
   {\text{$\frac{d}{k}$ odd}}\\ 
   k\equiv 0\,\mathrm{mod}\,8
   }} 2^{\frac{\frac{d}{k}-1}{2}} \Big[c_k(m+\tfrac{k(k+2)}{16})+c_k(m+\tfrac{k(k-2)}{16})\Big]+\nonumber\\
    &\hspace{2cm}+\frac{1}{d}\sum_{
   \substack{k|d\\ 
   {\text{$\frac{d}{k}$ odd}}\\ 
   k\equiv 2\,\mathrm{mod}\,8
   }} 2^{\frac{\frac{d}{k}-1}{2}} c_k(m+\tfrac{k(k-2)}{16})+\nonumber\\
       &\hspace{2cm}+\frac{1}{d}\sum_{
   \substack{k|d\\ 
   {\text{$\frac{d}{k}$ odd}}\\ 
   k\equiv 6\,\mathrm{mod}\,8
   }} 2^{\frac{\frac{d}{k}-1}{2}} c_k(m+\tfrac{k(k+2)}{16})\\
      M_d^{\varepsilon\varepsilon}(m) &= \frac{1}{d}\sum_{\substack{k|d\\ \text{$\frac{d}{k}$ even}\\ \text{$k$ even}}} 2^{\frac{d}{2k}} c_k(m)+
   \frac{1}{d}\sum_{\substack{k|d\\ \text{$\frac{d}{k}$ even}\\ \text{$k$ odd}}} 2^{\frac{d}{2k}-1} c_k(m)+\nonumber\\
   &\hspace{2cm}-\frac{1}{d}\sum_{
   \substack{k|d\\ 
   {\text{$\frac{d}{k}$ odd}}\\ 
   k\equiv 0\,\mathrm{mod}\,8
   }} 2^{\frac{\frac{d}{k}-1}{2}} \Big[c_k(m+\tfrac{k(k+2)}{16})+c_k(m+\tfrac{k(k-2)}{16})\Big]+\nonumber\\
    &\hspace{2cm}-\frac{1}{d}\sum_{
   \substack{k|d\\ 
   {\text{$\frac{d}{k}$ odd}}\\ 
   k\equiv 2\,\mathrm{mod}\,8
   }} 2^{\frac{\frac{d}{k}-1}{2}} c_k(m+\tfrac{k(k-2)}{16})+\nonumber\\
       &\hspace{2cm}-\frac{1}{d}\sum_{
   \substack{k|d\\ 
   {\text{$\frac{d}{k}$ odd}}\\ 
   k\equiv 6\,\mathrm{mod}\,8
   }} 2^{\frac{\frac{d}{k}-1}{2}} c_k(m+\tfrac{k(k+2)}{16})\\
      M_d^{\mathbb{I}\varepsilon}(m) &=
   \frac{1}{d}\sum_{\substack{k|d\\ \text{$\frac{d}{k}$ even}\\ \text{$k$ odd}}} 2^{\frac{d}{2k}-1} c_k(m+\tfrac{k-1}{2})+\nonumber\\
   &\hspace{2cm}+\frac{1}{d}\sum_{
   \substack{k|d\\ 
   {\text{$\frac{d}{k}$ odd}}\\ 
   k\equiv 4\,\mathrm{mod}\,8
   }} 2^{\frac{\frac{d}{k}-1}{2}} \Big[c_k(m+\tfrac{k(k+2)}{16}-\tfrac{1}{2})+c_k(m+\tfrac{k(k-2)}{16}-\tfrac{1}{2})\Big]+\nonumber\\
    &\hspace{2cm}+\frac{1}{d}\sum_{
   \substack{k|d\\ 
   {\text{$\frac{d}{k}$ odd}}\\ 
   k\equiv 2\,\mathrm{mod}\,8
   }} 2^{\frac{\frac{d}{k}-1}{2}} c_k(m+\tfrac{k(k+2)}{16}-\tfrac{1}{2})+\nonumber\\
       &\hspace{2cm}+\frac{1}{d}\sum_{
   \substack{k|d\\ 
   {\text{$\frac{d}{k}$ odd}}\\ 
   k\equiv 6\,\mathrm{mod}\,8
   }} 2^{\frac{\frac{d}{k}-1}{2}} c_k(m+\tfrac{k(k-2)}{16}-\tfrac{1}{2})\,.
\end{align}
\end{subequations}
All $M_d^{\mathbb{I}\mathbb{I}}(m)$, $M_d^{\varepsilon\varepsilon}(m)$ and $M_d^{\mathbb{I}\varepsilon}(m)$ can be tested with MATHEMATICA to be positive-integer valued functions on $\mathbb{Z}_d$ and satisfy
\begin{align}
    \frac{1}{2}\sum_{m\in\mathbb{Z}_d}M_d^{\sigma\sigma}(m)= \sum_{m\in\mathbb{Z}_d} M_d^{\mathbb{I}\mathbb{I}}(m) =   
 \sum_{m\in\mathbb{Z}_d}M_d^{\mathbb{I}\mathbb{I}}(m)= \sum_{m\in\mathbb{Z}_d} M_d^{\mathbb{I}\varepsilon}(m)=2^{\frac{d}{2}-1}\,,
\end{align}
so that the indices $(m,i)$ in \eqref{eq:deven} correctly resolve the degeneracies of the Virasoro modules in \eqref{eq:Heven}. Hence, also for $d$ even, we conclude that \eqref{eq:deven} defines a valid projector on $\mathcal{H}\big[(\mathcal{L}^{(\sigma)})^{d}\big]$ so that the conjecture \eqref{eq:spproj} holds.

\subsection[\texorpdfstring{$(G_2)_1$}{G21} WZW model]{\boldmath \texorpdfstring{$(G_2)_1$}{G21} WZW model}
\label{subsec:G21}

Let us now review an example of a simple rational CFT containing a defect which is non-invertible and its fusion cannot be expressed as a sum of invertible defects (that is, it is not a duality defect). The $(G_2)_1$ WZW model is a CFT with central charge $c = \frac{14}{5}$ and two irreps $\mathcal{V}_\mathbb{I}$ and $\mathcal{V}_\Phi$ with dimensions $h_\mathbb{I} = 0$, $h_\Phi = \frac{2}{5}$. The field $\Phi$ satisfies the Fibonacci fusion rule
\begin{align}
    \Phi\times\Phi = \mathbb{I} +\Phi\,,\label{eq:FibFusion}
\end{align}
which can be derived from the Verlinde formula using the modular S-matrix
\begin{align}
    \frac{1}{\sqrt{2+\varphi}} \begin{pmatrix}
        +1 & \varphi\\
        \varphi & -1
    \end{pmatrix}\,,
\end{align}
where
$\varphi = \frac{1+\sqrt{5}}{2}$ denotes the golden ratio. The diagonal Hilbert space of local bulk fields is encoded in the partition function
\begin{align}
    Z(\tau) = |\chi_\mathbb{I}(\tau)|^2+|\chi_\Phi(\tau)|^2\,.
\end{align}

\subsubsection{The defect \texorpdfstring{$\mathcal{L}^{(\Phi)}$}{LPhi}}

The non-invertible and non-duality defect $\mathcal{L}^{(\Phi)}$ corresponding to the representation $\Phi$ acts 
as
\begin{align}
    \mathcal{L}^{(\Phi)} = \varphi\Pi_{\mathbb{I}\otimes\bar{\mathbb{I}}} +(1-\varphi)\Pi_{\Phi\otimes\bar{\Phi}}\,.
\end{align}
Computing general powers of this defect using the fusion rules \eqref{eq:FibFusion}, one finds
\begin{align}
    (\mathcal{L}^{(\Phi)})^k = F_{k-1}+F_k \mathcal{L}^{(\Phi)}\,,
\end{align}
where $F_k$ are the Fibonacci numbers, satisfying $F_{n+1}=F_n+F_{n-1}$ with $F_0=0$ and $F_1=1$. For any $m,n\in\mathbb{Z}$, let us write $(m,n)=(g\mu,g\nu)$ where $g=\mathrm{gcd}(m,n)$ and $\mathrm{gcd}(\mu,\nu)=1$.
The generic torus correlators $Z_{m,n}^{(\Phi)}$ then satisfy
\begin{align}
    Z_{m,n}^{(\Phi)}(\tau) = F_{g-1}Z(\tau)+F_gZ^{(\Phi)}_{\mu,\nu}(\tau)\,,
\end{align}
where, for $\mu,\nu$ coprime, one can find (denoting $\omega = e^{-\frac{4\pi i}{5}}$)
\begin{align}
    Z^{(\Phi)}_{\mu,\nu} = \Bigg\{
\begin{array}{ll}
   2(-1)^{\mu+1} \cos\frac{\mu\pi}{5}|\chi_\mathbb{I}|^2+\big[1-2(-1)^{\mu+1} \cos\frac{\mu\pi}{5}\big]|\chi_\Phi|^2 &\quad   \nu = 0\,\mathrm{mod}\,5\\
  |\chi_\Phi|^2 +\big(\omega^{\mu\nu}\chi_\mathbb{I}\bar{\chi}_\Phi+\mathrm{h.c.}\big) &\quad  \nu \in \{1,4\}\,\mathrm{mod}\,5\\
   |\chi_\mathbb{I}|^2 -\big(\omega^{-\mu\nu}\chi_\mathbb{I}\bar{\chi}_\Phi+\mathrm{h.c.}\big) &\quad  \nu \in \{2,3\}\,\mathrm{mod}\,5
\end{array}\,.
\end{align}
In particular, the Hilbert space $\mathcal{H}[(\mathcal{L}^{(\Phi)})^d]$ of the $(\mathcal{L}^{(\Phi)})^d$ defect-ending fields is encoded by the partition function
\begin{align}
  Z_{0,d}^{(\Phi)} = F_{g-1} |\chi_\mathbb{I}|^2 +F_{g+1}|\chi_\Phi|^2+ \big(F_g\chi_\mathbb{I}\bar{\chi}_\Phi+\mathrm{h.c.}\big)\,.\label{eq:Z0dG1}
\end{align}

\subsubsection{Checking the cyclic projection}

It is then possible to verify that the conjecture \eqref{eq:spproj} holds with the projector
\begin{align}
    \Pi_d^{(\Phi)} &= \sum_{m\in\mathbb{Z}_d}\sum_{i=1}^{M_d^{\mathbb{I}\mathbb{I}}(m)}\delta_{\mathbb{Z}_d}(L_0-\bar{L}_0-m)\Pi^{(m,i)}_{\mathbb{I}\otimes\overline{\mathbb{I}}}+\nonumber\\  &\hspace{2cm}+\sum_{m\in\mathbb{Z}_d}\sum_{i=1}^{M_d^{\Phi\Phi}(m)}\delta_{\mathbb{Z}_d}(L_0-\bar{L}_0-m)\Pi^{(m,i)}_{\Phi\otimes\overline{\Phi}}+\nonumber\\
    &\hspace{2cm}+\bigg(\sum_{m\in\mathbb{Z}_d}\sum_{i=1}^{M_d^{\mathbb{I}\Phi}(m)}\delta_{\mathbb{Z}_d}(L_0-\bar{L}_0+\tfrac{2}{5}-m)\Pi^{(m,i)}_{\mathbb{I}\otimes\overline{\Phi}}+\mathrm{h.c.}\bigg)\,,
\end{align}
where $M_d^{\mathbb{I}\mathbb{I}}(m)$, $M_d^{\Phi\Phi}(m)$, $M_d^{\mathbb{I}\Phi}(m)$ are numbers which can be expressed in terms of the Ramanujan $c$-function as
\begin{subequations}
\begin{align}
    M_d^{\mathbb{I}\mathbb{I}}(m)  &= \frac{1}{d}\sum_{k|d} F_{\frac{d}{k}-1} c_k(m)+\frac{1}{d}\sum_{\substack{k|d\\ k\equiv \pm 2 \,\mathrm{mod}\,5}} F_{\frac{d}{k}} c_k(m)+\nonumber\\
    &\hspace{2cm}+\frac{1}{d}\sum_{\substack{k|d\\ k\equiv 0 \,\mathrm{mod}\,5}} F_{\frac{d}{k}} \big[c_k(m)+c_k(m+\tfrac{2}{5}k)+c_k(m+\tfrac{3}{5}k)\big]\,,\\
    M_d^{\Phi\Phi}(m)&= \frac{1}{d}\sum_{k|d} F_{\frac{d}{k}-1} c_k(m)+\frac{1}{d}\sum_{\substack{k|d\\ k\equiv \pm 1 \,\mathrm{mod}\,5}} F_{\frac{d}{k}} c_k(m)+\nonumber\\
    &\hspace{2cm}+\frac{1}{d}\sum_{\substack{k|d\\ k\equiv 0 \,\mathrm{mod}\,5}} F_{\frac{d}{k}} \big[c_k(m)+c_k(m+\tfrac{2}{5}k)+c_k(m+\tfrac{3}{5}k)\big]\,,\\
    M_d^{\mathbb{I}\otimes \bar{\Phi}}(m)&=\frac{1}{d}\sum_{\substack{k|d\\ k\equiv \pm 1\,\mathrm{mod}\,5}} F_{\frac{d}{k}} c_k(m +\tfrac{2}{5}(k^2-1))+\nonumber\\
    &\hspace{2cm}+\frac{1}{d}\sum_{\substack{k|d\\ k\equiv \pm 2\,\mathrm{mod}\,5}} F_{\frac{d}{k}} c_k(m -\tfrac{2}{5}(k^2-1))\,.
\end{align}
\end{subequations}
One can readily verify (for instance, with MATHEMATICA) that $M_d^{\mathbb{I}\mathbb{I}}(m)$, $M_d^{\Phi\Phi}(m)$, $M_d^{\mathbb{I}\Phi}(m)$ are indeed positive-integer valued functions on $\mathbb{Z}_d$ for all $d$ which, upon summing over $m$ yield the correct multiplicities of irreducible representations observed in \eqref{eq:Z0dG1}.

\subsection[\texorpdfstring{$SU(2)_k$}{SU(2)k} WZW model]{\boldmath \texorpdfstring{$SU(2)_k$}{SU(2)k} WZW model}
\label{subsec:su2}

The irreps $\mathcal{V}_j$ of the $SU(2)_k$ CFT for $k\geqslant 1$ an integer are labelled by a half-integer $j$ where $0\leqslant j\leqslant \frac{k}{2}$. Their conformal dimensions are
\begin{align}
    h_j = \frac{j(j+1)}{k+2}\,.
\end{align}
The elements of the modular S-matrix are given as
\begin{align}
    S_{j,j^\prime} = \sqrt{\frac{2}{k+2}} \sin \frac{\pi(2j+1)(2j^\prime+1)}{k+2}\,.
\end{align}
They obey the useful property
\begin{align}
    S_{j,\frac{k}{2}-j^\prime} = (-1)^{2j}S_{j,j^\prime}\,.
\end{align}
The Verlinde fusion rules are consistent with the tensor product of $SU(2)$ irreps and read
    \begin{align}
        \tensor{N}{_{j_1, j_2}^{j_3}} = \left\{
        \begin{array}{ll}
            1 & \qquad\text{if $|j_1-j_2|\leqslant j_3 \leqslant \mathrm{min}(j_1+j_2,k-j_1-j_2)$} \\
            & \qquad\text{and $j_1+j_2+j_3\in\mathbb{Z}$}\,,\\[2mm]
            0 & \qquad\text{otherwise}\,.
        \end{array}
        \right.\label{eq:su2_wzw}
    \end{align}
We will focus on the case of bulk theories with the diagonal modular invariant 
\begin{align}
    Z(\tau) = \sum_{j=0}^{\frac{k}{2}}\big|\chi_j\big|^2\label{eq:wzw_diagonal}
\end{align}
(corresponding to $A_{k+1}$ in the ADE classification).
The Verlinde TDLs are then given by the operators
\begin{align}
    {\mathcal{L}}^{(j)}=\sum_{j^\prime=0}^{\frac{k}{2}}\frac{\sin \frac{\pi(2j+1)(2j^\prime+1)}{k+2}}{\sin \frac{\pi(2j^\prime+1)}{k+2}} \Pi_{j^\prime\otimes \overline{j^\prime}}\,.
\end{align}
It follows that ${\mathcal{L}}^{(\frac{k}{2})}$ is group-like: indeed, from \eqref{eq:su2_wzw} we have
\begin{align}
    {\mathcal{L}}^{(\frac{k}{2})}{\mathcal{L}}^{(j)} = {\mathcal{L}}^{(\frac{k}{2}-j)}\,.
\end{align}
In particular, the defect $\mathcal{L}^{(\frac{k}{2})}$ squares to identity, meaning that it obeys $\mathbb{Z}_2$ fusion rules.
At the same time, only ${\mathcal{L}}^{(0)}$ and ${\mathcal{L}}^{(\frac{k}{2})}$ have defect $g$-function equal to 1, so the defects ${\mathcal{L}}^{(j)}$ for $j\neq 0,\frac{k}{2}$ cannot be group-like. 

\subsubsection{Torus correlators, gauging and anomalies}

Furthermore, we can express a generic torus correlator with the insertions of the defect ${\mathcal{L}}^{(\frac{k}{2})}$ as
\begin{align}  
Z_{m,n}^{(\frac{k}{2})}(\tau)=(-i)^{kmn}\sum_{j=0}^{\frac{k}{2}} (-1)^{2jm}\chi_{j}(\tau)\,\bar{\chi}_{\frac{k}{4}+(j-\frac{k}{4})(-1)^n}(\bar{\tau})\,.\label{eq:ZmnWZW}
\end{align}
Note that the spectrum of the defect-ending operators for ${\mathcal{L}}^{(\frac{k}{2})}$ reads
\begin{align}
    \mathcal{H}({\mathcal{L}}^{(\frac{k}{2})})=\bigoplus_{j=0}^{\frac{k}{2}} \mathcal{V}_{j}\otimes \overline{\mathcal{V}_{\frac{k}{2}-j}}\,.\label{eq:ZWZWodd}
\end{align}
Observe that for $k\in 2\mathbb{Z}$, the correlators \eqref{eq:ZmnWZW} depend only on the $\mathbb{Z}_2$-parity of $m,n$. 
As a consequence, the torus correlators \eqref{eq:ZmnWZW} can be recast as a pure trace with no prefactors, namely
\begin{align}
     Z_{m,n}^{(\frac{k}{2})}(\tau)=\mathrm{Tr}_{\mathcal{H}[({\mathcal{L}}^{(\frac{k}{2})})^n]}({\mathcal{L}}^{(\frac{k}{2})})^m q^{L_0-\frac{c}{24}}\bar{q}^{\bar{L}_0-\frac{c}{24}}\,,\label{eq:traceWZW}
\end{align}
where the 4-valent junction on the torus is being implicitly resolved into a pair of 3-valent junctions by the defect $({\mathcal{L}}^{(\frac{k}{2})})^{m+n}$. This means that the $\mathbb{Z}_2$ symmetry generated by the defect ${\mathcal{L}}^{(\frac{k}{2})}$ is non-anomalous. 
Gauging this $\mathbb{Z}_2$ symmetry, the modular-invariant sum
\begin{align}
   \frac{1}{|\mathbb{Z}_2|}\sum_{m,n\in\mathbb{Z}_2}Z^{(\frac{k}{2})}_{m,n}(\tau)= \frac{1}{|\mathbb{Z}_2|}\sum_{m,n\in\mathbb{Z}_2}\mathrm{Tr}_{\mathcal{H}[({\mathcal{L}}^{(\frac{k}{2})})^n]}({\mathcal{L}}^{(\frac{k}{2})})^m q^{L_0-\frac{c}{24}}\bar{q}^{\bar{L}_0-\frac{c}{24}}
\end{align}
then gives the partition function of the corresponding $\mathbb{Z}_2$ orbifold. For $k\in 4\mathbb{Z}$, this can be explicitly evaluated as
\begin{align}
        \sum_{\substack{j=0\\ \text{$2j$ even}}}^{\frac{k}{4}-1}\big|\chi_{j}+\chi_{\frac{k}{2}-j}\big|^2+2\big|\chi_{\frac{k}{4}}\big|^2\,,
\end{align}
which corresponds to the ${D}_{k/2+2}$ series in the ADE classification of $SU(2)_k$ modular invariants, while for $k\in 4\mathbb{Z}+2$, we have
\begin{align}
     \sum_{\substack{j=0\\ \text{$2j$ even}}}^{\frac{k}{2}}\big|\chi_{j}\big|^2
    +\sum_{\substack{j=0\\ \text{$2j$ odd}}}^{\frac{k}{2}}\chi_{j}(\tau)\,\bar{\chi}_{\frac{k}{2}-j}(\bar{\tau})\,,
\end{align}
which in turn corresponds to the ${D}_{k/2+1}$ series. On the other hand, for $k\in 2\mathbb{Z}+1$, the defect torus correlators depend on $m,n$ modulo 4 because the $\mathbb{Z}_2$ symmetry enacted by the defect ${\mathcal{L}}^{(\frac{k}{2})}$ suffers from a 't Hooft anomaly and cannot be gauged. The exceptional modular invariants at levels 10, 16 and 28 can be obtained by a generalized orbifolding procedure involving insertion of certain conformal nets on the torus \cite{Frohlich:2009gb}.

\subsubsection{Checking the cyclic projection}

Let us also test the validity of the conjecture \eqref{eq:spproj}. First, note that the Hilbert space $\mathcal{H}[(\mathcal{L}^{(\frac{k}{2})})^{d}]$ of the defect ending fields is simply given by the diagonal bulk CFT partition function \eqref{eq:wzw_diagonal} for $d$ even and by \eqref{eq:ZWZWodd} for $d$ odd.
Recalling the result \eqref{eq:ZmnWZW} for the defect $\mathcal{L}^{(\frac{k}{2})}$, we indeed obtain that for $d\in 2\mathbb{Z}$, the formula \eqref{eq:spproj} holds with
\begin{align}
    \Pi_d  = \sum_{j=0}^{\frac{k}{2}} \delta_{\mathbb{Z}_d}\big[L_0-\bar{L}_0+jd+\tfrac{k}{4}\,d^2\big]\,\Pi_{j\otimes {j}}\,,\label{eq:projWZW1}
\end{align}
while for $d\in 2\mathbb{Z}+1$, we have
\begin{align}
    \Pi_d  = \sum_{j=0}^{\frac{k}{2}} \delta_{\mathbb{Z}_d}\big[L_0-\bar{L}_0+j(d+1)+\tfrac{k}{4}\,(d^2-1)\big]\,\Pi_{j\otimes (\frac{k}{2}-{j})}\,.\label{eq:projWZW2}
\end{align}
Both \eqref{eq:projWZW1} and \eqref{eq:projWZW2} are manifest projectors on $\mathcal{H}\big[(\mathcal{L}^{(\frac{k}{2})})^d\big]$ for any values of $k$. This means that \eqref{eq:spproj} holds also for the anomalous defects which occur for odd values of $k$.

\section{Examples of topological defects in free CFTs}\label{defectsFree}

In this Appendix, we will collect examples of topological defects which will be relevant for the holographic duality between closed strings on the $\mathrm{AdS}_3$ space and the symmetric product orbifold CFTs. Our objective here will be to showcase simple examples of topological defects in these theories. See \cite{Fuchs:2007tx} for a classification of topological defects in $c=1$ free boson theories.

\subsection{Non-compact free boson}
\label{subapp:noncompact}

Let us start by considering a single non-compact free boson CFT ($c=1$). The irreps $\mathcal{V}_k$ with respect to the Virasoro algebra enlarged by the $U(1)$ free-boson current $\p X$ are labelled by a continuum $-\infty \leqslant k \leqslant +\infty$ of momenta. 
Setting $\alpha^\prime =1$, these have conformal dimensions $h_k = k^2 /4$ with respect to the stress-energy tensor
\begin{align}
    T(z) =\,\, :\!\p X \p X \!\!:\!(z)\,.\label{eq:Tb}
\end{align}
The modules $\mathcal{V}_k$ are irreducible also w.r.t.\ the chiral algebra generated by $T(z)$, except for specific values of $k$ such that $k^2/4 = j^2$, where $j=0,\frac{1}{2},1,\ldots$
The diagonal spectrum of the theory is encoded in the partition function
\begin{align}
    Z(\tau)=\frac{1}{|\eta(\tau)|^2}\int\limits_{-\infty}^\infty dk\,q^{\frac{k^2}{4}}\bar{q}^{\frac{k^2}{4}}\,.\label{eq:Zb}
\end{align}

\subsubsection{Shift-symmetry defects}

Let us first consider topological defects satisfying the identity gluing conditions
\begin{subequations}
\label{eq:trivial}
    \begin{align}
        \p X^{(1)}(z) &=\p X^{(2)}(z)\,,\\
        \bar{\p} X^{(1)}(\bar{z}) &=\bar{\p} X^{(2)}(\bar{z})\,,
    \end{align}
\end{subequations}
along the interface. The corresponding defect operators are labeled by a continuous parameter $\rho\in\mathbb{R}$ and read
\begin{align}
    {\mathcal{L}}^{(\rho)} = \int\limits_{-\infty}^\infty dk\, e^{2\pi i k\rho}\,\sum_{N,\bar{N}}|k,N,\bar{N}\rangle\,\langle k,N,\bar{N}|\,.
\end{align}
Note that the action of ${\mathcal{L}}^{(\rho)}$ on the momentum plane waves implements the translation symmetry $X\longrightarrow X+2\pi \rho$ as we clearly have ${\mathcal{L}}^{(\rho)}|k\rangle = e^{2\pi i k\rho}|k\rangle$. These defects are group-like with the properties
\begin{subequations}
\label{eq:transl}
\begin{align}
{\mathcal{L}}^{(\rho)}{\mathcal{L}}^{(\sigma)}&={\mathcal{L}}^{(\rho+\sigma)}\,,\\
    {\mathcal{L}}^{(0)}&=\mathrm{id}\,,\\
    {\mathcal{L}}^{(-\rho)}&=({\mathcal{L}}^{(\rho)})^{-1}=({\mathcal{L}}^{(\rho)})^\dagger\,.
\end{align}
\end{subequations}
The corresponding vacuum torus correlator for generic winding numbers $(m,w)$ of the defect can be readily found as
\begin{align}
    Z_{m,w}^{(\rho)}(\tau) = \frac{1}{|\eta(\tau)|^2}\int\limits_{-\infty}^\infty dk\, e^{2\pi i m k\rho}\,q^{\frac{1}{4}(k+w\rho )^2}\bar{q}^{\frac{1}{4}(k-w\rho )^2}\,.\label{eq:Zmwbos}
\end{align}
Note that by changing the integration variable from $k$ to $-k$, one can readily see that 
\begin{align}
    Z_{m,w}^{(\rho)}(\tau)=Z_{-m,-w}^{(\rho)}(\tau)=Z_{m,w}^{(-\rho)}(\tau)\,,
\end{align}
meaning that the defect ${\mathcal{L}}^{(\rho)}$ has the same torus correlators as its conjugate $({\mathcal{L}}^{(\rho)})^\dagger={\mathcal{L}}^{(-\rho)}$.
In particular, the Hilbert space twisted by $w$ insertions of the defect $\mathcal{L}^{(\rho)}$ reads
\begin{align}
    \mathcal{H}[(\mathcal{L}^{(\rho)})^w]=\int_{-\infty}^\infty dk\, \mathcal{V}_{k+w\rho}\otimes \overline{\mathcal{V}_{k-w\rho}}\,.
\end{align}
We then observe that it is possible to extend the action of the defect operator $\mathcal{L}^{(\rho)}$ also on $\mathcal{H}[(\mathcal{L}^{(\rho)})^w]$ by defining
\begin{align}
    {\mathcal{L}}^{(\rho)}|_{\mathcal{H}[(\mathcal{L}^{(\rho)})^w]} = \int\limits_{-\infty}^\infty dk\, e^{2\pi i k\rho}\,\sum_{N,\bar{N}}|k+w\rho,N,\bar{N}\rangle\,\langle k-w\rho,N,\bar{N}|\,,
\end{align}
and that these extended operators continue to satisfy the the algebra \eqref{eq:transl}. As a result, the torus correlation function \eqref{eq:Zmwbos} can be recast as a trace
\begin{align}
    Z_{m,w}^{(\rho)}(\tau) = \mathrm{Tr}_{\mathcal{H}[({\mathcal{L}}^{(\rho)})^w]}({\mathcal{L}}^{(\rho)})^m q^{L_0-\frac{c}{24}}\bar{q}^{\bar{L}_0-\frac{c}{24}}\,.\label{eq:traceBos}
\end{align}
This is a manifestation of the fact that the symmetry $X\longrightarrow X+2\pi \rho$ is non-anomalous and can be gauged by taking the modular sum
\begin{align}
    |\rho|\sum_{m,w\in\mathbb{Z}} Z_{m,w}^{(\rho)}(\tau) =\frac{1}{|\eta(\tau)|^2}\sum_{n,w\in\mathbb{Z}}q^{\frac{1}{4}(\frac{n}{\rho}+w\rho )^2}\bar{q}^{\frac{1}{4}(\frac{n}{\rho}-w\rho )^2} \equiv Z(|\rho|;\tau) \,.\label{eq:modsumbos}
\end{align}
It yields the spectrum of a free boson compactified on a circle with radius $R=|\rho|$.

\subsubsection{Reflection-symmetry defect}
\label{subsubapp:reflection}

Instead, we will now consider the gluing conditions
\begin{subequations}
\label{eq:refl}
    \begin{align}
        \p X^{(1)}(z) &=-\p X^{(2)}(z)\,,\\
        \bar{\p} X^{(1)}(\bar{z}) &=-\bar{\p} X^{(2)}(\bar{z})\,.
    \end{align}
\end{subequations}
These can be implemented by the (self-conjugate) defect operator
\begin{align}
        {\mathcal{L}}^{(-)} = \int\limits_{-\infty}^\infty dk\, \sum_{M,\bar{M}}(-1)^{M+\bar{M}}|\!-\! k,M,\bar{M}\rangle\,\langle k,M,\bar{M}|\,.
\end{align}
On the momenta, the gluing automorphism $\Omega$ acts as $\Omega(k)=-k$ and, while on the oscillators we have $\Omega(\alpha_n)=-\alpha_n$, $\Omega(\bar{\alpha}_n)=-\bar{\alpha}_n$. That is, the defect ${\mathcal{L}}^{(-)}$ enacts the $\mathbb{Z}_2$ reflection symmetry $X\longrightarrow -X$. Correspondingly, one has
\begin{align}
    {\mathcal{L}}^{(-)}{\mathcal{L}}^{(-)} = \mathrm{id}\,.
\end{align}
The symmetry is non-anomalous and correspondingly, the torus correlators $Z_{m,n}^{(-)}(\tau)$ are well-defined for $m,n$ modulo 2. In particular, one can find
\begin{subequations}
    \begin{align}
        Z_{0,0}^{(-)}(\tau) &= Z(\tau)\,,\\[1.5mm]
        Z_{1,0}^{(-)}(\tau) &= 2\Big|\frac{\eta}{\theta_2}\Big|=
         \Big|\frac{\theta_3\theta_4}{\eta^2}\Big|\,,\\
        Z_{0,1}^{(-)}(\tau) &= 2\Big|\frac{\eta}{\theta_4}\Big|=
         \Big|\frac{\theta_2\theta_3}{\eta^2}\Big|\,,\\
        Z_{1,1}^{(-)}(\tau) &= 2\Big|\frac{\eta}{\theta_3}\Big|=
         \Big|\frac{\theta_2\theta_4}{\eta^2}\Big|\,.
    \end{align}
\end{subequations}
The modular-invariant sum
\begin{align}
    \frac{1}{2}\big[Z_{0,0}^{(-)}+Z_{1,0}^{(-)}+Z_{0,1}^{(-)}+Z_{1,1}^{(-)}\big] = \frac{1}{2}\bigg(Z(\tau)
    +\Big|\frac{\theta_3\theta_4}{\eta^2}\Big|
    +\Big|\frac{\theta_2\theta_3}{\eta^2}\Big|
    +\Big|\frac{\theta_2\theta_4}{\eta^2}\Big|
    \bigg)\label{eq:orbifold}
\end{align}
can be recognized as the partition function of the $\mathbb{Z}_2$-orbifold of the non-compact free boson.

\subsection{Compactified free boson}
\label{subapp:compact}

Let us now consider various defects for a single free boson compactified on a circle with radius $R$. The irreps $\mathcal{V}_{n,w}$ are now labelled by two integers $n$ (momentum) and $w$ (winding). The spectrum is encoded in the modular invariant partition function (note that we are setting $\alpha^\prime=1$)
\begin{align}
    Z(R;\tau) =\frac{1}{|\eta(\tau)|^2}\sum_{n,w\in\mathbb{Z}}q^{\frac{1}{4R^2}(n+wR^2 )^2}\bar{q}^{\frac{1}{4R^2}(n-wR^2 )^2} \,.\label{eq:part_compact}
\end{align}
Defining 
\begin{subequations}
    \begin{align}
        k_\mathrm{L}&=n+wR^2\,,\\
        k_\mathrm{R}&=n-wR^2\,,
    \end{align}
\end{subequations}
the corresponding vertex operators can be represented as
\begin{align}
   V_{n,w}(z,\bar{z})= e^{ik_\mathrm{L} X_\mathrm{L}}(z)\,e^{ik_\mathrm{R} X_\mathrm{R}}(\bar{z}) = e^{i(nX+ w\tilde{X}) }(z,\bar{z})\,,
\end{align}
where we define
\begin{subequations}
    \begin{align}
        X(z,\bar{z})&=X_\mathrm{L}(z)+X_\mathrm{R}(\bar{z})\,,\\
        \tilde{X}(z,\bar{z}) &=R^2 \big[X_\mathrm{L}(z)-  X_\mathrm{R}(\bar{z})\big]\,.
    \end{align}
\end{subequations}

\subsubsection{Shift-symmetry defects}
\label{subsubapp:reflectionCompact}

Considering the trivial gluing conditions \eqref{eq:trivial}, one has a family of defects given by the operators
\begin{align} 
    {\mathcal{L}}^{(W,N)} = \sum_{n,w\in\mathbb{Z}} e^{2\pi i nW}e^{2\pi i wN}\,\sum_{M,\bar{M}}|n,w,M,\bar{M}\rangle\,\langle n,w,M,\bar{M}|
\end{align}
for any $0\leqslant W,N<1$. These implement the translations $X\longrightarrow X+2\pi W $ and $\tilde{X}\longrightarrow \tilde{X}+2\pi N$ along the circle and its dual. For a generic defect torus correlator, one obtains
\begin{align}
     Z_{r,s}^{(W,N)}(R;\tau) &=\frac{1}{|\eta(\tau)|^2}e^{2\pi irsWN}\sum_{n,w\in\mathbb{Z}}e^{2\pi i r(nW + wN) }q^{\frac{1}{4R^2}[{n+sN}+(w+sW)R^2 ]^2}\times\nonumber\\
     &\hspace{7cm}\times\bar{q}^{\frac{1}{4R^2}[{n+sN}-(w+sW)R^2 ]^2} \,.\label{eq:torus_compact}
\end{align}
Note that relabelling the summation indices from $(n,w)$ to $(-n,-w)$ shows that 
\begin{align}
    Z_{r,s}^{(W,N)}=Z_{-r,-s}^{(W,N)}=Z_{r,s}^{(-W,-N)}\,,
\end{align}
namely that the defect and its conjugate give rise to the same torus correlators. 
These can be used to perform an orbifold construction only if $W,N$ are rational: for irrational $W, N$, the corresponding shifts are incompatible with the compactification lattice of $X$ and $\tilde{X}$ and the modular sum over $r,s$ would not yield a good CFT partition function.
For instance, taking $W = \frac{a}{b}$ (with $a,b$ positive and coprime) and $N=0$, the defect operator ${\mathcal{L}}^{(W,N)}$ generates a cyclic group of order $b$ and 
one obtains
\begin{align}
    \frac{1}{b}\sum_{r,s\in \mathbb{Z}_b} Z_{r,s}^{(\frac{a}{b},0)}(R;\tau) =\frac{1}{|\eta(\tau)|^2}\sum_{n,w\in\mathbb{Z}}q^{\frac{1}{4(\frac{a}{b}R)^2}[{n}+w(\frac{a}{b}R)^2 ]^2}\bar{q}^{\frac{1}{4(\frac{a}{b}R)^2}[{n}-w(\frac{a}{b}R)^2 ]^2}\,,
\end{align}
namely, the partition function of a free boson compactified on the circle with radius $\frac{a}{b}R$. Such theory can therefore be thought of as a $\mathbb{Z}_b$-orbifold $X\longrightarrow X+2\pi \frac{a}{b} $ of the theory with compactification radius $R$. Similarly, when $N=\frac{a}{b}$ is non-zero, while $W=0$, orbifolding by the defect would correspond to the $\mathbb{Z}_b$ orbifold $\tilde{X}\longrightarrow \tilde{X}+2\pi \frac{a}{b}$ which results in a free-boson theory compactified on a circle with radius $\frac{b}{a}R$.

 \subsubsection{T-duality defects}
 \label{subsubapp:Tduality}

Unlike in the case of the non-compact free boson, in the compact case it becomes interesting to study defects with the mixed gluing conditions 
\begin{subequations}
\label{eq:td}
    \begin{align}
        \p X^{(1)}(z) &=+\p X^{(2)}(z)\,,\\
        \bar{\p} X^{(1)}(\bar{z}) &=-\bar{\p} X^{(2)}(\bar{z})\,.
    \end{align}
\end{subequations}
First, compactifying the boson at the self-dual radius $R=R_{\text{s.d.}}=1$, it turns out that it is possible to construct an invertible defect with these gluing conditions. The corresponding defect operator, which implements the T-duality symmetry on the self-dual Hilbert space, then reads
\begin{align}
        {\mathcal{L}}^{\mathrm{T}} = \sum_{n,w\in\mathbb{Z}} \sum_{N,\bar{N}}(-1)^{\bar{N}}(-1)^{wn}|w,n,N,\bar{N}\rangle\,\langle n,w,N,\bar{N}|\,.
\end{align}
Namely, it acts by swapping $w,n$ and flipping the anti-holomorphic oscillators $\bar{\alpha}_n\to -\bar{\alpha}_n$. Clearly, acting on the local states in the theory, the operator ${\mathcal{L}}^{\mathrm{T}}$ squares to identity, so that it implements an invertible $\mathbb{Z}_2$ symmetry. This should not be surprising, becasue at the self-dual point, T-duality is a true symmetry of the bulk CFT Hilbert space. Calculating the defect torus correlators, one would obtain
\begin{subequations}
\begin{align}
    Z_{0,0}^\mathrm{T}(\tau,\bar{\tau}) &= \frac{1}{|\eta|^2} \sum_{n,w\in\mathbb{Z}}q^{\frac{1}{4}(n+w)^2}\bar{q}^{\frac{1}{4}(n-w)^2}\,,\\
    Z_{1,0}^\mathrm{T}(\tau,\bar{\tau}) &= \frac{1}{\eta}\sum_{n\in\mathbb{Z}}(-1)^{n}q^{n^2}\times\frac{1}{\bar{\eta}}\sqrt{\bar{\theta}_3(\bar{\tau})\bar{\theta}_4(\bar{\tau})} =\frac{1}{|\eta|^2}|{\theta}_3(\tau){\theta}_4(\tau)|\,,\\
    Z_{0,1}^\mathrm{T}(\tau,\bar{\tau}) &= \frac{1}{|\eta|^2}|{\theta}_2(\tau){\theta}_3(\tau)| \,,\\[2mm]
    Z_{1,1}^\mathrm{T}(\tau,\bar{\tau}) &= \frac{1}{|\eta|^2}|{\theta}_2(\tau){\theta}_4(\tau)|\,.
\end{align}
\end{subequations}
These are the same defect torus correlators as in the case of the standard $\mathbb{Z}_2$-orbifold action $X\to -X$, $\tilde{X}\to -\tilde{X}$, which is anomaly-free. The T-duality transformation $\mathrm{T}$ can be therefore gauged and one would obtain a theory with partition function of the form \eqref{eq:orbifold}.

Away from the self-dual radius, one can still construct sensible topological defects with the mixed gluing conditions \eqref{eq:td} as long as it is possible to write $R^{2} = \frac{p}{q}$ for some coprime positive integers $p,q$. The corresponding T-duality defect then reads
\begin{align}
        {\mathcal{L}}^{\mathrm{T}} = \sqrt{pq}\sum_{{n\in p\mathbb{Z}}}\sum_{w\in q\mathbb{Z}} \sum_{N,\bar{N}}(-1)^{\bar{N}}|w,n,N,\bar{N}\rangle\,\langle n,w,N,\bar{N}|\,,\label{eq:LTbos}
\end{align}
where the constraints on the sum over $n$ and $w$ ensure that the action of ${\mathcal{L}}^{\mathrm{T}}$ remains in the spectrum of local bulk fields, while preserving the gluing conditions \eqref{eq:td}. We also note that the defect $g$-function is now equal to $\sqrt{pq}$, meaning that the defect ${\mathcal{L}}^{\mathrm{T}}$ is not invertible. To verify that this is the correct value for the $g$-function, one could compute the spectrum of the defect-ending fields as
\begin{align}
     Z_{0,1}^\mathrm{T}(\tau,\bar{\tau}) &= \frac{1}{|\eta|^2}\sqrt{\tfrac{1}{2}\big[\theta_3(\tfrac{\tau}{pq})^2+\theta_2(\tfrac{\tau}{pq})^2\big]\bar{\theta}_3(\bar{\tau})\bar{\theta}_2(\bar{\tau})}\label{eq:LTLT}
\end{align}
and check that the multiplicities are the smallest possible positive integers. One can also expand\footnote{Since $\mathcal{L}^\mathrm{T}$ is self-conjugate, one could equally well write $({\mathcal{L}}^{\mathrm{T}})^2$ instead of ${\mathcal{L}}^{\mathrm{T}}({\mathcal{L}}^{\mathrm{T}})^\dagger$.}
\begin{align}
    {\mathcal{L}}^{\mathrm{T}}({\mathcal{L}}^{\mathrm{T}})^\dagger=\sum_{\alpha\in\mathbb{Z}_p}\sum_{\beta\in\mathbb{Z}_q}\mathcal{L}^{(\frac{\alpha}{p},\frac{\beta}{q})}\,,
\end{align}
so that ${\mathcal{L}}^{\mathrm{T}}$ is a duality defect in the sense that only group-like defects appear in the fusion ${\mathcal{L}}^{\mathrm{T}}({\mathcal{L}}^{\mathrm{T}})^\dagger$.

\newpage

\bibliography{references}
\bibliographystyle{utphys.sty}

\end{document}